\documentclass[preprint,amsmath,amssymb,aps,tightenlines]{revtex4}

\usepackage{graphicx}
\usepackage{subfigure}

\newcommand{\be}{\begin{equation}}
\newcommand{\ee}{\end{equation}}
\newcommand{\bsmm}{B_s \rightarrow \mu^+ \mu^-}
\newcommand{\bea}{\begin{eqnarray}}
\newcommand{\eea}{\end{eqnarray}}

\begin{document}
%\preprint{SNUTP}
%\draft

\preprint{\begin{tabular}{l}
\hbox to\hsize{August, 2002 \hfill KAIST-TH 2002/13}\\
\hbox to\hsize{hep-ph/0208112 \hfill KIAS-P02065}\\
\hbox to\hsize{\hfill MCTP-02-42}\\
\end{tabular} }

\bigskip

\title{SUSY breaking mediation mechanisms
and $(g-2)_{\mu}$, $B\rightarrow X_s \gamma$, ~$B \rightarrow X_{s} l^+ l^-$
and $B_s \rightarrow \mu^+ \mu^-$ }

\author{Seungwon Baek$^a$, ~P. Ko$^{b,c}$ and ~Wan Young Song$^c$}
\affiliation{$^a$ School of Physics, KIAS, Cheongryangri-dong, 
    Seoul 130-012, Korea }
\affiliation{$^b$ Michigan Center for Theoretical Physics, 
  University of Michigan Ann Arbor, MI 48109, USA}
\affiliation{$^c$ Dep. of Physics, KAIST,  Daejeon 305-701, Korea }
\date{\today}

\thispagestyle{empty}

%\tighten
\begin{abstract}
We show that there are qualitative differences in correlations among
$(g-2)_{\mu}$, $B\rightarrow X_s \gamma$, $B \rightarrow X_{s} l^+ l^-$ and
$B_s \rightarrow \mu^+ \mu^-$ in various SUSY breaking mediation mechanisms:
minimal supergravity (mSUGRA), gauge mediation (GMSB), anomaly mediation
(AMSB), gaugino  mediation ($\tilde{g}$MSB), weakly and strongly interacting 
string theories, and $D$ brane models. After imposing the direct search 
limits on the Higgs boson and SUSY particle search limits and $B\rightarrow 
X_s \gamma$ branching ratio,  we find all the scenarios can accommodate 
the $a_\mu \equiv ( g-2)_\mu /2 $ in the range of 
(a few tens)$\times 10^{-10}$, and predict 
that the branching ratio for $B\rightarrow X_s l^+ l^-$ can differ from 
the standard model (SM) prediction by $\pm 20 \%$ but no more.     
On the other hand, the $B_s \rightarrow \mu^+ \mu^-$ is sensitive to the
SUSY breaking mediation mechanisms through the pseudoscalar and stop masses 
($m_A$ and $m_{\tilde{t}_1}$), and the stop mixing angle. In the GMSB with
a small messenger number, the AMSB, the $\tilde{g}$MSB and the noscale 
scenarios, one finds that $B( B_s \rightarrow \mu^+ \mu^- ) \lesssim 2 \times 
10^{-8}$, which is below the search limit at the Tevatron Run II. Only the 
mSUGRA or string inspired models can generate a large branching ratio for 
this decay. 
\end{abstract}

\pacs{PACS numbers:}

\maketitle

%\narrowtext
%\tighten

%%%%%%%%%%%%%%%%%%%%%%%%%%%%%%%%%%%%%%%%%%%%%%%%%%%%%%%%%%%%%%%%%%%%%%%%%%%%%

\section{Introduction}

The minimal supersymmetric standard model (MSSM) is widely regarded as the
leading candidate for the physics beyond the 
standard model (SM)~\cite{nilles}. Its
detailed phenomenology both in the flavor preserving and flavor changing
sectors are heavily dependent on soft SUSY breaking lagrangians which contain
105 new parameters (including CP violating phases) compared to the SM.
This indicates that our understanding of SUSY breaking and mechanisms
mediating SUSY breaking to our world is not complete yet, although many
suggestions have been put forward over the past 20 years or so.
There are various models on soft terms in the literatures :
gravity mediation (SUGRA)~\cite{mSUGRA}, gauge mediation (GMSB)~
\cite{gauge}, anomaly mediated SUSY breaking (AMSB)~\cite{amsb}, 
and no--scale~\cite{noscale} or 
gaugino mediation ($\tilde{g}$MSB)~\cite{ginomsb}, etc., 
to name a few representative mechanisms.
Each mechanism predicts sparticle spectra and the trilinear couplings
which could be qualitatively different from one to another scenarios.
It is most important to determine the soft parameters from various different
experiments, and compare the resulting soft SUSY breaking parameters with those
predicted in the aformentioned SUSY breaking mediation mechanisms. This
process will provide invaluable informations on the origin of SUSY breaking,
which may be intrinsically rooted in very high energy regimes such as
intermediate, GUT or Planck scales.

Direct productions of SUSY particles and measuring their properties are
the best ways for this purpose. However, the importance of indirect effects
of SUSY particles through quantum loop corrections cannot be underestimated
at all for the following reasons. First of all, the experimental errors in
many low energy processes are now already (or will be in the near future)
at the level of probing the loop effects from SUSY particles :
the muon $(g-2)$, $B\rightarrow X_s \gamma$, $ B\rightarrow X_s l^+ l^-$,
$B_s \rightarrow \mu^+ \mu^-$, etc., to name a few.
On the other hand, theoretical uncertainties, which are mainly from our poor
understanding of (non)perturbative QCD effects, are becoming larger than the
experimental errors, and it is very important to reduce these theoretical
uncertainties before one can say a definite thing about a possible presence
of some new physics beyond the SM from these low energy observables.
Assuming this is achieved at some satisfactory level, we can study the
indirect effect of SUSY particles within various SUSY breaking mediation
mechanisms. Secondly, the high energy processes available at colliders are
usually insensitive to the flavor structures of soft SUSY breaking parameters.
On the other hand, the low energy favor changing neutral current (FCNC)
processes such as $B - \overline{B} $ and $K - \overline{K}$ mixings as well
as $B$ and $K$ decays can be very sensitive to such nontrivial flavor
structures in soft terms. Therefore, once some observables are dominated by
short distance physics and thus can be reliably calculable in renormalization
group (RG) improved perturbations theory, these may give us some hints for
possible existence of new physics beyond the SM with nontrivial flavor
structures.

In this work, we consider the following low energy processes, the muon
$(g-2)_\mu$, $B\rightarrow X_s \gamma$, $ B\rightarrow X_s l^+ l^-$ and
$B_s \rightarrow \mu^+ \mu^-$  for various SUSY breaking mediation
mechanisms : SUGRA, GMSB, AMSB, $\tilde{g}$MSB and certain classes of
string theories where dilaton/T/M moduli are the mediators. It turns out
there are qualitative differences among various correlations for different
SUSY breaking mediation mechanisms, especially depending on the messenger
scale. This qualitative difference may help us to distinguish various SUSY
breaking mediation mechanisms from low energy processes, in addition to
the informations provided by high energy collider experiments.

Suppose that the positive $\mu$ is preferred by the BNL data on $a_\mu$.
Then, the Wilson coefficient $C_{7,\gamma}$ for the process
$B\rightarrow X_s \gamma$ in SUSY models (except for AMSB) turns out to
have the same sign as in the SM case. And the sign of $C_7$ can be (partly)
observed in $B\rightarrow X_s l^+ l^-$ (or some exclusive channels) and
FB asymmetry therein. Therefore the correlations between the $B\rightarrow
X_s \gamma$ and $B\rightarrow X_s l^+ l^-$ will depend on the sign $(\mu)$
and the SUSY breaking mediation mechanisms. Since the ongoing $B$ factory
experiments began to observe $B\rightarrow K l^+ l^-$, it would be timely to
include the decay $B\rightarrow X_s l^+ l^-$ into our analysis.
Furthermore, if $\tan\beta$ should be moderately large in order to fit the
BNL data on $(g-2)_{\mu}$, the SUSY QCD corrections to $h-b-\bar{b}$ can
change by a significant amount \cite{hall,hbb}, thereby affecting $\bsmm$ by 
a (potentially) significant amount \cite{dedes}. This decay may be observable
at the Tevatron Run II down to the level of $2 \times 10^{-8}$. Therefore
this decay can cover parameter space (large $\tan\beta$ region) which
is not accessible by direct search for SUSY particles at the Tevatron Run II.
All these correlations will depend on the sign of the $\mu$ parameter,
$\tan\beta$  and  the detailed spectra of SUSY particles which are 
determined by soft SUSY breaking parameters.
Still, we could foresee that there may be qualitative differences in some
correlations among these observables through different chargino, stop, 
pseudoscalar masses $(m_{\chi^+}, m_{\tilde{t_1}}, m_A)$ depending on
$\tan\beta$, $\mu$, $M_3$ and the messenger scale $M_{\rm mess}$. 
The partial results in the GMSB and AMSB were reported in Ref.~\cite{bks}.
Similar study on the Higgs boson physics has been recently reported in 
Ref.~\cite{heynemeyer}. Also the muon $g-2$ in various SUSY breaking 
mediation mechanisms were considered in Ref.~\cite{kchoi}, including the 
collier phenomenology~\cite{tata} 
(see refs.~\cite{kane} for the recent study).  
In our work, we include 
$B\rightarrow X_s l^+ l^-$ and $B_s \rightarrow \mu^+ \mu^-$, and 
find a qualitative difference among various SUSY breaking mediation mechanisms
in the $B_s \rightarrow \mu^+ \mu^-$ mode. 
%This is because the muon $g-2$ and Higgs mass depend on the gaugino mass 

This work is organized as follows. In Sec.~II, we discuss $a_\mu^{\rm SUSY}$,
$B\rightarrow X_s \gamma$, $B\rightarrow X_s l^+ l^-$ and $B_s \rightarrow
\mu^+ \mu^-$ in the SM and MSSM in brief. In Sec.~III, we review various
SUSY breaking mediation mechanisms, and present numerical analsyses for
various low energy processes. Then the results  are summarized in Sec.~IV.

%%%%%%%%%%%%%%%%%%%%%%%%%%%%%%%%%%%%%%%%%%%%%%%%%%%%%%%%%%%%%%%%%%%%%%%%%%%%%
\section{Relevant processes and analysis procedures}
%%%%%%%%%%%%%%%%%%%%%%%%%%%%%%%%%%%%%%%%%%%%%%%%%%%%%%%%%%%%%%%%%%%%%%%%%%%%
\subsection{Muon anomalous magnetic moment : $a_\mu$}
%%%%%%%%%%%%%%%%%%%%%%%%%%%%%%%%%%%%%%%%%%%%%%%%%%%%%%%%%%%%%%%%%%%%%%%%%%%%

Let us define the $l_i \rightarrow l_j \gamma$ form factor $F_{ji} (0)$
as follows:
\begin{equation}
{\cal L}_{\rm eff} (l_i \rightarrow l_j \gamma) 
= e \frac{m_{l_i}}{2}  ~\overline{l}_{j} \sigma^{\mu\nu} F_{\mu\nu}
\left[ F_{ji}^L (0) P_L + F_{ji}^R (0) P_R \right] l_i .
\end{equation}
Then, the muon $(g-2)$ or $a_\mu$ is related with $F_{22}^{L(R)}$ by 
\begin{equation}
a_{\mu} = {1\over 2}~( g_{\mu} - 2 ) 
= m_{\mu}^2 \left[ F^L_{22} (0) + F^R_{22} (0) \right], 
\end{equation}
The SM contribution to $a_\mu$ had been calculated up to $O( \alpha^5 )$ in 
QED, up to two loops in the electroweak gauge interactions. Hadronic 
contributions are composed of vacuum polarization and light-light scattering 
parts, and are the source of the largest theoretical uncertainties. 
The BNL data is $ 116\;591\;597(67)\times 10^{-11}$. 
%\times 10^{-10}$. 
Comparing the BNL data of the year 2001 
$ a_\mu = 116\;591\;597(67)\times 10^{-11}$ with the most 
recently updated SM prediction, one finds that \cite{g-2}
\begin{equation}
\Delta a_\mu \equiv a_\mu^{\rm SUSY} \equiv 
a_\mu^{\rm exp} - a_\mu^{\rm SM} = (26 \pm 16) \times 10^{-10}. 
\end{equation}  
We assume that this small deviation can be explained by SUSY effects.

The SUSY contributions to $a_\mu$ come from the chargino-sneutrino and
the neutralino-smuon loop, the former of which is dominant in most parameter
space. Schematically, the result can be written as~\cite{martin}
\begin{equation}
a_\mu^{\rm SUSY} =
{\tan\beta \over 48 \pi}{m_\mu^2 \over M_{\rm SUSY}^2}
( 5 \alpha_2 + \alpha_1 )
=  14 \times 10^{-10}~ \tan\beta~
\left( { 100~{\rm GeV} \over M_{\rm SUSY}} \right)
\end{equation}
in the limit where all the superparticles have the same mass $M_{\rm SUSY}$.
In particular, the positive $\mu$ parameter implies the positive
$a_\mu^{\rm SUSY}$ in our convention. The current value for the deviation
between the BNL data~\cite{bnl} and 
the most recently updated SM prediction~\cite{g-2},
$(26 \pm 16) \times 10^{-10}$, can not be taken as a strong indication for
new physics beyond the SM. Therefore, we do not use $a_\mu$ as a constraint
but give predictions for it. If the data is updated with smaller statistical
and systematic errors and theoretical uncertainties, $a_\mu^{\rm SUSY}$
could provide a useful constraint on SUSY parameter space.
If there is no strong indication for new physics from the upcoming BNL data
on the muon $(g-2)_\mu$, it would rule out light SUSY spectra and/or large
$\tan\beta$ region. Also effective SUSY models will be in more comfortable
situations than before (see however Refs.~\cite{ko}).

%%%%%%%%%%%%%%%%%%%%%%%%%%%%%%%%%%%%%%%%%%%%%%%%%%%%%%%%%%%%%%%%%%%%%%%%%%%%
\subsection{$B \rightarrow X_s \gamma$ and $B \rightarrow X_s l^+ l^-$}
%%%%%%%%%%%%%%%%%%%%%%%%%%%%%%%%%%%%%%%%%%%%%%%%%%%%%%%%%%%%%%%%%%%%%%%%%%%%

It is well known that the $B \rightarrow X_s \gamma$ branching  ratio puts
a severe constraint on new physics, especially SUSY models from early days
of SUSY phenomenology. The relevant effective Hamiltonian for this decay is
\begin{equation}
H_{\rm eff} (b \rightarrow s \gamma) = - {4 G_F \lambda_t \over 
\sqrt{2}} \left[ C_{7L} O_{7L} + C_{7R} O_{7R} \right], 
\end{equation}
where $\lambda_t = V_{ts}^* V_{tb} (= - A \lambda^2 )$ in the Wolfenstein 
parametrization \cite{wolfenstein})  and  
\begin{equation}
O_{7L}  = {e \over (4 \pi)^2 }~m_b \bar{s}_L^{\alpha} 
\sigma^{\mu\nu} b_R^{\alpha}~F_{\mu\nu}.
\end{equation}
The operator $O_{7R}$ is obtained from $O_{7L}$ by the exchange 
$( L \leftrightarrow R)$.  Similarly one can expect a new physics 
contribution to $b\rightarrow s g$ : 
\begin{equation}
H_{\rm eff} (b \rightarrow s  g) = - {4 G_F \lambda_t \over 
\sqrt{2}} \left[ C_{8L} O_{8L} + C_{8R} O_{8R} \right],
\end{equation}
where 
\begin{equation}
O_{8L}  = {g_s \over (4 \pi)^2 }~m_b \bar{s}_L^{\alpha} 
\sigma^{\mu\nu} T^a_{\alpha \beta} b_R^{\beta}~G^a_{\mu\nu},
\end{equation}
and $O_{8R}$ is obtained from $O_{8L}$ by the exchange $(L\leftrightarrow R)$.
These two processes $b\rightarrow s\gamma$ and $b\rightarrow s g$ are unique 
in the sense that they are described in terms of only two independent operators
$O_{7(8)L}$ and $O_{7(8)R}$ whatever new physics there are. This fact makes it
easy to study these decays in a model independent  manner \cite{kn98}. 
The SM predictions for the $C_{7,8}$ at the $M_W$ scale are 
(in the limit $m_s = 0$)
\begin{eqnarray}
C_{7L}^{\rm SM} (M_W) & \approx & -0.22, 
\nonumber  
\\
C_{7R}^{\rm SM} (M_W) & = & 0,
\nonumber  
\\
C_{8L}^{\rm SM} (M_W) & \approx & -0.12, 
\nonumber  
\\
C_{8R}^{\rm SM} (M_W) & = & 0.
\end{eqnarray}
Note that $C_{7(8)R}^{\rm SM}$ in the SM is suppressed compared to 
$C_{7(8)L}^{\rm SM}$ by $m_s/m_b$, because $W$ boson couples only to the 
left-handed fermions. Such terms proportional to $m_s$ will be  neglected in 
our work by setting $m_s = 0$ whenever they appear. 

The magnetic dipole coefficient $C_{7,\gamma}$ for
this decay receives contributions from SM, charged Higgs and SUSY particles
in the loop. The charged Higgs contributions always add up to the SM
contributions, thereby increasing the rate. On the other hand,  the 
chargino-stop loop can interfere with the SM and the charged Higgs 
contributions either in a constructive or destructive manner depending on 
the sign of $\mu$, and could provide an important constraint on the SUSY 
parameter space. The gluino loop contribution to $B\rightarrow X_s \gamma$ 
is always negligible in the models we consider in this work.  
The most recent data~\cite{belle1,cleo}
\[
Br ( B \rightarrow X_s \gamma )_{\rm exp} = ( 3.21 \pm 0.43_{(stat)}
\pm 0.27_{(sys) -0.10(th)}^{~~~~~~~+0.18} ) \times 10^{-4},
\]
is in good agreement with the SM prediction~\cite{kagan,bsgamma} :
\[
Br ( B \rightarrow X_s \gamma )^{\rm SM}_{E_\gamma >1.6 GeV} = 
( 3.57 \pm 0.30 ) \times 10^{-4}.
\]
Thus, there is very little room for new physics contributions, unless SUSY
contributions interfere destructively with the SM contributions.

The inclusive decay $B\rightarrow X_s l^+ l^-$ has been also considered
extensively in the context of two Higgs doublet model\cite{bsll2hd}, 
mSUGRA~\cite{bsllsugra} model, 
GMSB ~\cite{bsllgmsb} and SUSY models with minimal flavor violations~
\cite{bsllmfv} as well as in the SM~\cite{bsllsm}. 
Here the local $\bar{s}b\bar{l}l$ operators are also important as well as 
the nonlocal photon exchange diagram due to $C_{7,\gamma}$. 
In the presence of new physics contributions to $b\rightarrow s \gamma$, 
there should be also generic new physics contributions to $b\rightarrow s l^+ 
l^-$ through electromagnetic penguin diagrams. This effect will modify the 
Wilson coefficient $C_9$ of the dim-6 local operator $O_9$ :
\begin{eqnarray}
H_{\rm eff} (b\rightarrow s ll) \supset 
H_{\rm eff} (b\rightarrow s \gamma ) 
- {4 G_F \lambda_t \over \sqrt{2}} ~\left[ C_9 O_9 + C_{10} O_{10} \right] 
\nonumber  \\
 - {4 G_F \lambda_t \over \sqrt{2}} ~\left[ C_S O_S + C_P O_P 
+  C_S^{'} O_S^{'} + C_P^{'} O_P^{'} 
\right],
\end{eqnarray}   
where the operators $O_9$, $O_{10}$, $O_S$ and $O_P$ are defined by 
\begin{eqnarray}
O_9 = {e^2 \over (4 \pi)^2} \left( \overline{s}_L \gamma_{\mu} b_L \right)~
( \overline{l} \gamma^{\mu} l ),
&&
O_{10} = {e^2 \over (4 \pi)^2} \left( \overline{s}_L \gamma_{\mu} b_L \right)~
( \overline{l} \gamma^{\mu} \gamma_5 l ), 
\nonumber  \\
%\end{equation}
%\begin{equation}
O_S = {e^2 \over (4 \pi)^2} ~ (\overline{s}_L b_R )~( \overline{l} l ),
&&
O_P = {e^2 \over (4 \pi)^2} ~ (\overline{s}_L b_R )~
( \overline{l} \gamma_5 l ).
\end{eqnarray} 
The primed operators are obtained by the exchange $L\leftrightarrow R$ 
on the quark bilinear operators. 

In the SM, the Wilson coefficients $C_{9,10}$'s are given by
\begin{equation}
C_9^{\rm SM} ( M_W ) \approx 2.01,
~~~
C_{10}^{\rm SM} ( M_W ) \approx 4.55.
\end{equation}

In fact,  the latter
is important when the dilepton invariant mass $m_{ll}$ is low because of
the photon propagator effect $1/ m_{ll}^2$ in the amplitude. The SM predictions
in NLO for $l = e$ and $\mu$ are \cite{bsllsm}
\\
\begin{eqnarray}
 B ( B \rightarrow X_s e^+ e^- ) & = & ( 6.3^{+1.0}_{-0.9} ) \times 10^{-6},
\nonumber  \\
 B ( B \rightarrow X_s \mu^+ \mu^- ) & = & ( 5.7 \pm 0.8  ) \times 10^{-6},
\end{eqnarray}
respectively. Although the inclusive decays are difficult to measure with
high precision, the ongoing B factory experiments began to be sensitive to
exclusive decay modes although the experimental errors are large 
\cite{bsllexp}:
\begin{eqnarray*}
  B( B \rightarrow K l^+ l^- ) & < & 0.6 \times 10^{-6} ~~~
  [ 0.47 - 0.75 ]
\nonumber  \\
  B( B \rightarrow K \mu^+ \mu^- ) & = & ( 0.99^{+0.40+0.13}_{-0.32-0.14} )
\times 10^{-6} ~~~
  [ 0.47 - 0.75 ]
\nonumber  \\
  B( B \rightarrow K^* e^+ e^- ) & < & 5.0 \times 10^{-6} ~~~
  [ 1.4 - 3.0 ]
\nonumber  \\
  B( B \rightarrow K^* \mu^+ \mu^- ) & = & 3.6 \times 10^{-6} ~~~
  [ 0.9 - 2.4 ]
\end{eqnarray*}
The SM predictions shown in the brackets (in units of $10^{-6}$) suffer from
large theoretical uncertainties because the form factors are poorly known
\cite{bsllth}.
In view of these data, it is quite timely to consider this class of processes
in various SUSY breaking mediation mechanisms.

This implies that in the $C_7$ coefficient for $B \rightarrow X_s \gamma$
process, the chargino contribution interfere destructively with the SM and
the charged Higgs contributions. However, this does not imply that $C_7$
necessarily has the opposite sign to the $C_7^{\rm SM}$.  For sufficiently
large $\tan\beta$ (namely, for large $a_{\mu}^{SUSY}$), it is possible to
have $C_7^{\rm MSSM} \approx - C_7^{\rm SM}$. Then the branching ratio for
$B\rightarrow X_s l^+ l^-$ will be substantially larger than the SM case.
In other words, if the deviation in  $a_{\mu}$ is larger than the current
value, it is very likely that the branching  ratio for
$B\rightarrow X_s l^+ l^-$ should be significantly enhanced compared to the
SM predictions.

In order to avoid the hadronic uncertainties related with exclusive $B$ 
decays and the long distance contributions from charmonia and charmed meson
intermediate states, we consider inclusive $B\rightarrow X_s l^+ l^-$ below 
the $J/\psi$ resonance. Defining 
\begin{equation}
R_{\mu\mu} \equiv {B( B\rightarrow X_s \mu^+ \mu^- )_{\rm MSSM} \over 
B( B\rightarrow X_s \mu^+ \mu^- )_{\rm SM} } 
\end{equation}
in the region with 
\[
2 m_\mu \leq m_{\mu\mu} \leq ( m_{J/\psi} - 100~{\rm MeV}). 
\]
In this region, the nonlocal contributions from the  virtual photon exchange
involving $O_{7\gamma}$ is more important than the local four fermion 
operators $O_{9}$ and $O_{10}$. Therefore the ratio $R_{\mu\mu}$ is strongly
correlated with the branching ratio of $B\rightarrow X_s \gamma$ and the sign
of $C_{7\gamma} (m_b)$. 
The forward backward asymmetry of dilepton energy distributions in the 
rest frame of the parent $B$ meson can be a sensitive probe of new physics. 
However we postpone studying this observable for the future project, and 
will not consider in the present work.

%%%%%%%%%%%%%%%%%%%%%%%%%%%%%%%%%%%%%%%%%%%%%%%%%%%%%%%%%%%%%%%%%%%%%%%%%%%%
\subsection{Hall--Rattazzi--Sarid Effect}
%%%%%%%%%%%%%%%%%%%%%%%%%%%%%%%%%%%%%%%%%%%%%%%%%%%%%%%%%%%%%%%%%%%%%%%%%%%%

Another important effect in the large $\tan\beta$ limit is the nonholomorphic
SUSY QCD corrections to the $h-b-\bar{b}$ couplings, the so-called
Hall -- Rattazzi -- Sarid effect \cite{hall}. Also for large $A_t$ and $y_t$
couplings, the stop -- chargino loop could be quite important.  One can
summarize these effects as the following relation between the bottom quark
mass and the bottom Yukawa coupling $y_b$ \cite{hbb} :
\begin{equation}
  m_b = y_b {\sqrt{2} M_W \cos\beta \over g}~( 1 + \Delta_b )
\end{equation}
where
\begin{eqnarray}
  \Delta_b & \simeq &
{2 \alpha_s \over 3 \pi}~M_{\tilde{g}} \mu \tan\beta
I ( M_{\tilde{b}_1} , M_{\tilde{b}_2} , M_{\tilde{g}} )
\nonumber \\
& + & {\alpha_t \over 4 \pi }~A_t \mu \tan\beta
I ( M_{\tilde{t}_1} , M_{\tilde{t}_2} , \mu )
\end{eqnarray}
and the loop integral $I(a,b,c)$ is given by
\begin{eqnarray*}
I(a,b,c) & = & \left[ a^2 b^2 \log ( a^2 / b^2 ) + b^2 c^2 \log ( b^2 / c^2 )
%\right.
%\nonumber \\
%& & \left.
+ c^2 a^2 \log ( c^2 / a^2 ) \right] /
\\
& & \left[ ( a^2 - b^2 ) ( b^2 - c^2 ) ( a^2 - c^2 ) \right]
\nonumber
\end{eqnarray*}
Therefore in the large $\tan\beta$ limit, the SUSY loop correction $\Delta_b$
can be large as well with either sign, depending on the sign of the $\mu$
parameter and the gluino mass parameter $M_{\tilde{g}}$. Note that the
muon $(g-2)$ picks up $\mu > 0$, whereas the $B\rightarrow X_s \gamma$
prefers a positive $\mu M_{\tilde{g}}$.

%%%%%%%%%%%%%%%%%%%%%%%%%%%%%%%%%%%%%%%%%%%%%%%%%%%%%%%%%%%%%%%%%%%%%%%%%%%
\subsection{$B_s \rightarrow \mu^+ \mu^-$}
%%%%%%%%%%%%%%%%%%%%%%%%%%%%%%%%%%%%%%%%%%%%%%%%%%%%%%%%%%%%%%%%%%%%%%%%%%%

The effective Hamiltonian for $B_s \rightarrow l^+ l^-$ is already given
in Eqs.~(10) (11), and  
%involves more 
%operators beyond that given in Eq.~(10) for $b\rightarrow s l^+ l^-$:
%\begin{equation}
%H_{\rm eff} ( B_s \rightarrow l^+ l^- )  \supset 
%H_{\rm eff} ( b\rightarrow s l^+ l^- ) 
% - {4 G_F \lambda_t \over \sqrt{2}} ~\left[ C_S O_S + C_P O_P 
%+  C_S^{'} O_S^{'} + C_P^{'} O_P^{'} 
%\right],
%\end{equation}
%where the operators $O_S$ and $O_P$ are defined as 
%\begin{equation}
%O_S = {e^2 \over (4 \pi)^2} m_b~ (\overline{s}_L b_R )~( \overline{l} l ),~~
%O_P = {e^2 \over (4 \pi)^2} m_b~ (\overline{s}_L b_R )~
%( \overline{l} \gamma_5 l ).
%\end{equation}
%The primed operators are obtained by the exchange $L\leftrightarrow R$ 
%on the quark bilinear operators. 
the branching ratio for this decay is given by \cite{arnowitt} 
\begin{eqnarray}
B ( \bsmm ) & = &
{G_F^2 \alpha^2 \tau_B M_B^5 \over 64 \pi^3}~f_{B_s}^2 ~
\left| \lambda_t  \right|^2
\sqrt{ 1 - {4 m_\mu^2 \over M_{B_s}^2 }}
\nonumber   \\
& & \left[
\left( 1 - {4 m_\mu^2 \over M_{B_s}^2 } \right) ~
\left| { ( C_S - C_S^{'} ) \over ( m_b + m_s ) } \right|^2 +
\left|   { ( C_P - C_P^{'} ) \over ( m_b + m_s ) } +
2 {m_\mu \over M_{B_s}^2 } ( C_{10} - C_{10}^{'} ) \right|^2
\right].
\end{eqnarray}

The branching ratio for the decay $B_s \rightarrow \mu^+ \mu^-$ is very small
in the SM : $(3.7 \pm 1.2) \times 10^{-9}$~\cite{dedes}. 
The current upper limit from CDF during Tevatron Run I is set to 
$< 2.6 \times 10^{-6}$ at 95 \% C.L.~\cite{bsmmexp}
At Tevatron Run II, CDF aims at achieving a single event sensitivity down to
$10^{-8}$ for an integrated luminosity of 2 fb$^{-1}$.
In SUSY models, both $B\rightarrow X_s l^+ l^-$ 
\cite{huang,Choudhury:1998ze} and $B_d \rightarrow \mu^+ \mu^-$ 
\cite{babu,Huang:2000sm} can be significantly enhanced 
in the large $\tan\beta$ limit, due to the neutral Higgs boson exchange,
and similarly for the analogous process $B_s \rightarrow \mu^+ \mu^-$. 
The SUSY effects are encoded in the Wilson coefficients $C$'s. For large
$\tan\beta$, one has, for example,
\[
C_S \simeq %{G_F \alpha \over \sqrt{2} \pi}~V_{tb} V_{ts}^* ~
\left( { \tan^3 \beta \over 4 \sin^2 \theta_W } \right) ~
\left( { m_b m_\mu m_t \mu \over M_W^2 M_A^2 } \right) ~
{\sin 2\theta_{\tilde{t}} \over 2} ~
f ( m_{\tilde{t_1}}^2, m_{\tilde{t_2}}^2, \mu^2 ).
\]
Here,  $C_P = - C_S$,  $C_S^{'} = ( m_s / m_b ) C_S$ and $C_P^{'} = -
( m_s / m_b ) C_P$, and the loop function $f(x,y,z)$ can be found in
Refs.~\cite{dedes,arnowitt,babu,Chankowski,Urban,cgnw,isidori}. 
%\cite{arnowitt}. 
(See also Refs.~\cite{cpv} for the discussions for CP violations therein.)
Note that the branching ratio for this decay is proportional to 
$\tan^6 \beta$, and thus can be enhanced by a significant amount  
for large $\tan\beta$, light  pseudoscalar boson ($m_A$), light stop 
($m_{\tilde{t}_1}$) and the large $\tilde{t}_L - \tilde{t}_R$ mixing angle
$\theta_{\tilde{t}}$.  
%but quickly decreases as $\tan\beta$ gets smaller. Also light pseudoscalar 
%and light stop and the large $\tilde{t}_L - \tilde{t}_R$
%mixing can lead to the larger branching ratio for $\bsmm$.
The Hall -- Rattazzi -- Sarid effect can further modify the result in either
direction depending on the sign$(\mu)$. For $\mu > 0$, the enhancement
becomes less pronounced due to the Hall-Rattazzi-Sarid effect.

%%%%%%%%%%%%%%%%%%%%%%%%%%%%%%%%%%%%%%%%%%%%%%%%%%%%%%%%%%%%%%%%%%%%%%%%%%%%%
\subsection{ Constraints }
%%%%%%%%%%%%%%%%%%%%%%%%%%%%%%%%%%%%%%%%%%%%%%%%%%%%%%%%%%%%%%%%%%%%%%%%%%%%%

When we scan over SUSY parameter space, we impose the direct search limits
on Higgs and SUSY particles (except for the GMSB scenario) \cite{heynemeyer}:
\begin{eqnarray}
m_{\tilde{e}, \tilde{\mu}, \tilde{\tau},\tilde{\nu} }
& > & 95,~~85,~~71,~~43~{\rm GeV},
\nonumber  \\
m_{\tilde{t}, \tilde{b}, \tilde{g}} & > & 95,~~85,~~190~{\rm GeV},
\nonumber  \\
m_{\chi^+}^{\rm mSUGRA} & > & 103~{\rm GeV~ for}~
                                m_{\tilde{\nu}} > 300~{\rm GeV},
\nonumber  \\
m_{\chi^+}^{\rm mSUGRA} & > & 83.6~{\rm GeV~ for}~
                                m_{\tilde{\nu}} < 300~{\rm GeV},
\nonumber  \\
m_{\chi^+}^{\rm AMSB}   & > & 45~{\rm GeV},
\nonumber  \\
m_{\chi^0}^{\rm mSUGRA} & > & 36~{\rm GeV},
\nonumber  \\
m_{\chi^0}^{\rm AMSB} & > & 45~{\rm GeV}.
\end{eqnarray}
For the GMSB, the LSP is always very light gravitinos, and we impose
\begin{equation}
m_{\rm NLSP}^{\rm GMSB} > 100 ~{\rm GeV},
\end{equation}
which is stronger than other experimental bounds on SUSY particle masses.
It turns out that the stau mass bound is quite strong in a certain region
of parameter space.

In order to be as model independent as possible, we do not assume that
the LSP is color and charge neutral (except for the GMSB scenario where
the gravitino is the LSP), nor do we impose the color-charge breaking minima
or the unbounded from below constraints~\cite{munoz,casas,abel}, 
since these constraints can be evaded in nonstandard cosmology.

Also we impose the $B\rightarrow X_s \gamma$ branching ratio as a constraint.
Then, using the aforementioned constraints, we find that the sign of
$C_{7,\gamma}$ for $B\rightarrow X_s \gamma$ cannot flip relative
to the SM case, and the branching ratio for $B\rightarrow X_s l^+ l^-$
remains close to the SM prediction. (Note that the previous study in the
context of mSUGRA suggested that two branches would be possible for
$Br (B\rightarrow X_s \gamma)$ for large $\tan\beta$, because of both sign
of $C_{7,\gamma}$ were allowed.) Therefore there is no chance to observe
$B\rightarrow X_s l^+ l^-$ at the level of $70 \% - 80 \%$ enhanced over
to the SM. 
For the muon $(g-2)$, we do not use it as a constraint but give predictions
for it, since the current value for the deviation between the BNL data and
the most updated SM prediction, $(26 \pm 16) \times 10^{-10}$, can not be
taken as a serious indication for new physics beyond the SM. %The muon
%$(g-2)_\mu$ could be a useful constraint once the data is updated with
%smaller statistical and systematic uncertainties. If there is no strong
%indication for new physics from $(g-2)_\mu$, it would rule out light SUSY
%spectra and/or large $\tan\beta$ region. Also effective SUSY models will
%be in more comfortable situations than before (see however Refs.~\cite{}).

%%%%%%%%%%%%%%%%%%%%%%%%%%%%%%%%%%%%%%%%%%%%%%%%%%%%%%%%%%%%%%%%%%%%%%%%%%%%%
\subsection{Procedures}
%%%%%%%%%%%%%%%%%%%%%%%%%%%%%%%%%%%%%%%%%%%%%%%%%%%%%%%%%%%%%%%%%%%%%%%%%%%%%

First of all, we assume  the radiative electroweak symmetry breaking (REWSB)
to trade $\mu$ and $B\mu$ with $M_Z$ and $\tan\beta$ using the following 
relations:
\begin{eqnarray}
\mu^2 & = & {m_{H_d}^2 - m_{H_u}^2 \tan^2 \beta \over \tan^2 \beta - 1}
- {1\over 2} M_Z^2,
\nonumber  \\
B\mu & = & ( m_{H_d}^2 +  m_{H_u}^2 + 2 \mu^2 
)  \sin 2\beta, 
\end{eqnarray}
where $m_{H_d}^2$ and $ m_{H_u}^2$ are loop corrected running masses for 
two Higgses (which are soft SUSY breaking).
The sign of $\mu$ is fixed to be positive 
but we do not assume anything about $\tan\beta$. 
There is no problem to accommodate the $a_\mu \sim
+ O(10) \times 10^{-10}$ in SUSY models we consider in this work, except for 
the AMSB scenario. The decay $B\rightarrow X_s \gamma$ will be in good shape
for $\mu > 0$, since the chargino loop contribution can cancel the charged 
Higgs  contributions for the positive $\mu$. The data will constrain the 
absolute value of $C_{7\gamma} ( m_b )$. Then, for a small $\tan\beta$, the 
predicted branching ratio for $B\rightarrow X_s l^+ l^-$ is essentially the 
same as the SM prediction. On the other hand, for large $\tan\beta$, one can 
have either signs of $C_{7\gamma}$ so that the branching ratio for 
$B\rightarrow X_s l^+ l^-$ can take two values for a given 
$B\rightarrow X_s \gamma$ branching ratio. However, in the SUSY breaking 
mediation scenarios we consider, it turns out that the current lower bound
on the Higgs boson is too severe that the parameter space in which the 
branching ratio for $B\rightarrow X_s l^+ l^-$ becomes large with 
$C_{7\gamma} \approx - C_{7\gamma}^{\rm SM}$ is essentially excluded.
Therefore there is little hope to observe a large deviation in 
Br ($B\rightarrow X_s \gamma$). This is true for the minimal SUGRA scenario,
in particular, and this observation is newly made in the present work for 
the first time to our best knowledge. On the other hand, the decay 
$B_s \rightarrow \mu^+ \mu^-$ depends on the stop mass and the stop mixing 
angle, which are sensitive to the SUSY breaking mediation mechanisms and the
messenger scale. So we anticipate there are qualitative differences in the
predictions for $B( B_s \rightarrow \mu^+ \mu^- )$. In order to have a light
stop and large $\tilde{t}_L - \tilde{t}_R$ mixing, it is crucial to have
a large messenger scale and a lighter squark mass parameter at the messenger
scale. Then RG running will produce the stop mass and the $A_t$ parameter
which determine the stop mixing. This phenomenon will be most clearly seen
in the GMSB scenario with different messenger scale and different  number of
the messenger fields (see Sec.~III B).

%%%%%%%%%%%%%%%%%%%%%%%%%%%%%%%%%%%%%%%%%%%%%%%%%%%%%%%%%%%%%%%%%%%%%%%%%%%

\section{SUSY Breaking Mediation Mechanisms}

In this section, we review several SUSY breaking mediation mechanisms 
%we are going to investigate in the subsequent sections
: minimal SUGRA, gauge mediated SUSY breaking, anomaly mediated SUSY breaking,
gaugino mediated SUSY breaking ($\tilde{g}$MSB) (which includes the no--scale
supergravity scenario), weakly interacting superstring models with dilaton 
and moduli mediations, heterotic $M$ theory and $D-$brane models.
%: minimal SUGRA~\cite{mSUGRA}, gauge mediated SUSY breaking
%\cite{gauge}, anomaly mediated SUSY breaking \cite{amsb}, gaugino mediated
%SUSY breaking ($\tilde{g}$MSB) \cite{ginomsb} (which includes the no--scale
%supergravity scenario \cite{noscale}), weakly interacting superstring models
%with dilaton and moduli mediations \cite{string}, heterotic $M$ theory
%\cite{mtheory} and $D-$brane models \cite{dbrane}.
When we give expressions for the soft SUSY braking parameters, we assume
that all the parameters are real in order to avoid SUSY CP problem.
It would be straightforward to relax this assumption with substantial
complications in the numerical analysis, which we do not aim to do in
this work. Thus there is no new source of CP violations beyond the KM phase
in the CKM mixing matrix. Also scalar fermion masses are unversal in many
cases, so that the SUSY flavor problem is mitigated significantly.

\subsection{Minimal Supergravity (mSUGRA)}

Supergravity theories, which may be a low energy effective field theory of
more fundamental theories such as superstring or $M$ theories, are completely
specified by three objects:
\begin{itemize}
\item K\"{a}hler potential $K ( \Phi, \Phi^*, V, V^{\dagger} )$:
a real scalar function of  chiral and vector superfields $\Phi_i$ and $V$
in the visible sector, respectively
\item Gauge kinetic functions $f_{ab} (\Phi)$: a holomorphic function of
chiral superfields $\Phi$, where $a,b$ are gauge group indices
\item Superpotential $W( \Phi )$: a holomorphic function of chiral superfield
\end{itemize}
The holomorphic functions $f_{ab} (\Phi)$ and $W( \Phi )$ are protected
from the radiative corrections by nonrenormalization theorem, whereas the
K\"{a}hler potential will be renormalized in general.  All the couplings may
depend on the hidden sector fields or moduli (we denote these fields
collectively by $\Sigma$), although we suppressed this dependence.  From 
these three objects, one can derive the soft terms such as the sfermion
masses, trilinear couplings, gaugino masses.

If one assumes a simple form for the K\"{a}hler potential,
$K = \sum_i \Phi_i^* \Phi_i + \sum_j X_j^* X_j$ with $X_j$ hidden sector 
fields, %which is independent of hidden sector fields,
the soft parameters satisfy universal sfermion masses $m_0$.
If the Yukawa couplings in the superpotential $W(\Phi)$ is assumed to be
constant independent of the hidden fields or modulis ($\Sigma$), we get 
universal trilinear coupling $A_0$ with exact proportionality.
Assuming that the gauge kinetic function is independent of the gauge group,
one has the universal gaugino mass $M_{1/2}$.
Although these specific assumptions are {\it ad hoc} out of question,
it leads to a simple universality in the scalar mass and trilinear couplings
at the GUT scale so that SUSY flavor problem can be signifiantly mitigated.
Also a restricted set of mSUGRA models can be motivated in the string
inspired SUGRA models where a dilaton plays a dominant role in SUSY breaking
mediation (see Sec.~II E).
One can also relax the condition for the gaugino unification at GUT scale.
In this case, low energy phenomenology can be richer, and there could be
qualitative changes in our results. But we keep the gaugino unification
assumption in this work in order to reduce the number of parameters,
relegating the study of nonuniversal gaugino mass scenarios for the future
publication. Under these assumtions, the mSUGRA model is specified by the 
following five parameters :
\begin{equation}
  m_0,~~M_{1/2},~~A_0,~~\tan\beta,~~{\rm sign}(\mu).
\end{equation}
We scan these parameters over the following ranges :
\begin{eqnarray}
%  50~{\rm GeV} & \leq & m_0, M_{1/2} \leq 1~{\rm TeV},
50~{\rm GeV} & \leq & M_{1/2} \leq 1~{\rm TeV},
\nonumber  \\
%  -3~{\rm TeV} & \leq & A_0 \leq + 3~{\rm TeV}
\nonumber  \\
  1.5 & \leq & \tan\beta \leq 60,
\end{eqnarray}
with $A_0 = 0$, $m_0 = 300$ GeV and $\mu > 0$. 
For a negative $\mu$, we have $a_\mu^{\rm SUSY} < 0$ and 
also the $B\rightarrow X_s \gamma$ constraint becomes much more severe 
since the chargino-stop loop interferes constructively with the SM and 
the charged Higgs loop contributions. 
Earlier phenomenological analysis of mSUGRA scenarios can be found on the 
muon $(g-2)_\mu$, $B\rightarrow X_s \gamma$ and 
$B\rightarrow X_s l^+ l^-$, \cite{bsllsugra}, for example. 

\begin{figure}
\centering
\includegraphics[width=0.8\textwidth]{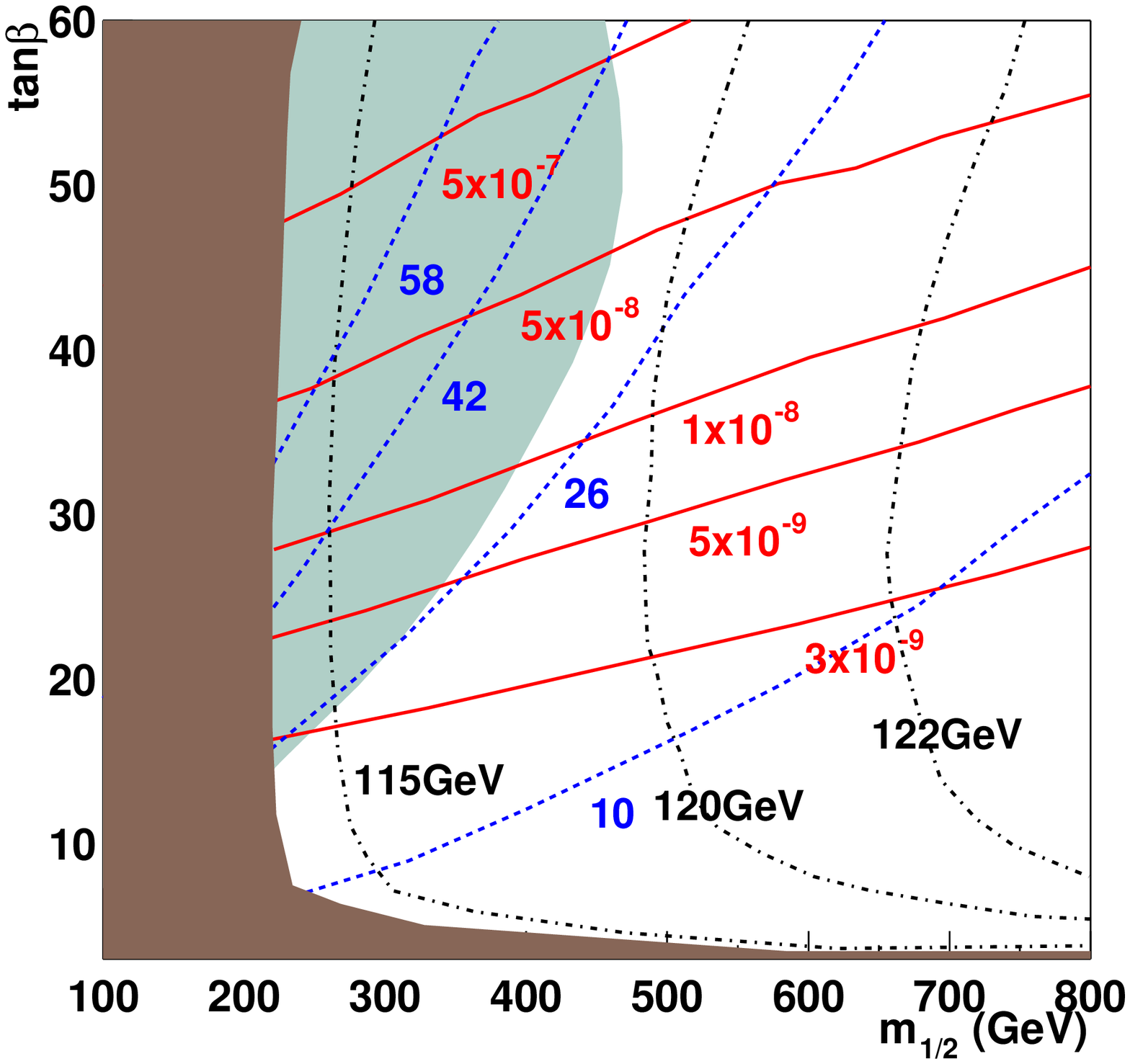}
\vspace{-1cm}
\caption{\label{fig:msugra1}
The contour plots for $a_\mu^{\rm SUSY}$ in unit of $10^{-10}$
(in the short dashed curves) and the Br ($B_s \rightarrow \mu^+ \mu^- $)
(in the solid curves) in $( m_{1/2}, \tan\beta)$ plane in the minimal SUGRA
model for $m_0 = 300$ GeV and $A_0 = 0$. The brown (dark) region is excluded
by the Higgs and SUSY particle mass bounds, and the green (light gray) region
is excluded by $B\rightarrow X_s \gamma$ branching ratio.
}
\end{figure}

In Fig.~\ref{fig:msugra1}, we show the constant contour plots for
$a_\mu^{\rm SUSY}$ in unit of $10^{-10}$ (in the short dashed curves) and
the Br ($B_s \rightarrow \mu^+ \mu^- $) (in the solid curves) in the
$( M_{1/2}, \tan\beta)$ plane for $m_0 = 300$ GeV and $A_0 = 0$.
The left dark region is excluded by direct search limits on SUSY
particles and Higgs boson masses, and the light gray region is excluded by
the lower bound on the $B\rightarrow X_s \gamma$.
The dot--dashed contours corresponds to $m_h = 115, 120, 122$ GeV's for
the future reference. The result for $ \bsmm $ is essentially the same
as the Fig.~2 of Dedes et al. \cite{dedes},  except that we did not assume
that the LSP should be color/charge neutral but did impose
$B\rightarrow X_s \gamma$ at 95 \% CL.

\begin{figure}
\centering
\hspace{-1.5cm}
\subfigure[]{
\includegraphics[width=0.5\textwidth]%{figs/msugra/br_bs-mug23.ps}}
{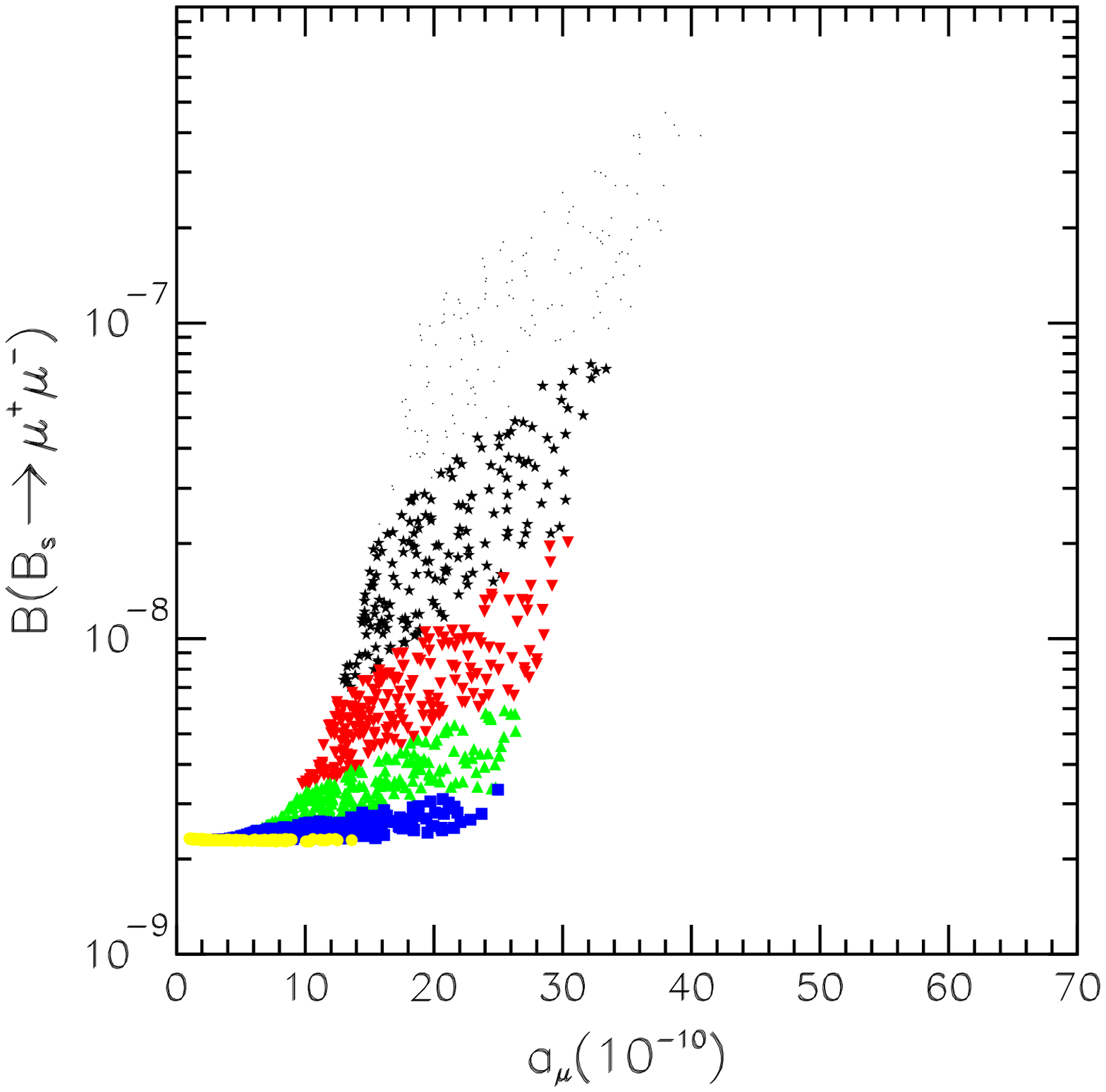}}
\subfigure[]{
%\hspace{-1.5cm}
\includegraphics[width=0.5\textwidth]%{figs/msugra/bsll-b2sr.ps}}
{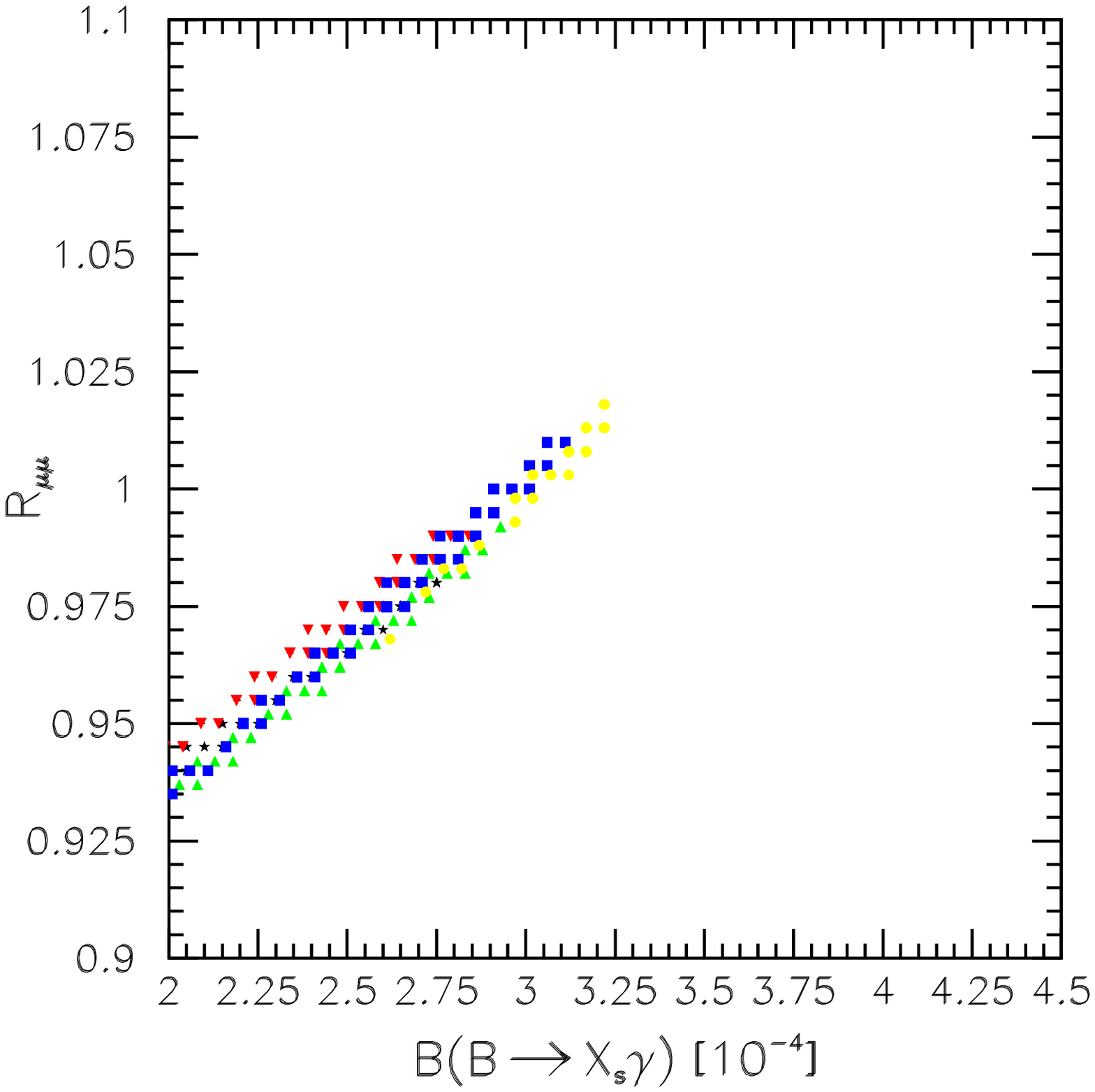}}
\caption{\label{fig:msugra2}
The correlations between
(a) the muon $a_\mu^{\rm SUSY}$ and $B(\bsmm)$, and
(b) $Br ( B \rightarrow X_s \gamma)$ and $R_{\mu\mu}$ in the mSUGRA model
with $A_0 = 0$ and $m_0 = 300$ GeV.
The regions 
Different colors represent different ranges of $\tan\beta$: 
$3 \leq \tan\beta \leq 10$ (yellow),
$10 \leq \tan\beta \leq 20$ (blue),
$20 \leq \tan\beta \leq 30$ (green),
$30 \leq \tan\beta \leq 40$ (red),
$40 \leq \tan\beta \leq 50$ (black star) and
$50 \leq \tan\beta$ (black dots). }
%\label{fig:msugra2}
\end{figure}

In Fig.~\ref{fig:msugra2} (a), we show the correlation between the muon
$a_\mu^{\rm SUSY}$ and $B (B_s \rightarrow \mu^+ \mu^- )$. For convenience,
we represent different $a_\mu^{\rm SUSY}$'s with different shapes
(also different colors). The regions $a_\mu^{\rm SUSY} < 10 \times 10^{-10}$,
$10 \times 10^{-10} < a_\mu^{\rm SUSY} < 26 \times 10^{-10}$,
$26\times 10^{-10} < a_\mu^{\rm SUSY} < 42\times 10^{-10}$,
$42 \times 10^{-10} < a_\mu^{\rm SUSY} < 58 \times 10^{-10}$, and
$a_\mu^{\rm SUSY} > 58\times 10^{-10}$ are represented by the stars (black),
the inverted triangles (red), the triangles (green), the squares (blue) and
the circles (yellow). The $B_s \rightarrow \mu^+ \mu^-$ branching ratio can
be enhanced up to $2\times 10^{-7}$ ($\sim 6 \times 10^{-7}$) for large
$\tan\beta$, if we impose (do not impose) $B\rightarrow X_s \gamma$
constraint. The current upper limit from CDF : $2.6 \times 10^{-6}$
(95 \% C.L.), and large $\tan\beta$ region of the mSUGRA model will be
within the reach of Tevatron Run II by searching the $\bsmm$ decay mode
down to the level of $\sim 2 \times 10^{-8}$. In the region where $\bsmm$
branching ratio is larger than $10^{-7}$, the $a_\mu^{\rm SUSY}$ is around
$(20-30) \times 10^{-10}$, which is much larger than the aimed experimental
uncertainties. On the other hand, this enhancement effect diminishes quickly
as $\tan\beta$ (and $a_\mu^{\rm SUSY}$) becomes smaller. If the new BNL data
on $a_\mu^{\rm SUSY}$ turns out small ($\lesssim 15 \times 10^{-10}$),
the $\bsmm$ branching ratio cannot be larger than $10^{-8}$ and there would
be no chance to observe this decay at the Tevatron Run II.

The correlation between the Br ($B\rightarrow X_s \gamma$) and $R_{\mu\mu}$
is an interesting quantity as well, since it can be useful to determine the
sign of the $C_{7\gamma}$ coefficient. In Fig.~\ref{fig:msugra2} (b), we show
this correlation in the mSUGRA model. The $R_{\mu\mu}$ can be
enhanced up to 13 \% compared to the SM prediction for large $\tan\beta$,
but no more. In particular the sign of $C_{7\gamma}$ in the mSUGRA model is
the same as the SM case, although there is some destructive interference
between the SM and charged Higgs contributions and the chargino-stop
contribution. In the previous comprehensive analyses by KEK group~
\cite{bsllsugra}, 
it was noted that there could be two branch for this correlation imposing 
the direct search limits available as of 1998. It was due to the possibility 
to have $C_{7,\gamma} \approx - C_{7,\gamma}^{\rm SUSY}$ for light chargino 
and stops for the positive $\mu (> 0)$. Now this is no longer true when the
direct search limits are updated. The lower limits on Higgs boson and other 
SUSY particles rule out the parameter space in which
$C_{7,\gamma} \approx - C_{7,\gamma}^{\rm SUSY}$. Also note that the large
$\tan\beta$ region allows a smaller branching ratio for $B\rightarrow X_s
\gamma$, because the chargino-stop contributions grows as $\tan\beta$ becomes
large and it interfere with the SM and the charged Higgs contributions in
a destructive manner. Considering experimental and theoretical uncertainties,
it would not be possible to use $R_{\mu\mu}$ to indirectly probe the mSUGRA
effects. This is also true for other scenarios we consider in this work.

\subsection{Gauge Mediated SUSY Breaking (GMSB)}

In the gauge mediated SUSY breaking (GMSB), SUSY breaking in the hidden
sector is mediated to the observable sector through SM gauge interactions
of $N_{\rm mess}$ messenger superfields $\Psi_i, \Psi_i^c$, which lie in
the vectorlike representation of the SM gauge group. The messenger fields
couple to a gauge singlet superfield $X$ through
\[
W = \lambda_i X \Psi_i \Psi_i^c.
\]
The vev of $X$ (both in the scalar and the $F$ components) will induce
SUSY breaking in the messenger sector, which in turn induce the following
set of SUSY breaking soft parameters in the MSSM sector at the messenger
scale $M_{\rm mess}$ :
\begin{eqnarray}
  M_a (M_{\rm mess}) & = &
%N {\alpha_a \over 4 \pi}~\Lambda,
           N_{\rm mess} \Lambda ~g \left( { \Lambda \over M^2_{\rm mess} }
\right)~{\alpha_a \over 4 \pi},
\nonumber
\\
  m_{ij}^2 (M_{\rm mess}) & = &
%2 N \delta_{ij} \sum_a C_a^i ~\left( {\alpha_a \over 4 \pi }
%\right)^2 \Lambda^2,
2 N_{\rm mess} \Lambda^2 ~f \left( { \Lambda \over M^2_{\rm mess}} \right)~
\sum_a \left( {\alpha_a \over 4 \pi } \right)^2~C_a
\nonumber
\\
  A_{ijk} (M_{\rm mess}) & = & 0.
\end{eqnarray}
Here $\alpha_a$ (with $a = 3,2,1$) are the SM gauge couplings of
$SU(3)_c \times SU(2)_L \times U(1)_Y$, $C_a^i$'s are the quadratic Casimir
invariant of the MSSM matter fields, and $f(x)$ and $g(x)$ are loop functions
whose explicit form can be found in Ref.~\cite{gauge}. In the limit 
$\Lambda << M^2_{\rm messenger}$, these loop  functions $g(x)$ and $f(x)$ 
are well approximated to one: $f(x) \approx g(x) \approx 1$ for $x <<1$. 
We have normalized the $U(1)_Y$ charge to a GUT group such as $SU(5)$.
Also we have ignored the nonvanishing results for $A_{ijk}$ which arise
from two-loop diagrams, since they are suppressed by loop factors.
Therefore the free parameters in GMSB are
\[
M_{\rm mess},~~N_{\rm mess},~~\Lambda,~~\tan\beta,~~ {\rm sign}(\mu),
\]
where $N$ is the number of messenger superfields, $M$ is the messenger
scale, and the $\Lambda$ is SUSY breaking scale :
\[
\Lambda \approx \langle F_X \rangle / \langle X \rangle.
\]
In practice, we trade $\Lambda$ for the bino mass parameter $M_1$, and
we scan these parameters over the following ranges :
\begin{eqnarray}
10^4 ~{\rm GeV} & \leq & \Lambda \leq 2 \times 10^5 ~{\rm GeV},
\nonumber  \\
N_{\rm mess} & = & 1,~~5
\nonumber  \\
M_{\rm mess} & = & 10^6 {\rm GeV},~~{\rm and}~~10^{15}~{\rm GeV}.
\end{eqnarray}
Earlier phenomenological analysis of GMSB scenarios can be found on the muon
$(g-2)_\mu$ \cite{gmsb:amu}, $B\rightarrow X_s \gamma$ and 
$B\rightarrow X_s l^+ l^-$, \cite{bsllgmsb}. 
The discussion of $B_s \rightarrow \mu^+ \mu^-$ in the GMSB scenarios is 
given in this work for the first time.

\begin{figure}
\centering
\includegraphics[width=0.8\textwidth]{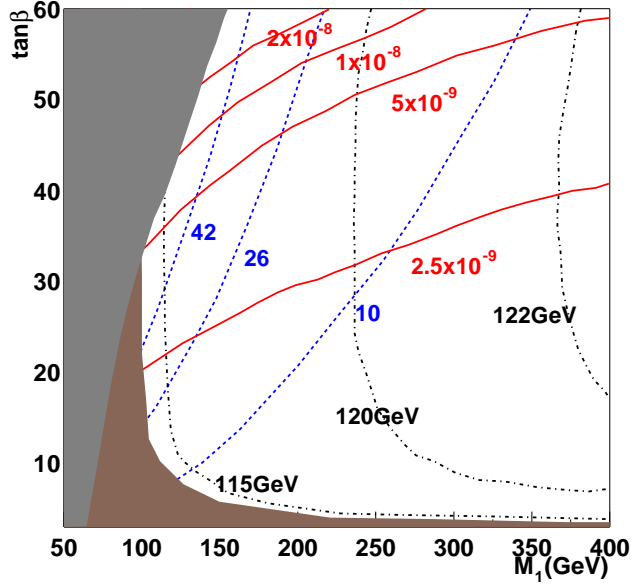}
\caption{\label{fig:gmsb1}
The contour plots for $a_\mu^{\rm SUSY}$ in unit of $10^{-10}$
(in the short dashed curves) and the Br ($B_s \rightarrow \mu^+ \mu^- $)
(in the solid curves) in the $( M_1, \tan\beta)$ plane in the GMSB model with
$N_{\rm mess} =1$ and $M_{\rm mess} = 10^6$ GeV. The light gray region is
excluded by the light stau mass bound, and the brown (dark) region is 
excluded by the lower bounds on the masses of Higgs bosons.}
%$B \rightarrow X_s \gamma$ branching ratio.}
%\label{fig:gmsb1}
\end{figure}

\begin{figure}
\centering
\hspace{-1.5cm}
\subfigure[]{
\includegraphics[width=0.5\textwidth]%{figs/gmsb-1-6/br_bs-mug23.ps}}
{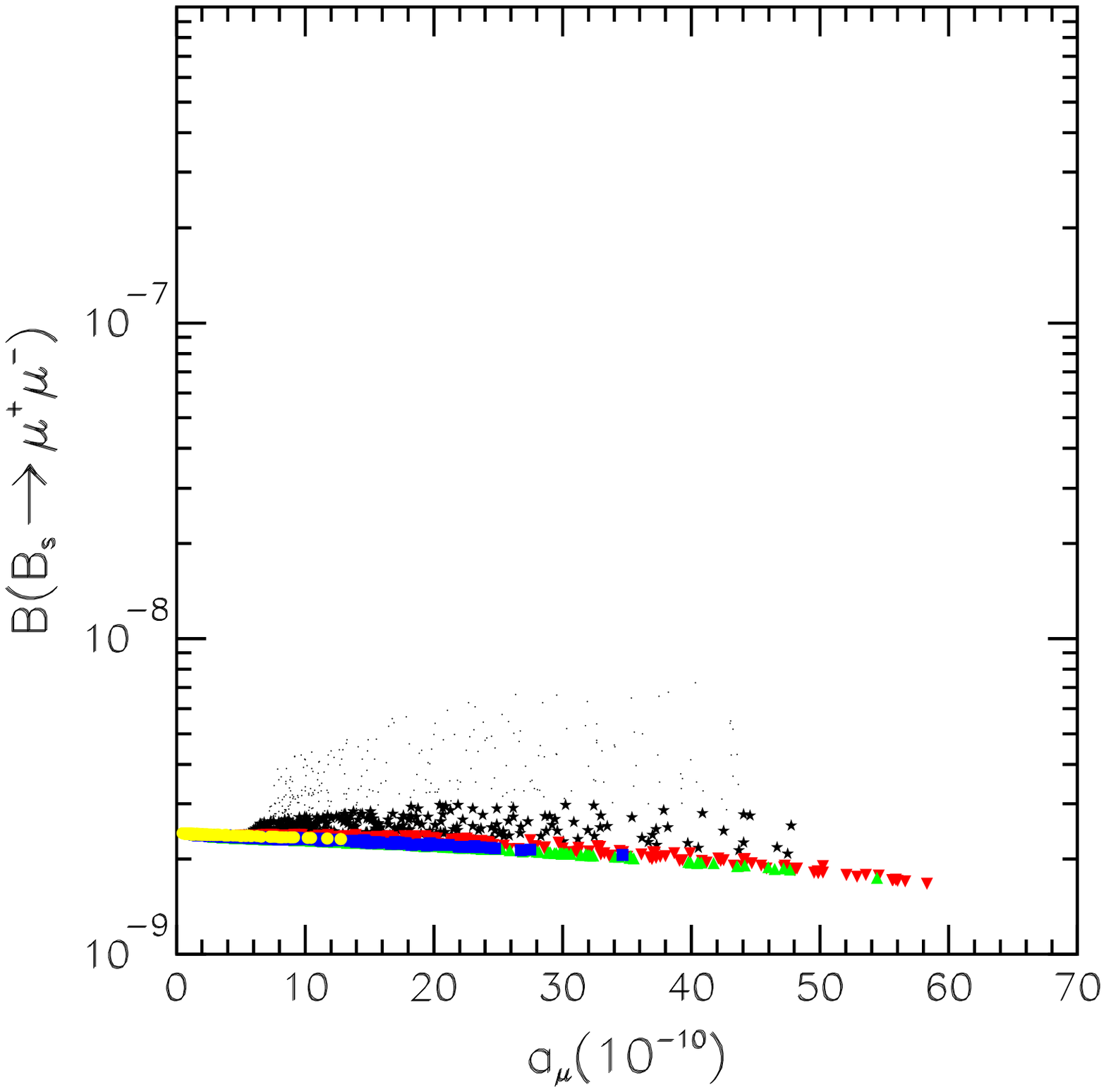}}
%\hspace{-1.5cm}
\subfigure[]{
\includegraphics[width=0.5\textwidth]%{figs/gmsb-1-6/bsll-b2sr.ps}}
{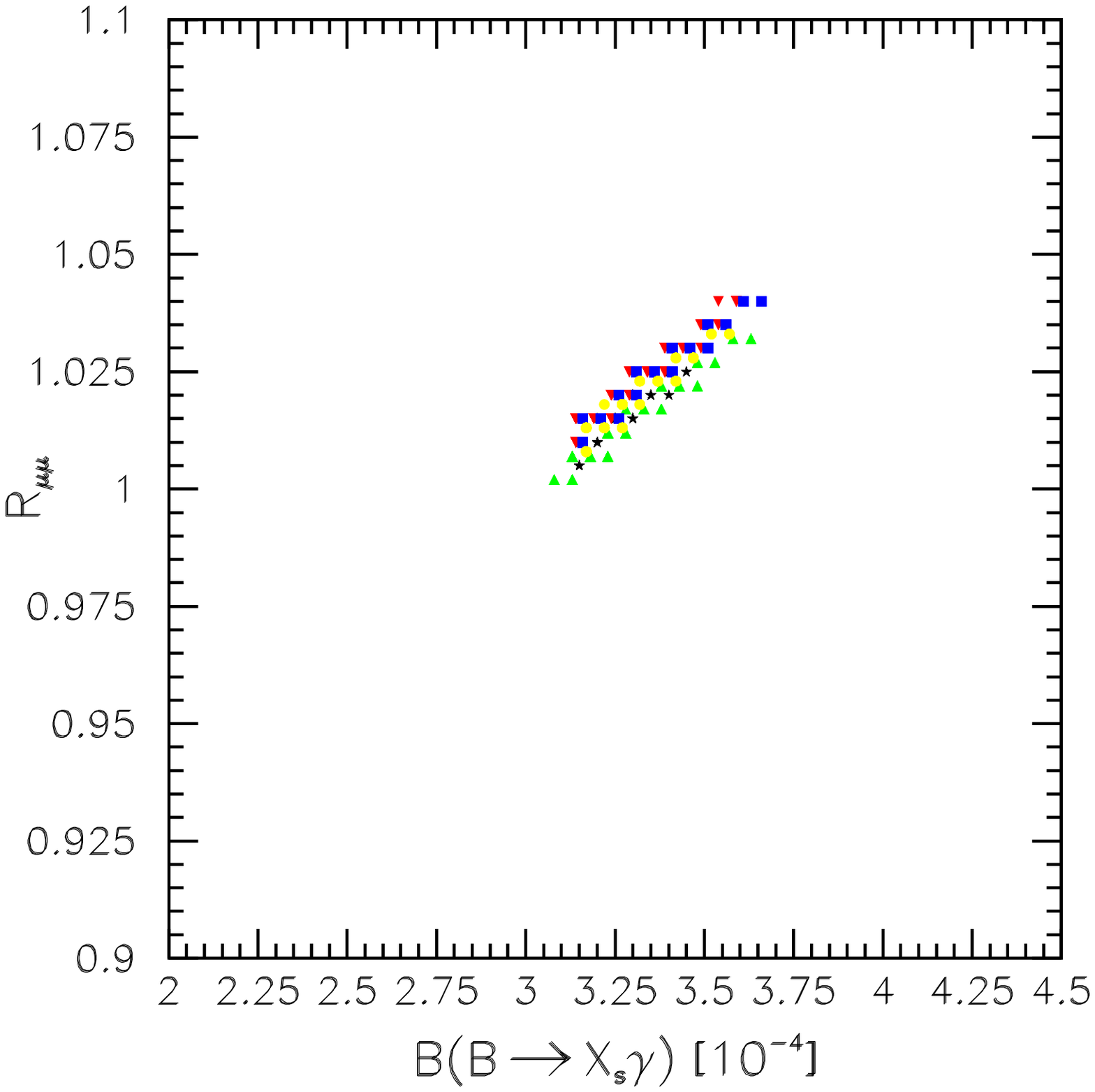}}
\caption{\label{fig:gmsb2}
The correlations of the $Br ( B \rightarrow X_s \gamma)$ with
(a) $R_{\mu\mu}$, and (b) $a_\mu^{\rm SUSY}$ in the GMSB scenario with
$N_{\rm mess} = 1$ and $M_{\rm mess} = 10^6 $ GeV.
The legends are the same as Fig.~\ref{fig:msugra2}.}
%\label{fig:gmsb2}
\end{figure}

In Fig.~\ref{fig:gmsb1}, we show the contour plots for the $a_\mu^{\rm SUSY}$
and $B ( B_s \rightarrow \mu^+ \mu^- )$ in the $( M_1 , \tan\beta)$ plane
for $N_{\rm mess} = 1$ and $M_{\rm mess} = 10^6$ GeV.  The left dark region
is excluded by direct search limit on Higgs boson mass, and the gray region
is excluded by the lower bound on the NLSP mass, which is quite significant.
For a low $M_{\rm mess}$ scale, the RG runs only for a short distance and
its effects are not very large. The resulting $A_t$ parameter at the
electroweak scale is very small, leading to negligible
$\tilde{t}_L - \tilde{t}_R$ mixing. Also the stop mass is relatively large
in this case.
Therefore both the chargino-stop and the charged Higgs - top contributions
to $B\rightarrow X_s \gamma$ are not that important, and there is no strong
constraint  from $B\rightarrow X_s \gamma$. By the same token, the branching
ratio for $\bsmm$ is always smaller than $10^{-8}$, and this becomes
unobservable at the Tevatron Run II.  Therefore if the $a_\mu^{\rm SUSY}$
turns out to be positive and the decay $\bsmm$ is observed at the Tevatron 
Run II, the GMSB scenarios with low messenger scales would be excluded.
$R_{\mu\mu}$ tends to decrease down to 0.9, but this is no significant 
deviation from the SM prediction, and it would not be possible to observe
indirect SUSY signals from $B \rightarrow X_s \mu^+ \mu^-$
(See Figs.~\ref{fig:gmsb2} (a) and (b)).

\begin{figure}
\includegraphics[width=0.8\textwidth]{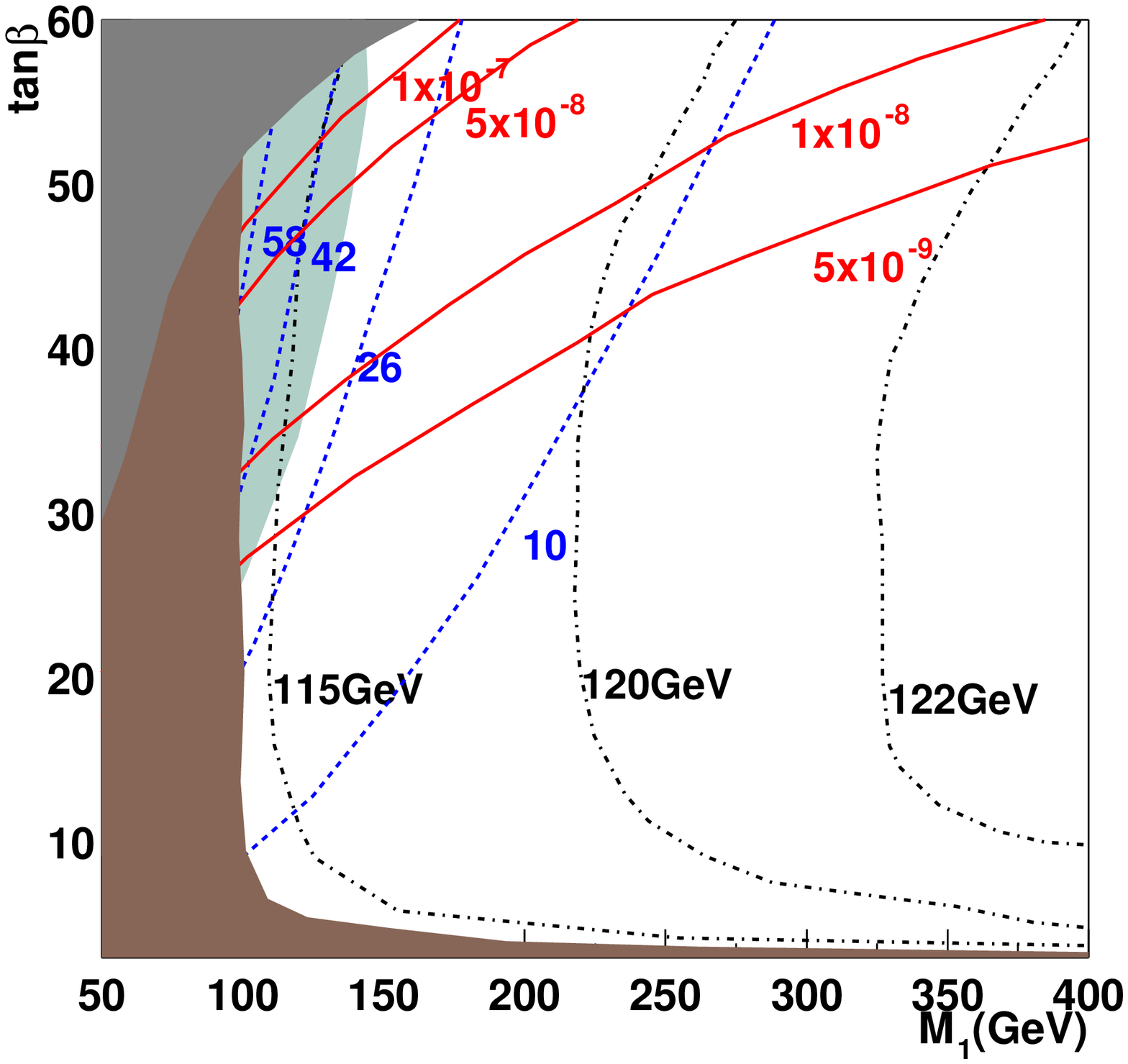}
\caption{\label{fig:gmsb3}
The contour plots for $a_\mu^{\rm SUSY}$ in unit of $10^{-10}$
(in the short dashed curves) and the Br ($B_s \rightarrow \mu^+ \mu^- $)
(in the solid curves) in the $( M_1, \tan\beta)$ plane for the GMSB model
with $N_{\rm mess} =1$ and $M_{\rm mess} = 10^{15}$ GeV.
The gray region is excluded by the NLSP mass bound, and the green region is 
excluded by $B\rightarrow X_s \gamma$ branching ratio.}
%\label{fig:gmsb3}
\end{figure}

\begin{figure}
\centering
\hspace{-1.5cm}
\subfigure[]{
\includegraphics[width=0.5\textwidth]%{figs/gmsb-1-15/br_bs-mug23.ps}}
{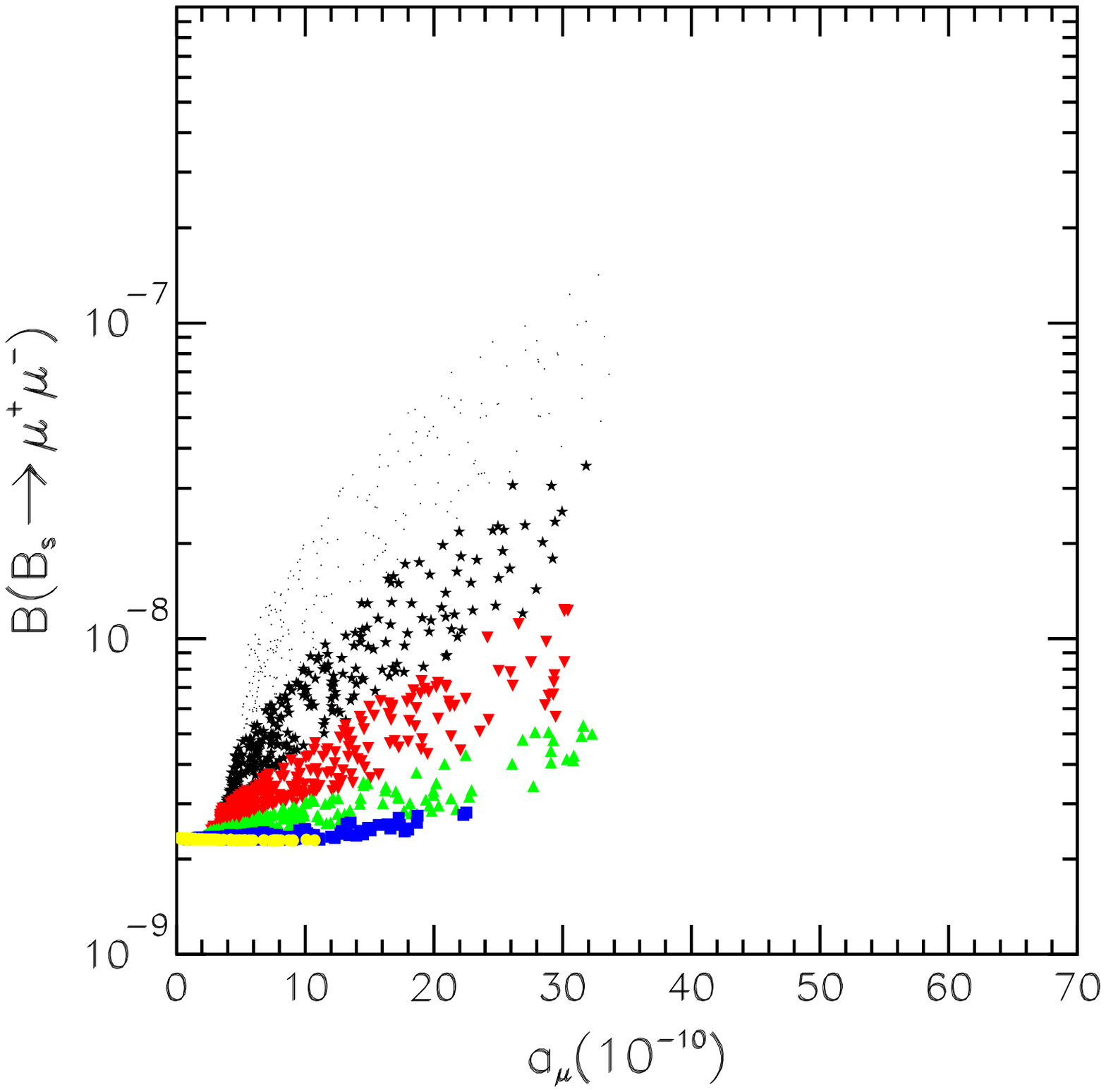}}
%\hspace{-1.5cm}
\subfigure[]{
\includegraphics[width=0.5\textwidth]%{figs/gmsb-1-15/bsll-b2sr.ps}}
{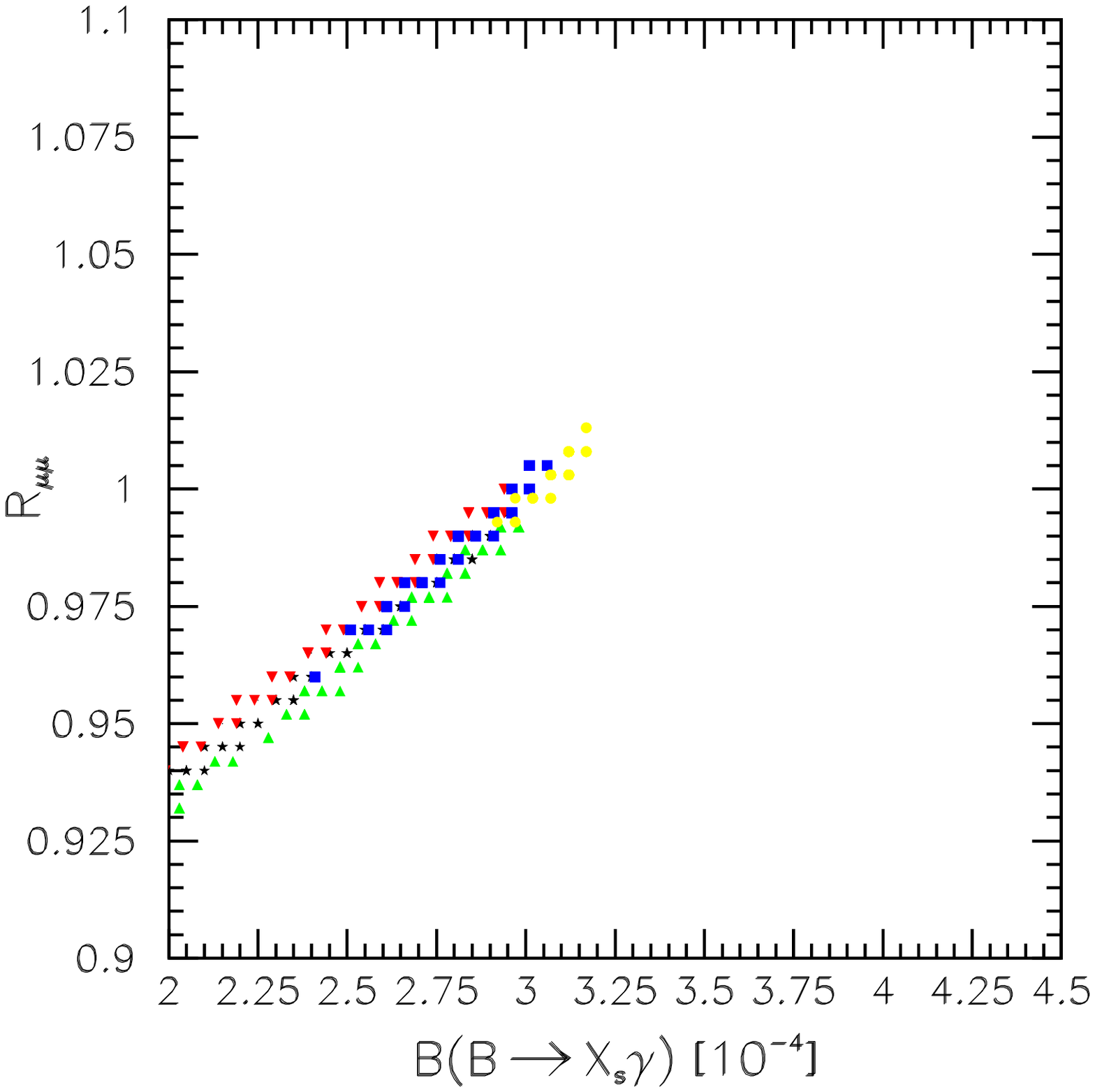}}
\caption{\label{fig:gmsb4}
The correlations of the $Br ( B \rightarrow X_s \gamma)$ with
(a) $R_{\mu\mu}$, and (b) $a_\mu^{\rm SUSY}$ in the GMSB scenario with
$N_{\rm mess} = 1$ and $M_{\rm mess} = 10^{15} $ GeV.
The legends are the same as Fig.~\ref{fig:msugra2}.}
%\label{fig:gmsb4}
\end{figure}

If the messenger scale becomes as high as the GUT scale, the RG effects
become important. The $A_t$ parameter at the electroweak scale becomes larger,
leading to large $\tilde{t}_L - \tilde{t}_R$ mixing. Therefore, the
chargino-stop contribution begins to compensate the SM and charged
Higgs - top contributions to $B\rightarrow X_s \gamma$.
The overall features look alike the mSUGRA or the dilaton dominated case
(see Figs.~\ref{fig:gmsb3} and \ref{fig:gmsb4}).
Still the resulting branching ratio for $\bsmm$
is fairly small, and can be as large as $2\times 10^{-8}$ for very large
$\tan\beta \sim 60$, and much smaller for $\tan\beta \lesssim 50$. So one can
safely assert that the GMSB with $N_{\rm mess} = 1$ is excluded if the decay
$\bsmm$ is discovered at the Tevatron Run II. (This is also true for the case
of the minimal AMSB scenario and noscale scenario as discussed
in the following subsection.)

\begin{figure}
\includegraphics[width=0.8\textwidth]{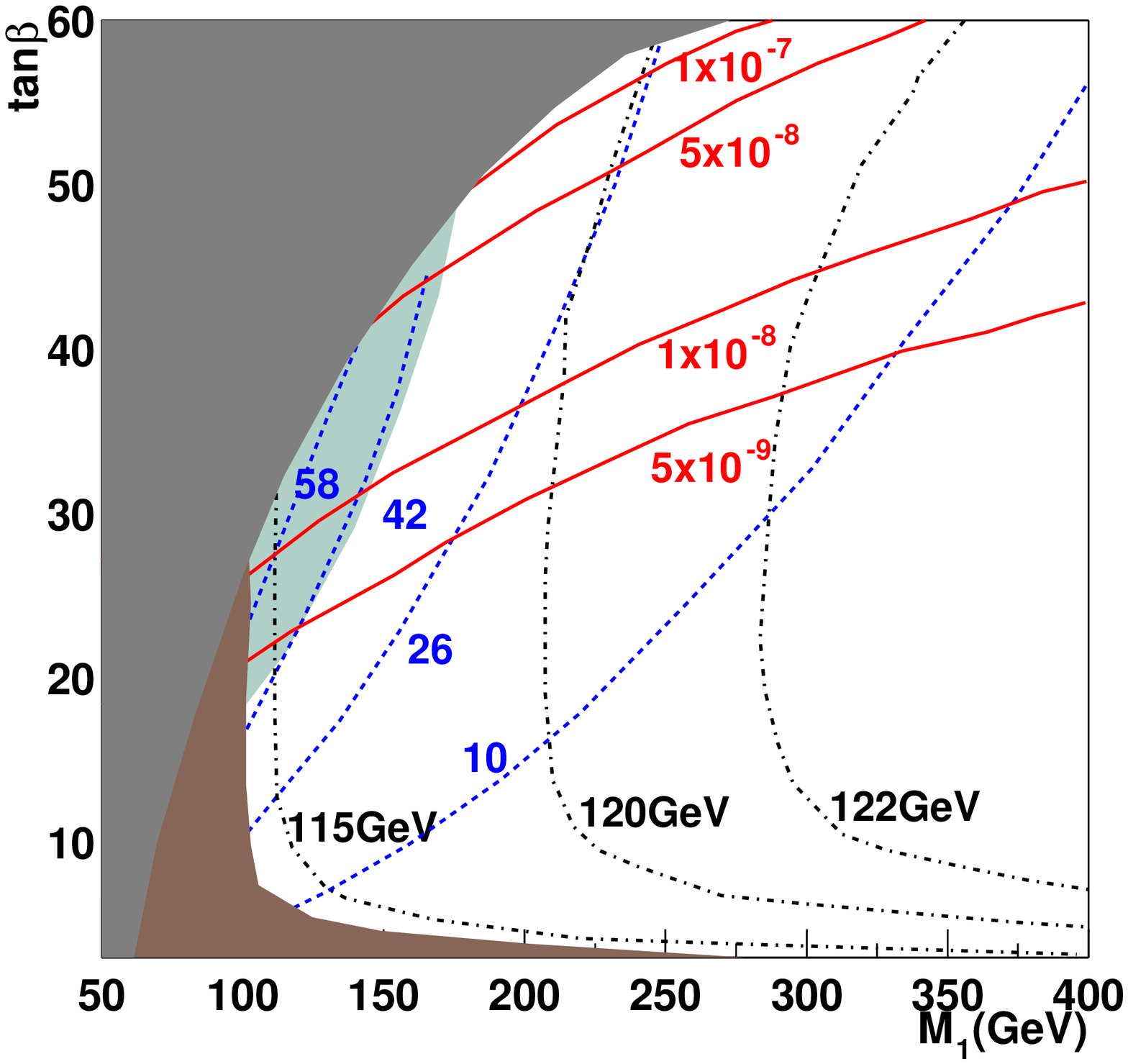}
\caption{\label{fig:gmsb5}
The contour plots for $a_\mu^{\rm SUSY}$ in unit of $10^{-10}$
(in the short dashed curves) and the Br ($B_s \rightarrow \mu^+ \mu^- $)
(in the solid curves) in the $( M_1, \tan\beta)$ plane for the GMSB model
with $N_{\rm mess} =5$ and $M_{\rm mess} = 10^{15}$ GeV.
The gray region is excluded by the NLSP mass bound, and the green region is 
excluded by $B\rightarrow X_s \gamma$ branching ratio. }
%$B \rightarrow X_s \gamma$ branching ratio.}
%\label{fig:gmsb5}
\end{figure}

As the number of the messenger fields $N_{\rm mess}$ increases from 1 to 5,
scalar fermions get lighter compared to the lower $N_{\rm mess}$ case for
unified gaugino masses. Therefore, the chargino-stop contributions to
$B\rightarrow X_s \gamma$ and $\bsmm$ become more important than the lower
$N_{\rm mess}$ case. Still $B\rightarrow X_s \gamma$ is not constraining,
but the $\bsmm$ branching ratio can be enhanced significantly, like in the
mSUGRA model (see Fig.~\ref{fig:gmsb5}).
The muon $a_\mu^{\rm SUSY}$ can be up to  $48 \times 10^{-10}$, and the
$\bsmm$ branching ratio can be enhanced up to  $2 \times 10^{-7}$.
The messenger scale dependence is similar to the previous case,
and will not be repeated here.

In summary, the lighter stop mass in the GMSB scenario with $N_{\rm mess} =1$
is generically heavy, although it gets lighter if the messenger scale becomes
higher and the RG effects become more important. Still the resulting $\bsmm$
branching ratio is smaller than $2\times 10^{-8}$, and only for very large
$\tan\beta \approx 60$ this upper limit can be achieved.
In most parameter space, it is much smaller, and there would no chance to
observe  it at Tevatron Run II.  On the other hand, if the number of messenger
fields and the messenger scale increases, the pseudoscalar and the stop get
lighter and the $A_t$ parameter gets larger leading to the large stop mixing. 
Thus the branching ratio for $\bsmm$ can be enhanced within the reach of the 
Tevatron Run II.

\subsection{Anomaly Mediated SUSY Breaking (AMSB)}

In the AMSB scenario, it is assumed that the hidden sector SUSY breaking
is mediated to our world only through the auxiliary component of the
supergravity multiplet. This is possible if the K\"{a}hler potential has
the so-called sequestered form :
\begin{equation}
K =-3 \log \left[ \xi ( \Phi, \Phi^{\dagger} ) + \zeta ( z, z^{\dagger} ) 
\right],
\label{eq:seq}
\end{equation}
where $\Phi$ and $z$ are the observable and the hidden fields, respectively.
In this case, the compensator field $X$ will take a VEV of the form :
\begin{equation}
 \langle X \rangle = 1 + F_X \theta^2
\end{equation}
Here $F_X$ is an auxiliary field in the gravity supermultiplet,
whose VEV is given by (assuming the vanishing cosmological constant)
\begin{equation}
F_X = {1 \over M_*^2}~ \left( W + {1\over 3} {\partial K \over \partial z}
F_z \right) ,
\end{equation}
where $M_* \approx 2.4 \times 10^{18}$ GeV is the reduced Planck scale.

The soft terms can be extracted by expanding the supergravity lagrangian
in the background with nonvanishing $F_\Phi$.
The results are the following :
\begin{eqnarray}
  M_a & = & - {b_a \alpha_a \over 4 \pi}~M_{\rm aux},
\nonumber \\
  m_{ij}^2 & = & \left( - {\dot{\gamma} \over 4}~M_{\rm aux}^2 + m_0^2 \right)
                  ~\delta_{ij},
\nonumber \\
  A_{ijk} & = & {1\over 2}~( \gamma_i + \gamma_j + \gamma_k ) M_{\rm aux}.
\end{eqnarray}
Here $b_a = ( 3, -1, -33/5)$ (with $a = 3,2,1$) are the one-loop beta function
coefficients for the SM gauge group $SU(3)_c \times SU(2)_L \times U(1)_Y$,
$\gamma_i \equiv - d \ln Z_i / d \ln \mu$ is the anomalous dimension
of the field $\Phi_i$, and the dot acting on $\gamma_i$ denotes the
differentiation with respect to $\ln \mu$.
We have simply added $m_0^2$ to the scalar fermion mass parameters of the
original AMSB model in order to avoid the tachyon problem in the slepton
sector, and will assume that the above set of equations make initial
conditions at the GUT scale for the RG equations.  Note that in the pure AMSB
case ($m_0^2 = 0$) the soft terms are scale invariant so that they are valid
for arbitrary scale and are completely fixed by a single overall scale
$M_{\rm aux}$ and the gauge couplings at low energy. However this nicety is
lost when we add $m_0^2$ to the scalar fermion masses.  Thus, the minimal
AMSB model is specified by the following four parameters :
\[
\tan\beta, ~~{\rm sign}(\mu), ~~m_0,~~ M_{\rm aux}.
\]
We scan these parameters over the following ranges :
\begin{eqnarray}
  20~{\rm TeV} & \leq & m_{\rm aux} \leq 100~{\rm TeV},
\nonumber  \\
             0 & \leq & m_0 \leq 2~{\rm TeV},
\nonumber  \\
         1.5   & \leq & \tan\beta \leq 60,
%\nonumber  \\
%{\rm sign} (\mu) & = & -1.
\end{eqnarray}
Earlier phenomenological analysis of the minimal AMSB scenarios can be 
found on the muon $(g-2)_\mu$ and $B\rightarrow X_s \gamma$~
\cite{amsb:amu,kchoi,martin}, and  
on $B_s \rightarrow \mu^+ \mu^-$~\cite{bks}.
The discussion on $B\rightarrow X_s l^+ l^-$ in the AMSB scenarios is given 
in this work for the first time.

In the brane world scenarios which became popular during recent years, the
K\"{a}hler potential takes a sequestered form Eq.~(\ref{eq:seq}) in a natural
way. The resulting scalar fermion masses take the above form (flavor
independent) so that the SUSY flavor problem is solved in the AMSB model.
However, Anisimov {\it et al.} recently argued that this form is not generic 
in the brane world SUSY breaking scenario \cite{dine}. The bulk supergravity
effects generate tree level scalar fermion masses which are generically flavor
dependent. Only a certain special class of  models have zero tree level
scalar masses and thus become genuine AMSB models 
(see, for example,~\cite{harnik}).  In this work, we consider
this class of models where the above expressions for the  soft terms make
good descriptions. This general remark is also true of the gaugino mediation
(and no-scale supergravity) scenario(s) to be discussed in the subsequent
subsection.

\begin{figure}
\includegraphics[width=0.8\textwidth]{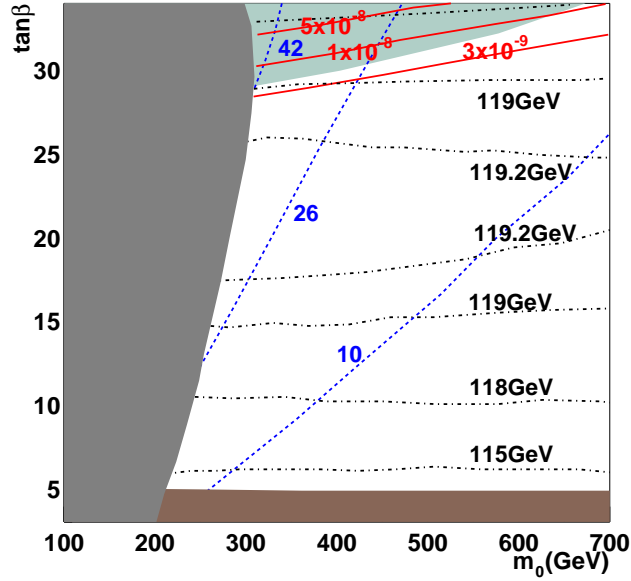}
\caption{\label{fig:amsb1}
The contour plots for $a_\mu^{\rm SUSY}$ in unit of $10^{-10}$
(in the short dashed curves) and the Br ($B_s \rightarrow \mu^+ \mu^- $)
(in the solid curves) in the $( m_0, \tan\beta)$ plane for the minimal
AMSB scenarios with $M_{\rm aux} = 50$ TeV and $\mu > 0$.
The brown (dark), the gray and the green regions are excluded by the Higgs 
mass bound, the SUSY particle search, and $B\rightarrow X_s \gamma$ 
branching ratio, respectively.
}
%\label{fig:amsb1}
\end{figure}

\begin{figure}
\centering
\hspace{-1.5cm}
\subfigure[]{
\includegraphics[width=0.5\textwidth]%{figs/amsb/br_bs-mug23.ps}}
{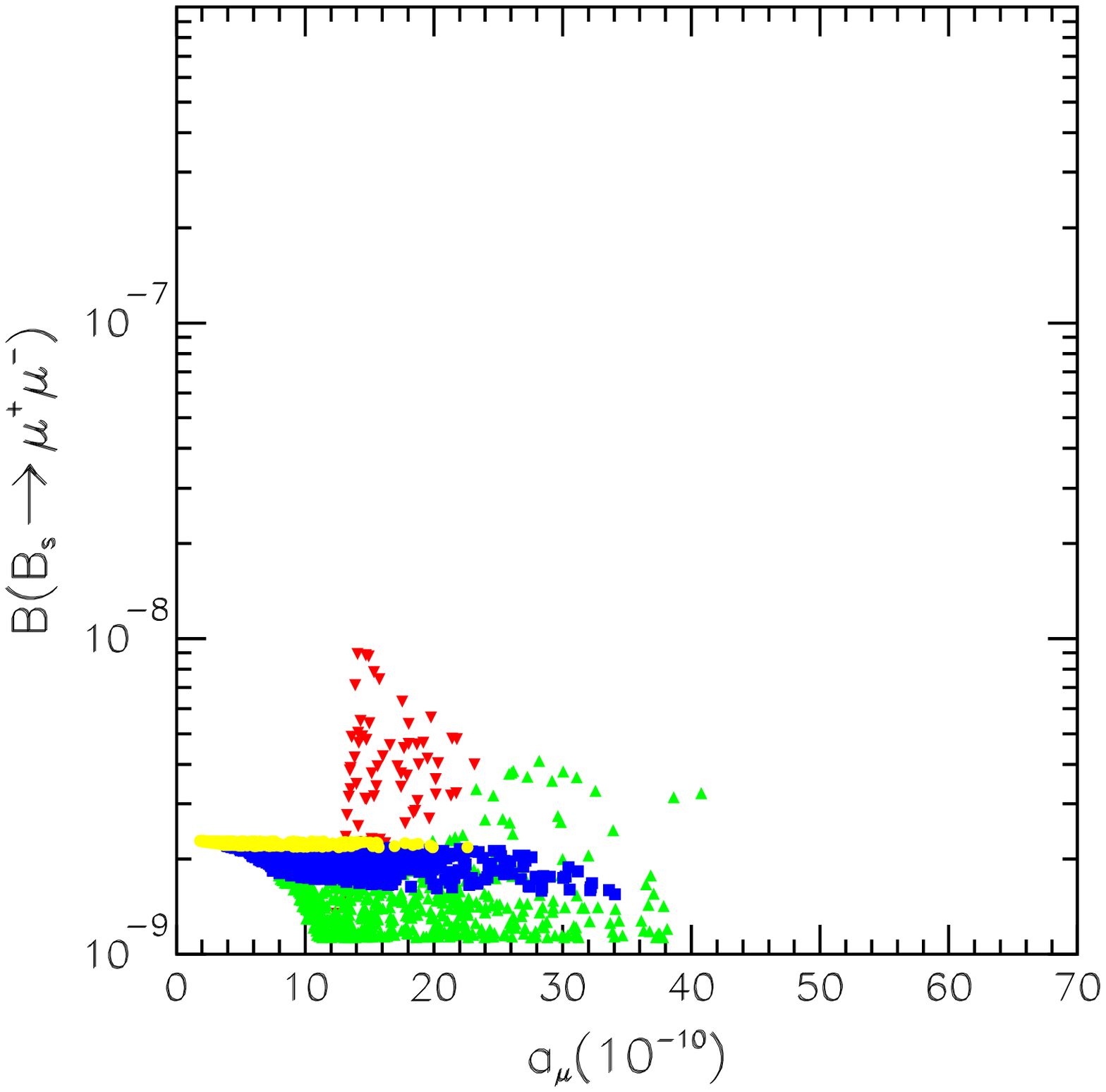}}
%\hspace{-1.5cm}
\subfigure[]{
\includegraphics[width=0.5\textwidth]%{figs/amsb/bsll-b2sr.ps}}
{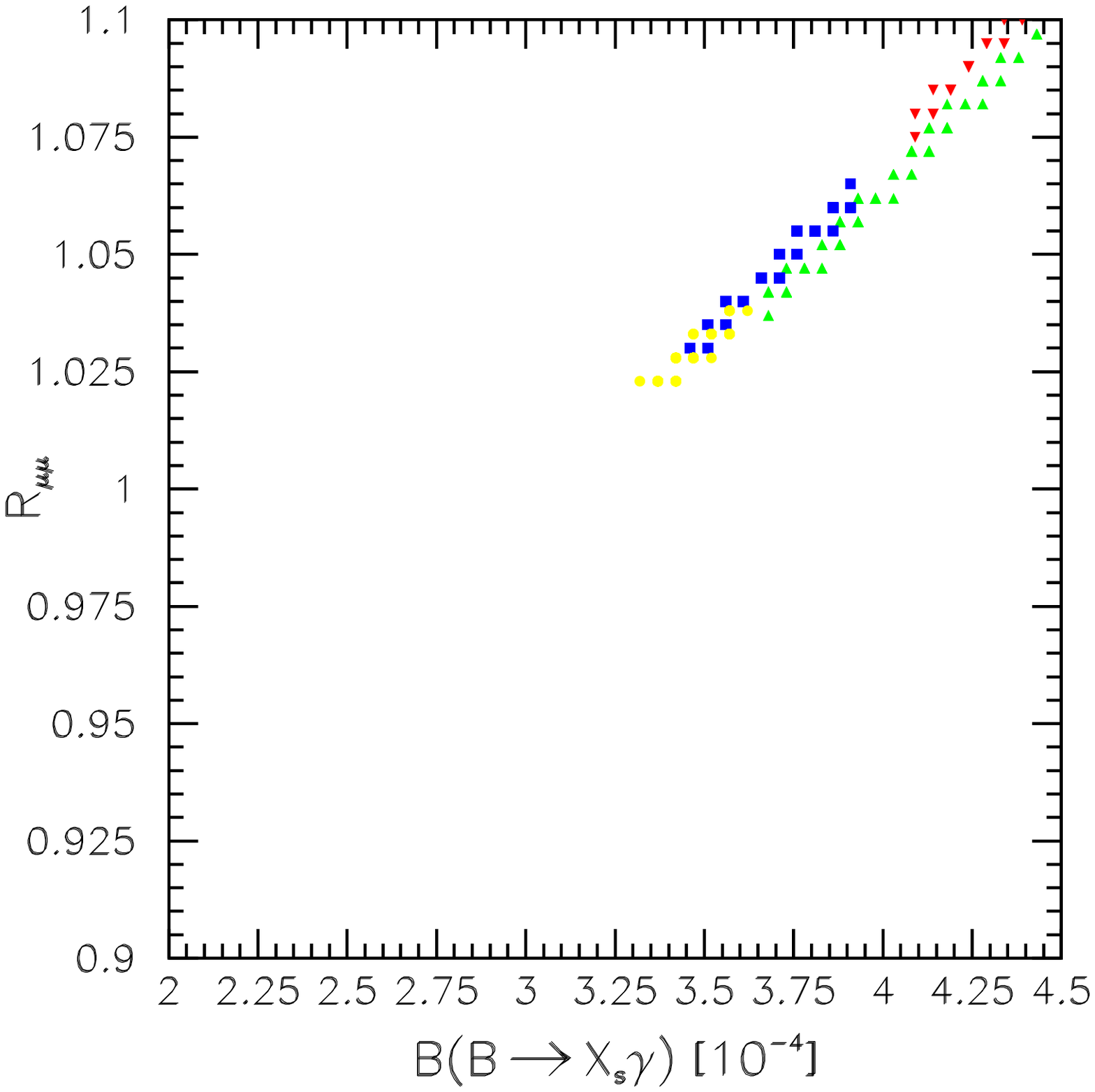}}
\hspace{-1.5cm}
\caption{\label{fig:amsb2}
The correlations of 
(a) $Br ( B_s \rightarrow \mu^+ \mu^- )$ with $a_\mu$, and 
(b) $Br ( B \rightarrow X_s \gamma)$ with $R_{\mu\mu}$ 
in the minimal AMSB scenario for $M_{\rm aux} = 50$ TeV.
The legends are the same as Fig.~\ref{fig:msugra2}.}
%\label{fig:amsb2}
\end{figure}

In Fig.~\ref{fig:amsb1}, we show the contour  plots for the $a_\mu^{\rm SUSY}$
and $B ( B_s \rightarrow \mu^+ \mu^- )$ in the $( m_0 , \tan\beta)$ plane for
$M_{\rm aux} = 50$ TeV. The low $\tan\beta$ region is excluded by the lower
limit on the neutral Higgs boson (the dark region), and the small $m_0$
region is strongly constrained by the stau mass bound (the gray region).
In the case of the AMSB scenario with $\mu > 0$, the
$B\rightarrow X_s \gamma$ constraint is even stronger compared to other
scenarios, since the chargino-stop contribution is additive to the SM and
the charged Higgs contribution  because of $\mu M_3 < 0$ in the AMSB scenario .
This is represented by the green (light gray) region in Fig.~\ref{fig:amsb1}.
Almost all the parameter space with large $\tan\beta > 30$ is excluded by the
upper limit on $B\rightarrow X_s \gamma$. Also stop mass becomes much heavier
in the AMSB scenario compared to the mSUGRA or noscale scenarios.
This makes the decay $\bsmm$ unobservable at the Tevatron Run II, since its
branching ratio cannot be larger than $4 \times 10^{-9}$.  Note that the
reason for the small $\bsmm$ branching ratio in the GMSB with low
$N_{\rm mess}$ or in the AMSB scenarios is heavy stop masses so that
chargino-stop loop contribution is suppressed. In the no scale scenario,
on the other hand, lighter stop can be much lighter but this
region of parameter space is excluded by Higgs and SUSY particle mass bounds.
Therefore, if the $a_\mu^{\rm SUSY}$ turns out to be positive and the decay
$\bsmm$ is observed at the Tevatron Run II, the minimal AMSB scenario would
be excluded. Also there is no significant  deviation in $R_{\mu\mu}$ from 1,
and it would not be possible to observe indirect SUSY signals from
$B \rightarrow X_s \mu^+ \mu^-$ (See Figs.~\ref{fig:amsb2} (a) and (b)).

\begin{figure}
\includegraphics[width=0.8\textwidth]{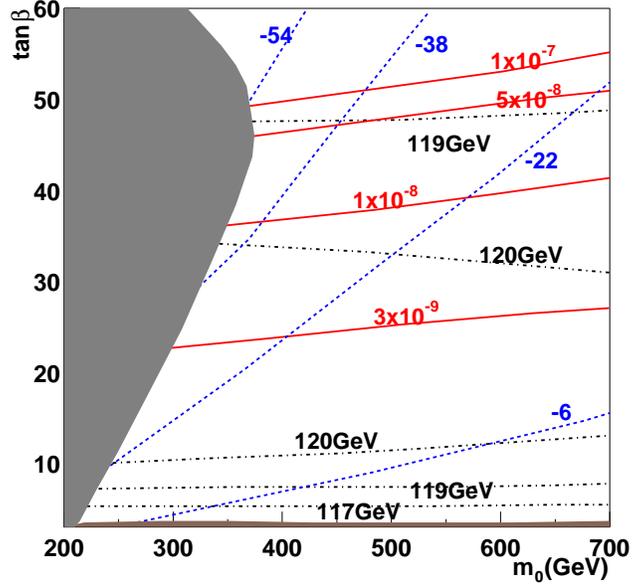}
\caption{\label{fig:amsb3}
The contour plots for $a_\mu^{\rm SUSY}$ in unit of $10^{-10}$
(in the short dashed curves) and the Br ($B_s \rightarrow \mu^+ \mu^- $)
(in the solid curves) in the $( m_0, \tan\beta)$ plane for the minimal
AMSB scenarios with $M_{\rm aux} = 50$ TeV and $\mu < 0$.}
%\label{fig:amsb3}
\end{figure}

In the AMSB model, the $B\rightarrow X_s \gamma$ is less constraining for
the negative $\mu <0$. In this case, the $a_\mu$ is also negative, which is
marginally consistent with the current data on the muon $(g-2)$. For large
$\tan\beta$, the $B_s \rightarrow \mu^+ \mu^-$ can be enhanced  up to  $6\times
10^{-7}$, for which the $a_\mu^{\rm SUSY}$ should be also large with the 
negative sign. All these features can be observed in Fig.~\ref{fig:amsb3}, 
where we show the contour  plots for the $a_\mu^{\rm SUSY}$ and 
$B ( B_s \rightarrow \mu^+ \mu^- )$ in the $( m_0 , \tan\beta)$ plane for
$M_{\rm aux} = 50$ TeV and the negative $\mu < 0$. 
%There is no longer the $B\rightarrow X_s \gamma$ constraint. 

\subsection{No-scale and Gaugino Mediated SUSY Breaking}

If we assume the following nonminimal K\"{a}hler potential and the gauge
kinetic function in supergravity models,
\begin{equation}
K = - 3 \ln ( T + T^* - \Phi_i^* \Phi_i ),~~~
f_{ab} = \delta_{ab} T/ 4 \pi ~(a=1,2,3),
\end{equation}
we get
\begin{equation}
M_a = M_{\rm aux},~~~m_{ij}^2 = 0,~~~A_{ijk} = 0,
\end{equation}
at the messenger scale close  to the GUT scale. Therefore, only gauginos
become massive, and other soft parameters are simply zero including the
gravitino masses. Thus the name ``no-scale SUGRA'' naturally arises
\cite{noscale}. Since the scalar fermion masses and trilinear couplings
take the simplest form to be flavor conserving, namely zero, at the
messenger scale, SUSY flavor problem is significantly mitigated up to  
corrections due to the RG effects when we run the above parameters down
to the electroweak scale. This no-scale scenario was a popular alternative
to the mSUGRA scenario discussed in the Sec.~II A. However both
scenarios assumed very specific and add hoc forms for the K\"{a}hler
potential and the gauge kinetic functions, and thus were not justified well
from deeper theoretical  frameworks.

After the role of branes began to be understood better and included into
the particle physics model building, it was realized that the no-scale
scenario could be naturally realized in the higher dimensional spacetime.
Suppose that the SUSY breaking occurs on a hidden brane, the MSSM matter
fields are confined to the visible brane which is distinct from the hidden
brane where SUSY is broken, and gauge fields live in the bulk.
Then SUSY breaking can be felt by the bulk gauge supermultiplets, thereby
generating soft masses for gauginos. Due to the locality in the extra
dimension, the soft terms for the MSSM matter fields on the visible brane
has to vanish. Only the gaugino can develop nonzero masses at the
compactification scale $M_c$. The scalar fermions get SUSY breaking masses
only through loop effects involving gauginos. This scenario is called the
gaugino$(\tilde{g})$ mediation \cite{ginomsb}. In the gaugino mediated SUSY
breaking scenario ($\tilde{g}$MSB), the model parameters are
\[
\tan\beta, ~~{\rm sign}(\mu), ~~M_{\rm aux}~~, B = m_{ij}^2 = A_{ijk} = 0
\]
at the compactification scale $M_c$. If we relax $B=0$ condition, the
gaugino mediation model becomes the so-called no-scale supergravity with the
corresponding K\"{a}hler potential being the same as Eq.~(32).
Earlier phenomenological analysis of $\tilde{g}$MSB scenarios and noscale 
scenario can be found on the muon 
$(g-2)_\mu$ and $B\rightarrow X_s \gamma$~\cite{kchoi,tata}. 
The discussions of $B\rightarrow X_s l^+ l^-$ and 
$B_s \rightarrow \mu^+ \mu^-$ in the no scale scenarios including the 
$\tilde{g}$MSB scenarios is given in this work for the first time. 

\begin{figure}
\includegraphics[width=0.8\textwidth]{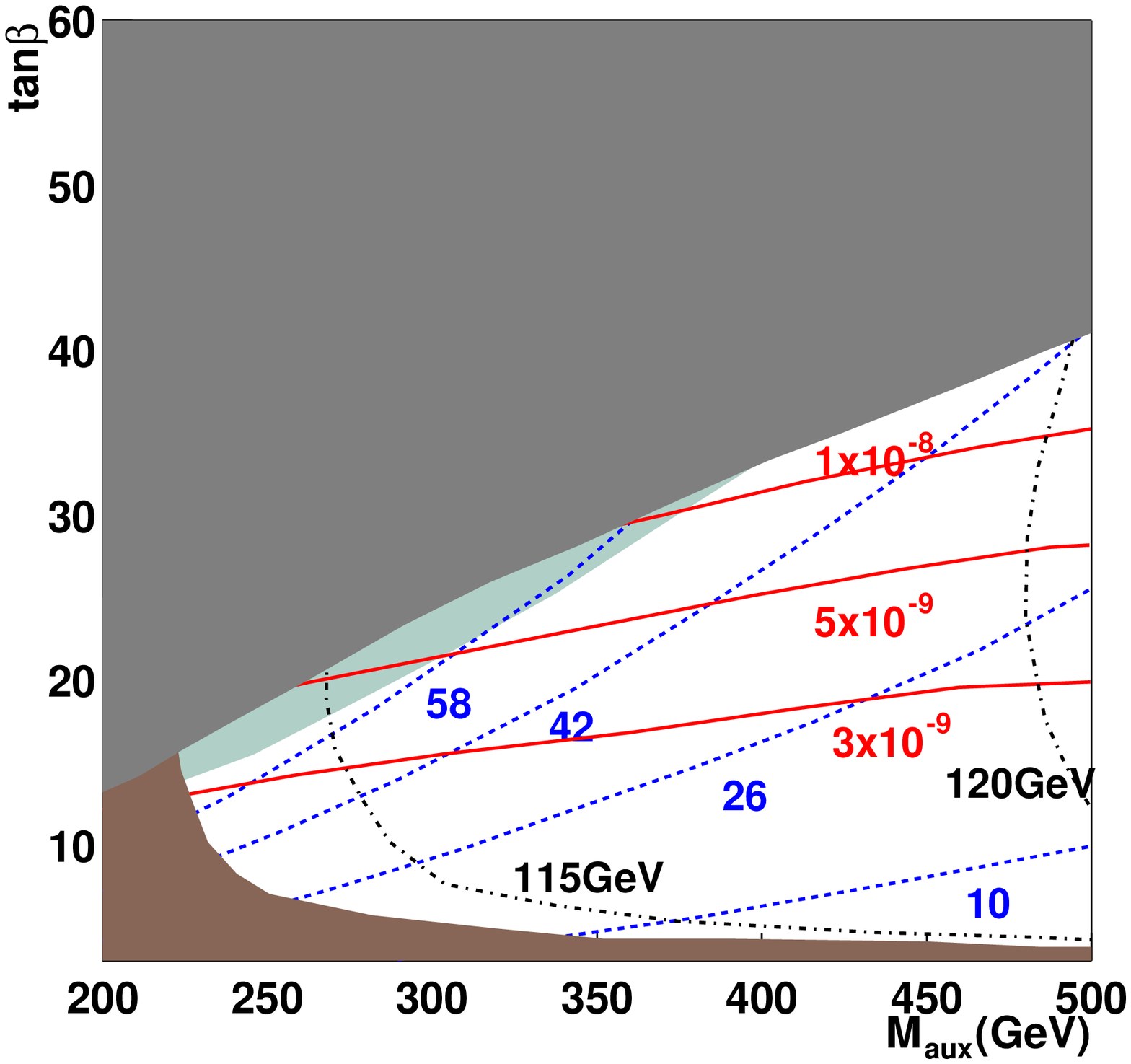}
\caption{\label{fig:noscale}
The contour plots for $a_\mu^{\rm SUSY}$ in unit of $10^{-10}$
(in the short dashed curves) and the Br ($B_s \rightarrow \mu^+ \mu^- $)
(in the solid curves) in the $( M_{\rm aux} , \tan\beta)$ plane for the
noscale scenario. The light gray region is excluded by the light stau mass
bound, and the green region is excluded by the lower bound to the
$B\rightarrow X_s \gamma$ branching ratio.}
%\label{fig:noscale}
\end{figure}

In Fig.~\ref{fig:noscale}, we show the contour  plots for the
$a_\mu^{\rm SUSY}$ and $B ( B_s \rightarrow \mu^+ \mu^- )$ in the
$( M_{\rm aux} , \tan\beta)$ plane. The black region is excluded by
direct search limits on SUSY and Higgs particles, and the green (light gray) 
denote the region excluded by the $B\rightarrow X_s \gamma$ constraint.
The dark gray region is excluded by the stau/smuon limit.
In the allowed parameter space, the $a_\mu^{\rm SUSY}$ can easily become
up to  $\sim 70 \times 10^{-10}$ and one can easily accommodate the BNL data.
On the other hand, the branching ratio for $\bsmm$ is always smaller than
$2 \times 10^{-8}$ and becomes unobservable at the Tevatron Run II, as in
the AMSB models. This is because the large $\tan\beta$ region, where the
branching ratio for $\bsmm$ can be much enhanced, is significantly
constrained by lighter stau and smuon mass bounds and the lower bound of
$B\rightarrow X_s \gamma$. Therefore if the $a_\mu^{\rm SUSY}$ turns out to
be positive and the decay $\bsmm$ is observed at the Tevatron Run II, the
noscale scenario would be excluded. Also there is an anticorrelation between
$R_{\mu\mu}$ and $B ( B \rightarrow X_s \gamma)$ and varies between 0.95
and 1.14. Thus it would not be possible to observe indirect SUSY signals
from $B \rightarrow X_s \mu^+ \mu^-$. Noscale SUGRA models with non universal
gaugino masses are discussed in Ref.~\cite{noscale2}.

%\subsection{Variations on AMSB scenarios}
%
%The pure AMSB scenario, despite of its simplicity, suffers from the 
%tachyonic slepton problem. There have been several suggestions to solve this
%problem in a qualitatively different manners. 
%In this subsection, we consider three different variations to the pure AMSB
%scenario. 
%
%One can consider a hybrid model of AMSB and GMSB \cite{Pomarol:1999ie}

%%%%%%%%%%%%%%%%%%%%%%%%%%%%%%%%%%%%%%%%%%%%
\subsection{Deflected anomaly mediation}
%%%%%%%%%%%%%%%%%%%%%%%%%%%%%%%%%%%%%%%%%%%%

The deflected anomaly mediation~\cite{damsb} is a kind of combination 
of pure anomaly mediation  with gauge mediation scenario. 
If a heavy threshold arises from SUSY breaking effects, integrating out the 
heavy degree of freedoms would kick the low energy SUSY breaking parameters 
off the pure AMSB trajectories and solve the tachyonic slepton problem. 
The model contains a light singlet $X$ which describes a flat direction 
in supersymmetric limit as well as $N$ flavors of gauge-charged messengers 
$\Psi_i,\Psi^c_i$ which are coupled to $X$ in the superpotential
\bea
W=\lambda_i X\Psi_i\Psi^c_i 
\eea\par
as in the GMSB scenario. 
If the VEV of $X$ is determined by the SUSY breaking effects, not by SUSY 
conserving dynamics, one has
\bea
\frac{F_X}{X}=\rho \frac{F_{\phi}}{\phi}\,,
\eea
where $F_{\phi}$ is the $F$ component of the Weyl compensator and 
$\rho$ depends on the details of how $X$ is stabilized, but $\rho\neq 1$ 
in general.

At energy scales below $M\approx \lambda_i\langle X \rangle$,
the heavy thresholds effects of $\Psi_i,\Psi^c_i$ make all soft 
parameters to leave the RG trajectory of pure AMSB. We then have
\bea
&& M_a(M) = (-b_a + N(1-\rho)){\alpha_a (M)\over 4\pi} M_{\rm aux}, 
\nonumber  \\
&& A_{ijk}(M) =\tilde{A}_{ijk} (M), 
\nonumber \\
&& m_{ij}^2(M) =\tilde{m}^2_{ij}(M)
    -2N(1-\rho)\delta_{ij}\sum_a{C_a^i}\left(
{\alpha_a(M)\over 4\pi}\right)^2|M_{\rm aux}|^2\,,
\label{eq:damsb}
\eea
where 
%$b_a=(33/5,1,-3)$ ($a=1,2,3$), and 
$\tilde{A}_{ijk}$, $\tilde{m}^2_{ij}$ 
are the pure AMSB soft parameters in the MSSM, as given in Eq.~(28).
Then the deflected anomaly mediation is described by  six input parameters,
\bea
M_{\rm aux}, \quad M, \quad \rho, \quad \tan\beta,
\quad N, \quad {\rm sign}(\mu).
\eea
For numerical analysis, we take $\rho\approx 0$ which corresponds assuming 
that $X$ is stabilized by the Coleman-Weinberg mechanism~\cite{damsb}.
We scan over the following parameter space:
\begin{eqnarray*}
3 \leq \tan\beta \leq 50,~~~10~{\rm TeV} \leq M_{\rm aux} \leq 80~{\rm TeV},
\\
M = 10^{12}~{\rm GeV},~~~N = 6,~~~\mu > 0. 
\end{eqnarray*}
As is clear from Eq.~(\ref{eq:damsb}), $B \to X_s \gamma$ constraint can 
be weakened by the sign flip of $M_3$, unlike the pure AMSB case.
  
\begin{figure}
\includegraphics[width=0.8\textwidth]{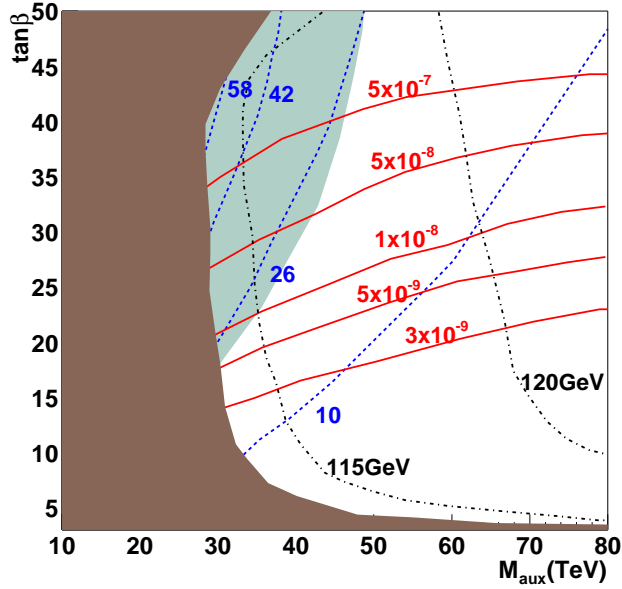}
\caption{\label{fig:damsb1}
The contour plots for $a_\mu^{\rm SUSY}$ in unit of $10^{-10}$
(in the short dashed curves) and the Br ($B_s \rightarrow \mu^+ \mu^- $)
(in the solid curves) in the $( M_{\rm aux}, \tan\beta)$ plane for the 
deflected AMSB scenarios with $M = 10^{12}$ GeV, 
$N=6$, $\rho = 0$ and $\mu > 0$.}
%\label{fig:damsb1}
\end{figure}

\begin{figure}
\centering
\hspace{-1.5cm}
\subfigure[]{
\includegraphics[width=0.5\textwidth]%{figs/hetero_M_05/br_bs-mug23.ps}}
{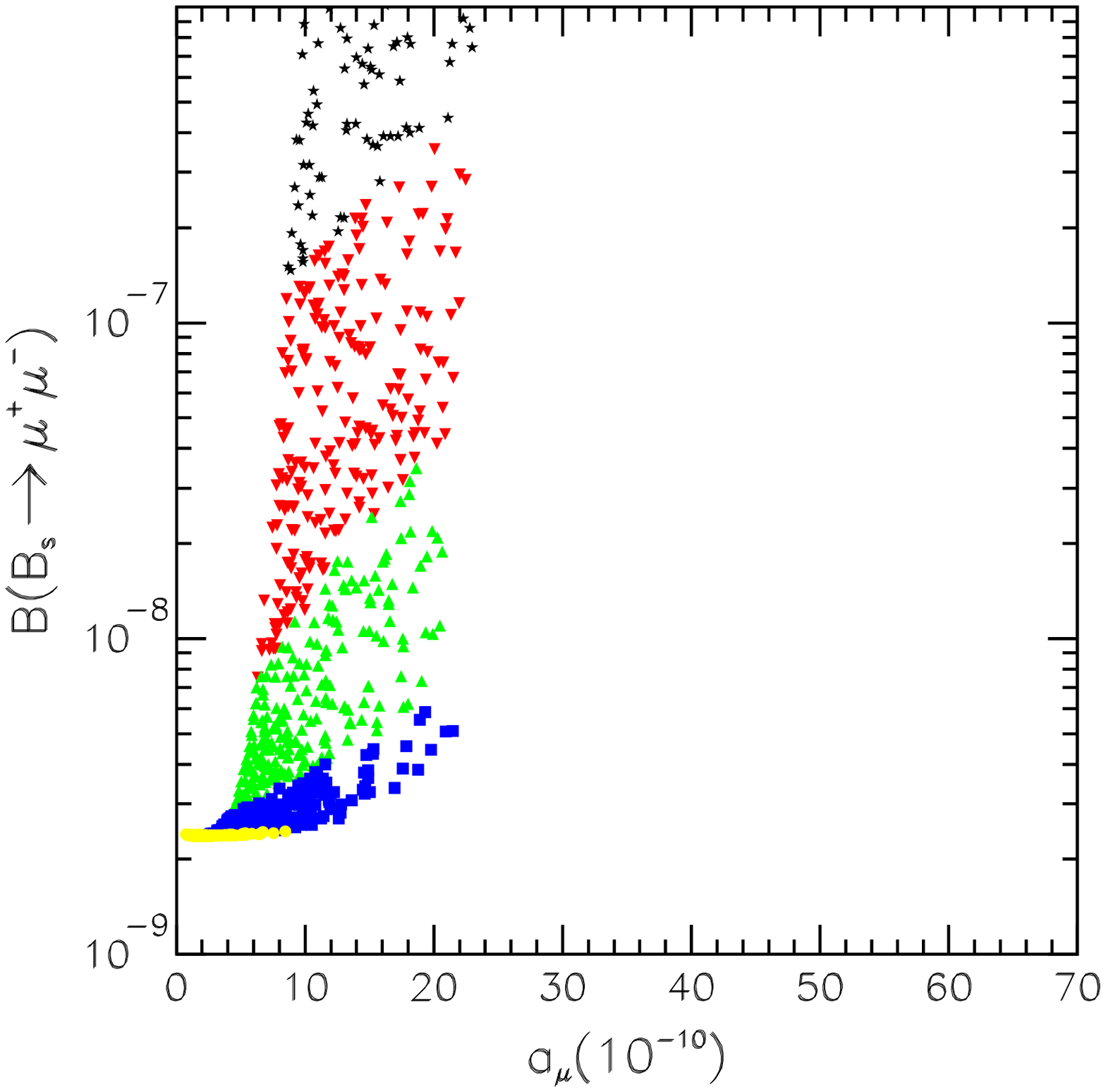}}
%\hspace{-1.5cm}
\subfigure[]{
\includegraphics[width=0.5\textwidth]%{figs/hetero_M_05/bsll-b2sr.ps}}
{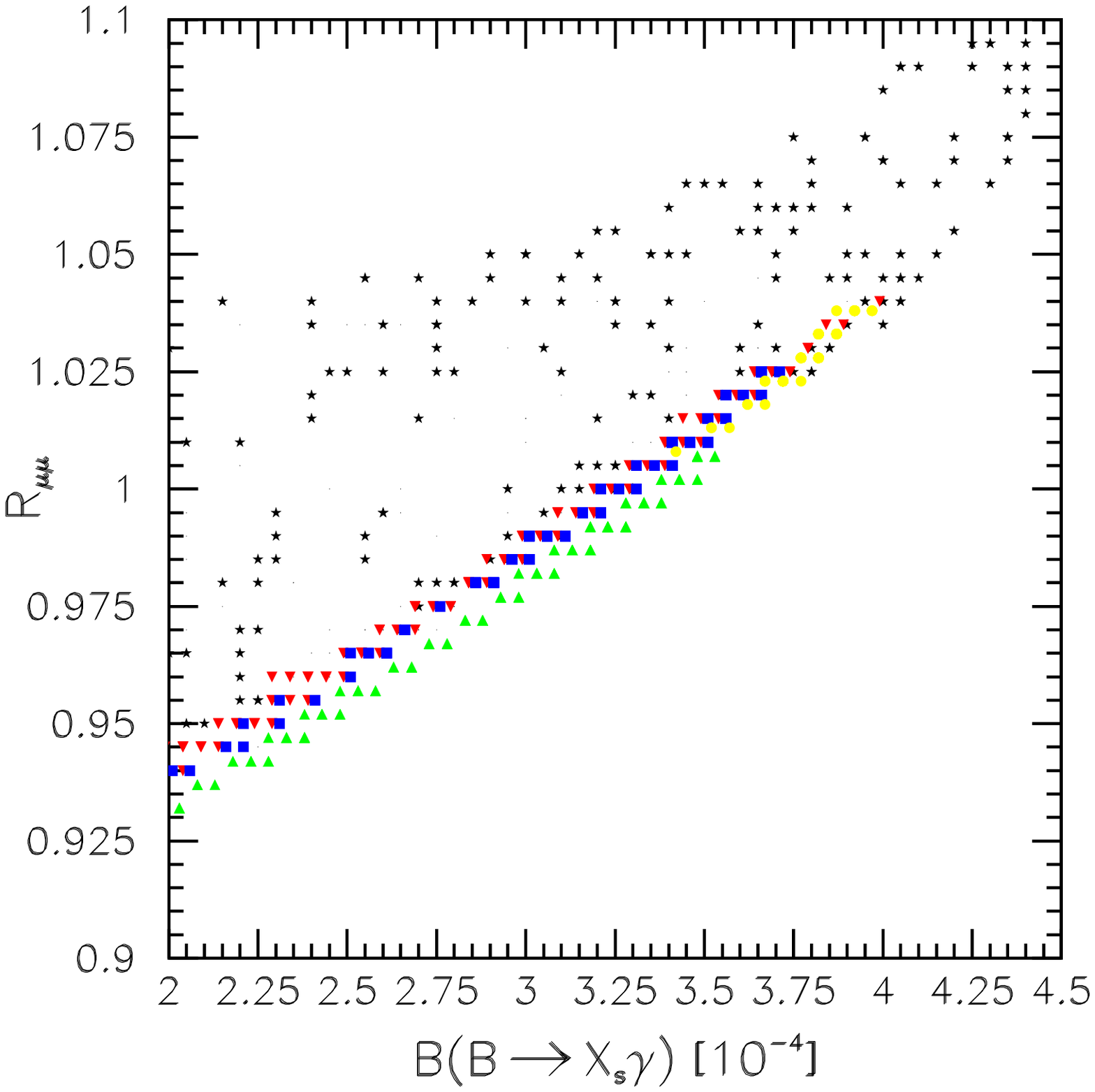}}
\caption{\label{fig:damsb2}
The correlations of 
(a) $Br ( B_s \rightarrow \mu^+ \mu^- )$ with $a_\mu$, and 
(b) $Br ( B \rightarrow X_s \gamma)$ with $R_{\mu\mu}$ 
in the deflected AMSB scenario with $M = 10^{12}$ GeV, 
$N=6$, $\rho = 0$ and $\mu > 0$.
The legends are the same as Fig.~\ref{fig:msugra2}.}
%\label{fig:damsb2}
\end{figure}

In Fig.~\ref{fig:damsb1}, we show the contour plots in the 
$( M_{\rm aux}, \tan\beta)$ plane with other parameters fixed to 
aforementioned  values.  In Figs.~\ref{fig:damsb2} (a) and (b),  the 
correlations between (a) $R_{\mu\mu}$ and $B(B\rightarrow X_s \gamma)$ and
(b) $B( B_s \rightarrow \mu^+ \mu^- )$ and $a_\mu$ in unit of $10^{-10}$ are  
shown. As expected, the sign flip of $M_3$ makes the model more consistent
with $B\rightarrow X_s \gamma$. Note that the muon anomalous MDM cannot be
greater than $22 \times 10^{-10}$, and $0.93 \leq R_{\mu\mu} \leq 1.10$.
On the other hand, $B ( B_s \rightarrow \mu^+ \mu^- )$ becomes as large as
$\sim 10^{-6}$, because

%%%%%%%%%%%%%%%%%%%%%%%%%%%%%%%%%%%%%%%%%%%\par
\subsection{Gaugino--assisted AMSB}
%%%%%%%%%%%%%%%%%%%%%%%%%%%%%%%%%%%%%%%%%%%\par

In the line of minimal AMSB, eventually it is inevitable to clarify how one 
could generate the universal soft scalar mass squared $m_0^2$ put in the 
minimal AMSB by hand. 
The gaugino--assisted anomaly mediation (gAMSB) gives a simple origin of 
$m_0^2$ without additional fields or symmetries below the Planck 
scale~\cite{gamsb}. The setup is keeping the model of AMSB in its original 
form, but placing the MSSM gauge and gauginos in the bulk. 
Under the assumption of no singlet in the hidden sector boundary, 
the gauginos get masses via AMSB dominantly whereas scalar masses get 
contributions from both AMSB and a tiny hard breaking of SUSY 
by some operators on the hidden sector. 
These operators contribute to scalar masses at 1-loop, and being dominant 
in most of parameter space, and the tachyonic sleptons are cured. 
In this regard, it is a hybrid of gaugino mediation and anomaly mediation. 

The soft terms at input scale are similar with mAMSB, but now the scalar 
masses get additional contribution which is not universal, but proportional 
to the matter gauge charges. Explicitly, we have
\bea
m^2 = \tilde{m}^2 + 
         2 \zeta(3)\Gamma(4) C(i) \frac{g^2}{16\pi^2}
         ~{1\over ( M_* L)^2}~m^2_{3/2} \,,
\eea
where $C(i)$ is the quadratic Casimir for the $i$ matter scalar 
representation, and the $\tilde{m}^2$ is the scalar masses in the pure 
AMSB scenario. The second term play the role of $m_0^2$ in the pure 
AMSB scenario where the tachyonic slepton problem is solved by adding
$m_0^2$. 

\begin{figure}
\includegraphics[width=0.8\textwidth]{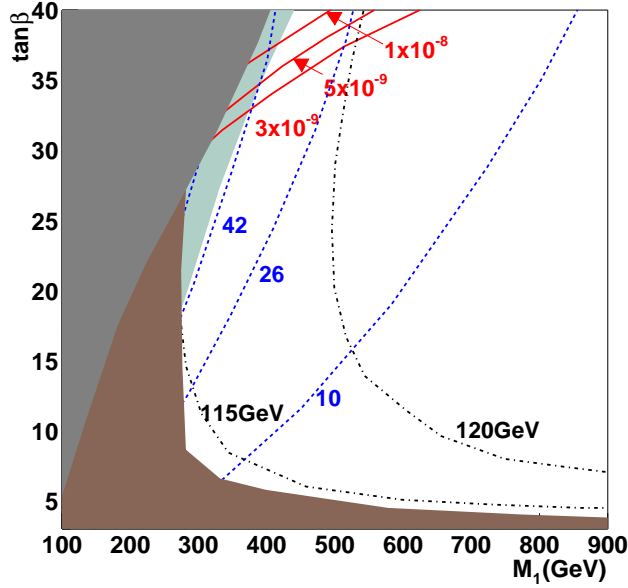}
\caption{\label{fig:gamsb1}
The contour plots for $a_\mu^{\rm SUSY}$ in unit of $10^{-10}$
(in the short dashed curves) and the Br ($B_s \rightarrow \mu^+ \mu^- $)
(in the solid curves) in the $( M_1, \tan\beta)$ plane for the 
gaugino--assisted AMSB scenarios with $\mu >0$.}
%\label{fig:gamsb1}
\end{figure}

\begin{figure}
\centering
\hspace{-1.5cm}
\subfigure[]{
\includegraphics[width=0.5\textwidth]%{figs/hetero_M_05/br_bs-mug23.ps}}
{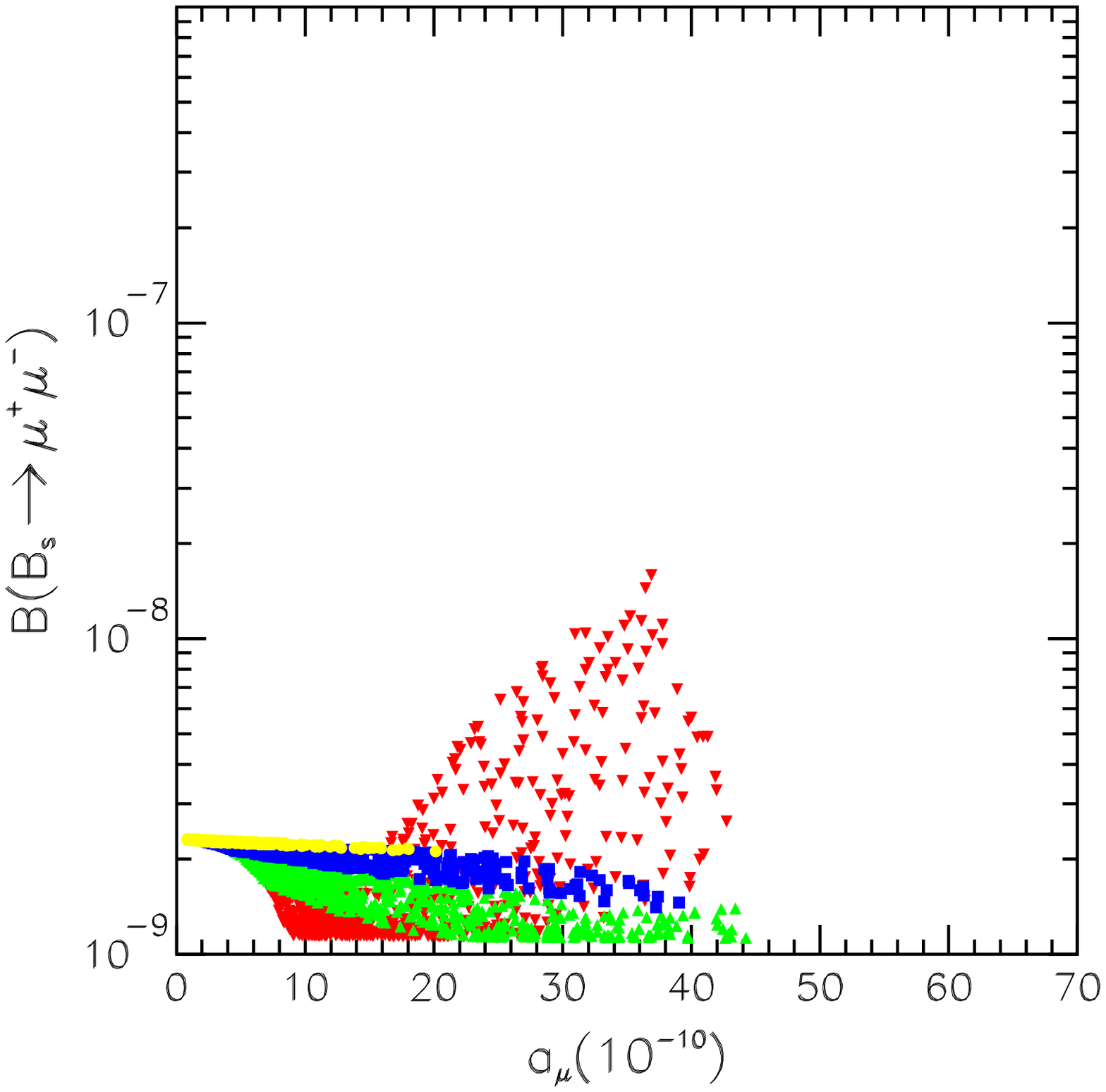}}
%\hspace{-1.5cm}
\subfigure[]{
\includegraphics[width=0.5\textwidth]%{figs/hetero_M_05/bsll-b2sr.ps}}
{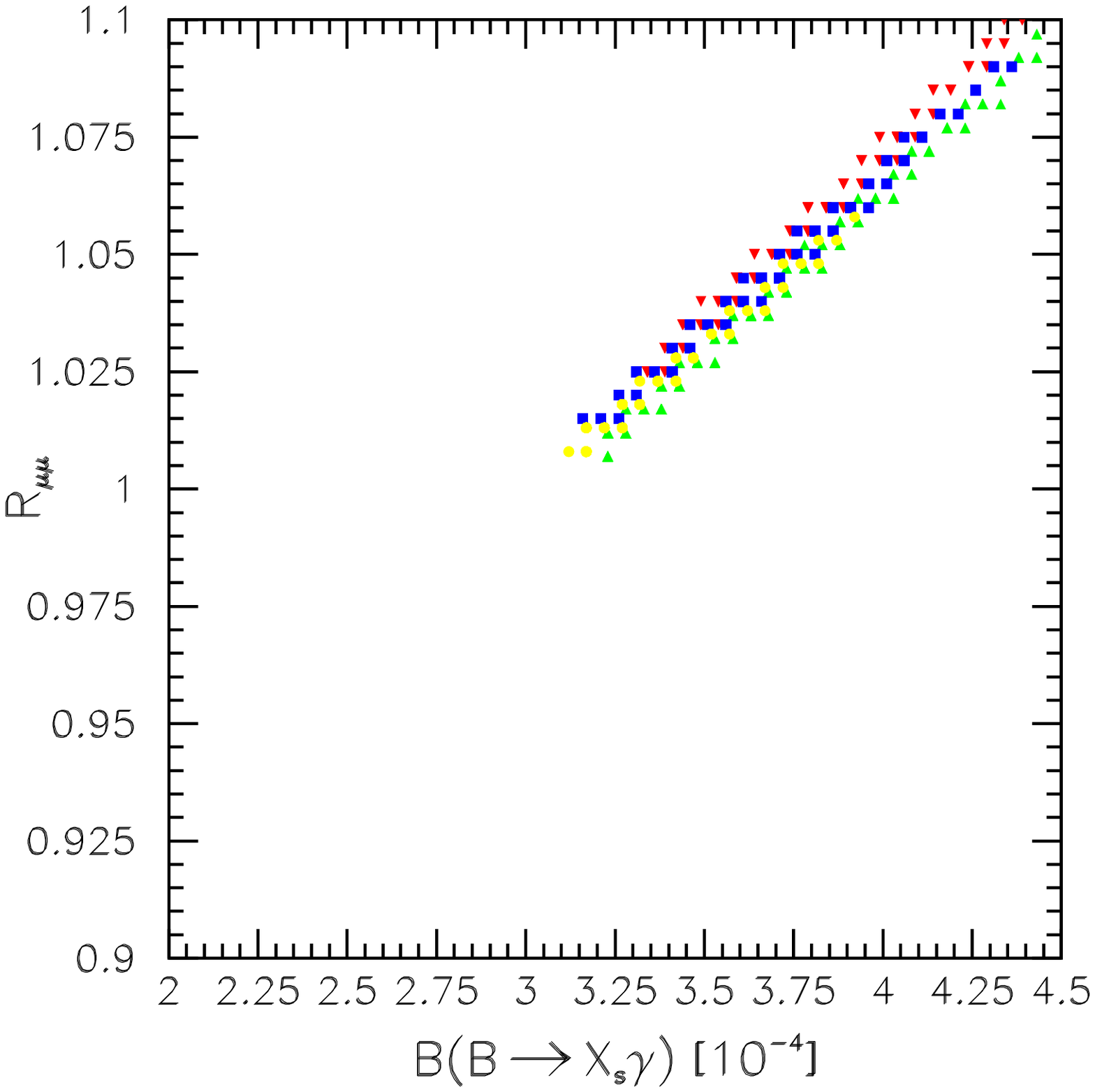}}
\caption{\label{fig:gamsb2}
The correlations of 
(a) $Br ( B_s \rightarrow \mu^+ \mu^- )$ with $a_\mu$, and 
(b) $Br ( B \rightarrow X_s \gamma)$ with $R_{\mu\mu}$ in the 
gaugino-assisted AMSB scenario. 
The legends are the same as Fig.~\ref{fig:msugra2}.}
%\label{fig:gamsb2}
\end{figure}

The parameter of the gaugino assisted AMSB scenario is the same as the mAMSB:
\[
m_{3/2},~~~\eta,~~~\tan\beta,~~~{\rm sign} (\mu) .
\]
We scan over 
\[
3 \leq \tan\beta \leq,~~~ 0 \leq m_{3/2} \leq 100~{\rm TeV},~~~
\eta = 1,~~~\mu > 0 .
\]
The qualitative features of the predictions in the gaugino-assisted AMSB 
scenario are similar to the mAMSB.
$B\rightarrow X_s \gamma$ gives a strong constraint, especially for large 
$\tan\beta$. We find $B ( B_s \rightarrow \mu^+ \mu^- ) \lesssim 1.6 \times 
10^{-8}$ scanning over the parameter space. The results are depicted in 
Fig.~\ref{fig:gamsb1} and Figs.~\ref{fig:gamsb2} (a) and (b).
The squark masses are generically large and their contributions to the decay
$B_s \rightarrow \mu^+ \mu^-$ is small. On the other hand, the charginos
and the sleptons are relatively light and can contribute to $a_\mu$ up to 
$45 \times 10^{-10}$. Finally we find $1.0 \leq R_{\mu\mu} \leq 1.1$ and
there is no large deviation from the SM prediction.

\subsection{Weakly interacting string models with dilaton/moduli mediations}

In the string theory, SUSY breaking is parametrized in terms of the nonzero
values of the auxiliary components of dilaton and overall modulus superfields
($S$ and $T$, respectively) \cite{brignole}:
\begin{eqnarray}
F^S & = & \sqrt{3} ( S + S^* ) m_{3/2}~\sin\theta,
\nonumber  \\
F^T & = & \sqrt{3} ( T + T^* ) m_{3/2}~\cos\theta.
\end{eqnarray}
Then universality of scalar fermion masses naturally follows in the dilaton
dominated SUSY breaking mechanism.
For weakly interacting heterotic string theories, the K\"{a}hler potential
and the gauge kinetic function of the 4-dimensional low energy effective
supergravity theory are given by \cite{brignole}
\begin{eqnarray}
  K & = & - \ln ( S + S^*) - 3 \ln ( T + T^* ) +
( T + T^* )^{n_i} \Phi_i \Phi_i^* ,
\nonumber
\\
f_{ab} & = & {\delta_{ab}  \over 4 \pi}~S.
\end{eqnarray}
Here $n_i$ is the modular weight of the MSSM superfield $\Phi_i$.
The soft terms at string scale can be derived from the above functions by
well known formulae \cite{brignole}. For $n_i = -1$ as an example, we have
\begin{equation}
  M_a = \sqrt{3} M_{\rm aux} = - A_{ijk},~~~
m_{ij}^2 = M_{\rm aux}^2 \delta_{ij}.
\end{equation}
Here $M_{\rm aux} = m_{3/2} \sin\theta$ where $\theta$ is the Goldstino angle
defined as $\tan\theta = F_S / F_T$.
This model is specified by three independent parameters :
\[
M_{\rm aux},~~~\tan\beta,~~~{\rm sign}(\mu).
\]
Note that the Goldstino angle $\theta$ does not appear as an observable at
this level.

In the dilaton domination scenario, one encounters the color charge breaking
minima and the unbounded from below directions in the effective potential,
if one starts the RG running from the usual GUT scale \cite{munoz,casas,abel}.
On the other hand, this problem can be evaded if one starts the RG running
from the lower scale, for example, from the intermediate scale
$M_{X} \sim 10^{11}$ GeV \cite{allanach}. The detailed phenomenology on 
$a_\mu^{\rm SUSY}$, $B\rightarrow X_s \gamma$ and the neutralino-nucleus 
scattering in the limit of the dilaton domination has been already 
discussed by two of us in Ref.~\cite{kolee} both for $M_{\rm string}$ equal 
to the usual GUT scale and the intermediate string scale. 
In this work, we ignore the CCB and UFB problems
and assume that the soft parameters are given at the conventional GUT scale
$M_{\rm GUT} = 2 \times 10^{16}$ GeV, and will discuss other processes
$B\rightarrow X_s l^+ l^-$ and $B_s \rightarrow \mu^+ \mu^-$.
Earlier phenomenological analysis of weakly interacting string theories with
dilaton domination scenarios can be found on the muon
$(g-2)_\mu$ \cite{kchoi,martin}, $B\rightarrow X_s \gamma$. 
Discussions on $B\rightarrow X_s l^+ l^-$ and 
$B_s \rightarrow \mu^+ \mu^-$ in the weakly interacting string theories are
given in this work for the first time.

\begin{figure}
\includegraphics[width=0.8\textwidth]{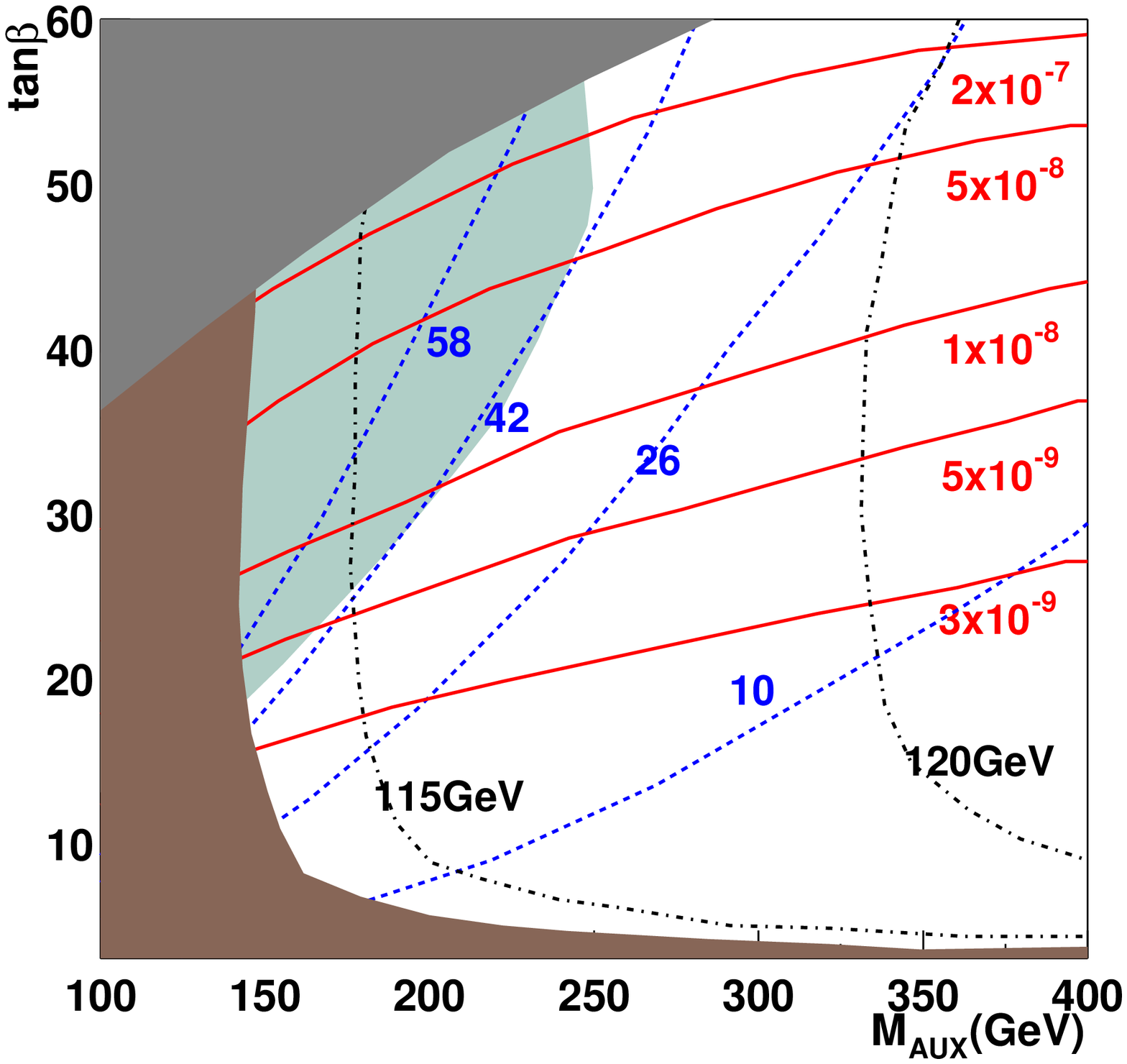}
\caption{\label{fig:dilaton1}
The contour plots for $a_\mu^{\rm SUSY}$ in unit of $10^{-10}$
(in the short dashed curves) and the Br ($B_s \rightarrow \mu^+ \mu^- $)
(in the solid curves) in the $( M_{\rm aux} , \tan\beta)$ plane for the
dilaton dominated scenario.
The light gray region is excluded by the light stau mass
bound, and the green region is excluded by the lower bound to the
$B\rightarrow X_s \gamma$ branching ratio.}
%\label{fig:dilaton1}
\end{figure}

In Fig.~\ref{fig:dilaton1}, we show the constant contour plots for
$a_\mu^{\rm SUSY}$ in unit of $10^{-10}$ (in the short dashed curves) and
the Br ($B_s \rightarrow \mu^+ \mu^- $) (in the solid curves) in the
$( M_{\rm aux}, \tan\beta)$ plane. In this scenario, the $a_\mu^{\rm SUSY}$
can be as large as $50 \times 10^{-10}$ without any conflict with other
constraints. The branching ratio for the decay $\bsmm$ can be as large
as $2 \times 10^{-7}$.
Therefore the upcoming Tevatron Run II can probe a large portion of the
parameter space of this scenario (down to $\tan\beta \sim 30$).
We also find that $R_{\mu\mu}$ can vary between 0.95 and 1.15, which has a
correlation with $B\rightarrow X_s \gamma $ similar to Fig.~\ref{fig:msugra2}.
Other comments are similar to the mSUGRA case.

\subsection{Heterotic $M$ theory with dilaton/moduli mediations}

Following the pioneering works of Witten and Horava and Witten \cite{horava}, 
five different perturbative string theories are
now regarded as different facets of one fundamental theory called $M$ theory,
which describe the string theory in the strong coupling limit.
The low energy limit of the $M$ theory is believed to be the 11-D SUGRA theory.
Compactified on the orbifold $S^1 / Z_2$ of the length $\pi\rho$ with two
10-dim. branes at the orbifold fixed points with two $E_8$ gauge groups
living on each brane, this theory can accommodate the unification of three
gauge coupling and Newton's constant for gravity by adjusting the length of
the 11-th dimensional orbifold. Further compactifying the 10-dim branes to
4 dimensional Minkowski space and Calabi-Yau or orbifolds with volume $V$,
one can derive 4-dim low energy effective SUGRA from Horava-Witten theory.
Note that there are three independent scales: $\kappa^2 = M_{11}^{-9}$ (where
$\kappa^2$ and $M_{11}$ being the Newton's constant and the 11-dim. Planck
constant), $\pi\rho$ (the length of the $S^1 / Z_2$ orbifold interval) and
$V$ (the volume of the 6-dim. internal space). There will be two model
independent moduli superfields $S$ and $T$, whose scalar components satisfy
\begin{eqnarray}
{\rm Re} (S) & = & {1 \over 2 \pi (4 \pi)^{2/3} }~M_{11}^6 V,
\nonumber  \\
{\rm Re} (T) & = & {6^{1/3} \over 2 ( 4 \pi)^{4/3} } ~M_{11}^3 V^{1/3}
\pi\rho.
\end{eqnarray}
In case there are more $T$ moduli, the scalar masses become nonuniversal 
in general, and SUSY flavor problem may get worse.  The RG running effects 
involving the gluino mass parameter can mitigate this problem to some extent. 
In the following we take a simple framework in which the scalar mass term 
is universal from the outset. This assumption is well justified if we 
consider the compactified space is the Calabi-Yau manifold with Hodge-Betty 
number $h_{1,1} = 1$. 

For such strongly interacting string models (or M theories), the K\"{a}hler
potential and the gauge kinetic function of the low energy effective 
supergravity theories are given by \cite{mtheory}
\begin{eqnarray}
  K & = & - \ln ( S + S^*) - 3 \ln ( T + T^* ) + \left( {3 \over T + T^*}
 + {\alpha \over S + S^* } \right) ~\Phi_i \Phi_i^* ,
\nonumber
\\
f_{ab} & = & { \delta_{ab} \over 4 \pi}~( S + \alpha T ).
\end{eqnarray}
The soft terms at the string scale are derived from the above functions
as follows \cite{ckm}:
\begin{eqnarray}
  M_a & = & {\sqrt{3} m_{3/2} \over 1 + \epsilon}~\left[ \sin\theta
+ { \epsilon \over \sqrt{3} } \cos\theta \right],
\nonumber  \\
 A_{ijk} & = & - {\sqrt{3} m_{3/2} \over 3 + \epsilon }~
\left[ ( 3 - 2 \epsilon) \sin\theta + \sqrt{3} \epsilon \cos\theta \right],
\nonumber  \\
m_{ij}^2 & = & m_{3/2}^2 \delta_{ij} \left[ 1 - {3 \over (3 + \epsilon )^2 }~
\left( \epsilon ( 6 + \epsilon ) \sin^2 \theta + ( 3 + 2 \epsilon ) \cos^2
\theta - 2 \sqrt{3} \epsilon \sin\theta \cos\theta \right) \right].
\end{eqnarray}
Here $\theta$ is the Goldstino angle as before, and
\[
\epsilon \equiv \alpha ( T + T^* ) / ( S + S^* ).
\]
Therefore there are five independent input parameters in the heterotic $M$
theory :
\[
m_{3/2},~~~ \sin\theta,~~~\epsilon,~~~ \tan\beta,~~~{\rm sign}(\mu).
\]
Note that the universality in the scalar masses and gaugino masses as well
as trilinear couplings are realized in this scenario, which are functions
of the Goldstino angle $\theta$ and the parameter $\epsilon$. In most
parameter space, one has $M_a^2 > m_{ij}^2$ at the string scale. 
In the heterotic $M$ theory, one recovers the dilaton domination scenario
in the limit of $\epsilon \rightarrow 0$, namely $( T+T^* ) << ( S + S^* )$.
The parameter $\epsilon$ lies in the range $0 < \epsilon < 1$ for the
standard embedding of the spin connection, but it can take a negative value 
for the nonstandard embedding. 
For the latter the gaugino mass is even larger than the scalar masses.
%(???) and the allowed parameter space becomes larger/smaller  (????). 
Overall phenomenology of this scenario for low energy processes is more or
less the same as the mSUGRA or the dilaton domination scenarios.
Let us make a comment that the problem of CCB and UFB are solved in
this scenario in a wide region of parameter space $\epsilon$ and $\theta$
\cite{mtheory}.

Let us make two comments on the phenomenological niceties of the heterotic 
$M$ theory compared to the weakly interacting heterotic string theory other
than the unification of the gauge coupling and Newton's constant: 
\begin{itemize}
\item Although we do not care about the CCB and the UFB problems in this 
work, it is worthwhile to note that the case with $0 < \epsilon < 1$ (the 
standard embedding) has the CCB and UFB problems and there is no parameter
space left if the top mass is to be reproduced, whereas the nonstandard 
embedding (for which $-1 < \epsilon < 0$) has no such a problem. 
\item The limit $\theta \rightarrow 0 $ exists in the heterotic 
$M$ theory (except for $\epsilon=0$) with  soft masses remaining finite, 
%\begin{eqnarray}
%  M & \rightarrow & m_{3/2}~{\epsilon \over 1 + \epsilon},
%\nonumber  \\
%  m & \rightarrow & m_{3/2}~{\epsilon \over 3 + \epsilon},
%\nonumber  \\
 % A & \rightarrow & - m_{3/2}~{3 \epsilon \over 3 + \epsilon},
%\end{eqnarray}
all of which are order of $\epsilon~ m_{3/2}$. 
On the other hand, the soft terms in the weakly interacting heterotic string 
theory considered in the previous subsection vanish in the limit 
$\theta \rightarrow 0$. Therefore one has to include the string one loop 
corrections or the sigma model one loop corrections to the K\"{a}hler 
potential and the gauge kinetic function.
\end{itemize}
Note that this scenario is a special case of mSUGRA scenario except that
the gaugino mass parameter can change the sign depending on $\epsilon$ and
$\theta$. 
Earlier phenomenological analysis of heterotic $M$ theories can be found 
on the muon $(g-2)_\mu$ \cite{kchoi,cerdeno}, $B\rightarrow X_s \gamma$ 
\cite{Huang:1998tn,kchoi}, and 
and $B\rightarrow X_s \tau^+ \tau^-$ \cite{Huang:1998tn}. 
The discussion on %$B\rightarrow X_s l^+ l^-$ and 
$B_s \rightarrow \mu^+ \mu^-$ in the heterotic $M$ theories is given in 
this work for the first time.

In the heterotic $M$ theory, the universal gaugino mass $M_a$ is dominant 
over the common scalar mass $|m_{ij}|$ at the messenger scale. Then at the 
electroweak scale,  $m_{\tilde{\tau}}^2 < 0$ in most region of parameter 
space for large $\tan\beta$ except for very narrow range of $\epsilon$ and 
$\theta$. Also, for $\theta > \pi$, the gaugino mass parameter changes 
its sign as in the AMSB scenario, and the HRS effect comes into play for 
positive $\mu$.  With this general comment in mind, we fix 
$\theta = 0.15 \pi$ and show the contour plots for $\epsilon = +0.5$ and 
$-0.8$ in the $( M_1 , \tan \beta )$ plane, and various correlations 
in Figs.~17--20.  
Note that the common gaugino mass parameter $M_a$ can be negative altogether 
for certain range of $\theta$ for a fixed $\epsilon$. Then the situation 
would be similar to the AMSB scenario where $M_3 < 0$. 
However, in the heterotic $M$ theory, all the three gaugino mass parameter 
changes the signs unlike the AMSB case where only $M_3$ change its sign. 
Therefore, $a_\mu^{\rm SUSY} >0$ implies $\mu M_2 >0$ and 
$B\rightarrow X_s \gamma$ prefers $\mu M_3 > 0$. 
There is no problem with satisfying both constraints if we flip the sign of 
$\mu$ for negative gaugino mass parameter. %For the negative $\mu$, the 
%allowed parameter space will be reflected with respect to $\theta = ???$.

\subsection{$D$ brane models}

Advances in understanding the role of $D-$branes in superstring theories
brought new ideas in particle physics model buildings. Several attempts
have been made to obtain (semi)realistic 4-dimensional models (SM, MSSM or
their variations) \cite{ibanez,dbrane}. 
SM gauge groups and matters can be put on the same or
different branes, according to which the patterns of the resulting soft SUSY
breaking terms can differ.  In this subsection, we choose a specific 
$D$ brane model where the SM gauge groups and 3 generations live on 
different $Dp$ branes \cite{dbrane}.
In this model, scalar fermion masses are not completely universal and
gaugino mass unification can be relaxed. Also the string scale is around
$10^{12}$ GeV (the intermediate scale) rather than GUT scale.

Since there are now three moduli ($T_i$) fields and one dilaton superfield
in this scenario, we modify the parametrization appropriate for several $T_i$
moduli as follows:
\begin{eqnarray}
  F^S & = & \sqrt{3}~( S + S^* )~ m_{3/2} \sin\theta,
\nonumber   \\
  F^i & = & \sqrt{3}~( T_i + T_i^* )~ m_{3/2} \cos\theta~\Theta_i
\end{eqnarray}
where $\theta$ and $\Theta_i ~(i = 1,2,3)$ with $| \Theta_i |^2 = 1$
parametrize the directions of the goldstinos in the $S, T_i$ field space.
Then, the gaugino masses are given by
\begin{eqnarray}
M_3 & = & \sqrt{3} m_{3/2} \sin\theta ,
\nonumber   \\
M_2 & = & \sqrt{3} m_{3/2} \Theta_1 \cos\theta ,
\nonumber   \\
M_Y & = & \sqrt{3} m_{3/2} \alpha_Y ( M_I ) ~\left( { 2 \Theta_3 \cos\theta
\over \alpha_1 ( M_I ) } + {\Theta_1 \cos\theta \over \alpha_2 ( M_I )}
+ {2 \sin\theta \over 3 \alpha_3 ( M_I ) } \right) ,
\end{eqnarray}
where
\begin{equation}
  {1\over \alpha_Y ( M_I ) } = { 2 \over \alpha_1 ( M_I ) }
+ { 1 \over \alpha_2 ( M_I ) } + {2 \over 3 \alpha_3 ( M_I ) }.
\end{equation}
The string scale $M_I$ is determined to be
$M_I = 10^{12} ~(5 \times 10^{14} )$ GeV from the $U(1)_1$ gauge coupling
$\alpha_1 ( M_I ) = 0.1 (1)$ \cite{dbrane}.  
Note that the gaugino masses are non universal in a natural way in this 
scenario, unlike other scenarios studied in the previous subsections.

The soft masses for the sfermions and Higgs fields are given by
\begin{eqnarray}
  m_Q^2     & = & m_{3/2}^2~ \left[~ 1 - {3\over 2} \left( 1 - \Theta_1^2
\right)~ \cos^2 \theta~  \right],
\nonumber   \\
  m_{u^c}^2 & = & m_{3/2}^2~ \left[~ 1 - {3\over 2} \left( 1 - \Theta_3^2
\right)~ \cos^2 \theta~  \right],
\nonumber   \\
  m_{d^c}^2 & = & m_{3/2}^2~ \left[~ 1 - {3\over 2} \left( 1 - \Theta_2^2
\right)~ \cos^2 \theta~  \right],
\nonumber   \\
  m_L^2     & = & m_{3/2}^2~ \left[~ 1 - {3\over 2} \left( \sin^2 \theta
+ \Theta_3^2 ~\cos^2 \theta \right) \right],
\nonumber   \\
  m_{e^c}^2 & = & m_{3/2}^2~ \left[~ 1 - {3\over 2} \left( \sin^2 \theta
+ \Theta_1^2 ~\cos^2 \theta \right) \right],
\nonumber   \\
  m_{H_2}^2 & = &  m_{3/2}^2~ \left[~ 1 - {3\over 2} \left( \sin^2 \theta
+ \Theta_2^2 ~\cos^2 \theta \right) \right],
\nonumber   \\
  m_{H_1}^2 & = & m_L^2.
\end{eqnarray}
Note that the scalar mass universality in the sfermion masses and Higgs
masses is achieved when
\begin{equation}
\sin^2 \theta = {1\over 4} ~~~{\rm and}~~~ \Theta_i^2 = {1 \over 3}~~~
{\rm for}~i = 1,2,3.
\end{equation}
And in this case the gaugino masses becomes also universal, when we take
only positive numbers for the solutions. For other choices of goldstino
angles, the scalar and the gaugino masses become nonuniversal, and there
could be larger flavor violations in the low energy processes as well as
enhanced SUSY contributions to the $a_\mu^{\rm SUSY}$.

The trilinear couplings are given by
\begin{eqnarray}
  A_u & = & {\sqrt{3} \over 2}~m_{3/2}~\left[
( \Theta_2 - \Theta_1 - \Theta_3 ) \cos\theta - \sin\theta \right],
\nonumber   \\
  A_d & = & {\sqrt{3} \over 2}~m_{3/2}~\left[
( \Theta_3 - \Theta_1 - \Theta_2 ) \cos\theta - \sin\theta \right],
\nonumber   \\
  A_e & = & 0.
\end{eqnarray}
Therefore the $D$ brane model we consider in this work is specified by
following six parameters :
\[
m_{3/2},~~\tan\beta,~~\theta,~~\Theta_{i=1,2},~~ {\rm sign} (\mu).
\]
Earlier phenomenological analysis of $D$ brane models can be found on 
the muon $(g-2)_\mu$~\cite{munoz2}.%  $B\rightarrow X_s \gamma$.
The discussion on  $B\rightarrow X_s \gamma$, $B\rightarrow X_s l^+ l^-$ and 
$B_s \rightarrow \mu^+ \mu^-$ in this scenario is given in the present work 
for the first time.

For numerical analysis, we fix $\Theta_i = + 1/\sqrt{3}$ for all $i=1,2,3$,
(the overall modulus limit) and we scan over the following parameter space :
$-\pi/4 \leq \theta \leq 4/ \pi$, $m_{3/2} \leq 300$ GeV, and
$\tan\beta \leq 50$.  In this case, the universality in sfermion and Higgs
masses parameters at the string scale is moderately broken. Still there
remain certain degrees of degeneracy: the squark masses are universal at
string scale $M_I$, the sleptons and the down type Higgs ($H_2$)
masses are the same, and the up type ($H_1$) Higgs are degenerate.
The gaugino masses are nonuniversal for this choice of parameters.
The point corresponding to the universality in the scalar and gaugino
masses is denoted by the filled triangle. Another interesting aspect of this
model is that the gluino mass parameter $M_3$ can have either sign for
$- \pi/4 < \theta <0$ as in the AMSB model, and the correlation between
$a_\mu^{\rm SUSY}$ and $B\rightarrow X_s \gamma$ resembles that of the AMSB
scenario.

In Figs.~\ref{fig:dbrane} (a) and (b), we show the correlations between
(a) $a_\mu^{\rm SUSY}$ and $B( \bsmm )$ and (b) $R_{\mu\mu}$ and
$B ( B \rightarrow X_s \gamma )$, respectively. Both $a_\mu^{\rm SUSY}$ and
$B( \bsmm )$ can be large for large $\tan\beta$, as in the mSUGRA.
In particular, $a_\mu^{\rm SUSY}$ can be as large as $70 \times 10^{-10}$,
unlike the minimal SUGRA model with $m_0 = 300$ GeV for which
$a_\mu^{\rm SUSY}$ is limited only to $32 \times 10^{-10}$. Also
$B ( \bsmm ) > 2 \times 10^{-8}$ is possible in an ample region of the
parameter space as in the mSUGRA. The fact that $M_3$ can change its sign
shows itself in the correlation in Fig.~\ref{fig:dbrane} (b). $R_{\mu\mu}$
can either decrease down to 0.86 or increase up to 1.15, depending on the
sign of $M_3$. Still the deviation from the SM is not significant, and
it would not be easy to observe this effect from $R_{\mu\mu}$.

On the other hand, one may assume that all the SM gauge groups are  embedded
within the same set of $D_p$ branes \cite{dbrane}. 
For this case, a salient feature is that there appear Higgs doublets 
$H_1^i$ and $H_2^i$ come in three generations.
Therefore there could be large FCNC contributions due to scalar exchanges,
unless one removes flavor changing neutral current interactions by imposing
some discrete symmetry. Although the soft terms for these models are known,
it is too premature to study the detailed phenomenology of this class of
models, before we know well enough how to handle this flavor changing
scalar interactions.

%%%%%%%%%%%%%%%%%%%%%%%%%%%%%%%%%%%%%%%%%%%%%%%%%%%%%%%%%%%%%%%%%%%%%%%%%%%%

\section{Conclusions}

In conclusion, we considered four low energy processes $(g-2)_{\mu}$,
$B\rightarrow X_s \gamma$, $B\rightarrow X_s l^+ l^-$ and
$B_s \rightarrow \mu^+ \mu^-$ in various models for SUSY breaking mediation
mechanisms which are theoretically well motivated. Since many models predict
universal scalar masses at the messenger scale, the RG running induces very
important features for the stop mass and the $A_t$ parameter depending on
the location of the messenger scale.
If the messenger scale is high around GUT scale, then the lower
bound on the $B\rightarrow X_s \gamma$ branching ratio is constraining for
the $\mu > 0$ case, because the chargino-stop loop contribution interfere
destructively with the SM and the charged Higgs contributions (except for
the AMSB scenario). The $A_t$ term relevant to the stop-chargino loop
contribution is generated mainly by the gluino loop by RG running effects.
On the other hand, in the GMSB scenario with low messenger scales
($M_{\rm mess} \sim 10^6 $ GeV or so), the stop mass is relatively heavy,
and the $A_t$ parameter is very small so that the stop - chargino
contribution to $B \rightarrow X_s \gamma$ is negligible. This is the reason
why the $\bsmm$ branching ratio is much suppressed in the GMSB with low
$M_{\rm mess}$ and $N_{\rm mess}$. This decay is also suppressed in the AMSB
scenario because the stop is relatively heavy in this scenario.
In fact, the branching ratio for $B_s \rightarrow \mu^+ \mu^-$ cannot be
larger than $2 \times 10^{-8}$ for these low messenger scale GMSB scenarios
with $N_{\rm mess} = 1$ or in the AMSB scenario. On the other hand, its
branching ratio can be much larger for mSUGRA or string inspired models where
the messenger scale is around the GUT scale. The Tevatron Run II can
probe the $B_s \rightarrow \mu^+ \mu^-$ decay mode down to 
$ \sim 2 \times 10^{-8}$ level in the branching ratio.
Therefore if $\bsmm$ is discovered at the Tevatron Run II, then the AMSB or
the GMSB with small $N_{\rm mess}$ will be definitely
excluded independent of the direct searches of SUSY particles. With the new
lower limits on Higgs (and SUSY particles), there is little chance to expect
large deviations in $R_{\mu\mu}$ from its SM prediction $R_{\mu\mu} =1$.
If any significant deviation in $R_{\mu\mu}$ is observed at $B$ factories,
it would reject all the SUSY breaking mediation scenarios we have considered
in this work.

%%%%%%%%%%%%%%%%%%%%%%%%%%%%%%%%%%%%%%%%%%%%%%%%%%%%%%%%%%%%%%%%%%%%%%%%%%

\begin{acknowledgments}
This work is supported in part 
by KOSEF through CHEP at Kyungpook National University (PK) and 
by KRF PBRG grant KRF-2002-070-C00022 (WS). 
\end{acknowledgments}

\vspace{1.5cm}

{\it Note Added}

While this work is being finished, a new result on the muon $(g-2)$ was 
reported by BNL Muon $(g-2)$ Collaboration \cite{bnl02}, and a few related 
works \cite{nath02,davier02} thereafter. 
The new result implies that \cite{davier02}
\[
a_\mu^{\rm exp} - a_\mu^{\rm SM} =  33.9 (11.2) ~~{\rm to}~~ 16.7 (10.7)   
\times 10^{-10} \equiv a_\mu^{\rm SUSY}, 
\]
depending on how the hadronic contributions are treated: the first and the
second numbers are based on the $e^+ e^- \rightarrow hadrons$ and 
the hadronic $\tau$ decays, respectively.  
Another calculation by Hagiwara et al. \cite{hagiwara} also indicates that
the SM prediction is 2.7 $\sigma$ below the experimental value.
This new data do not affect the conclusons of the present work very much.
In particular, there is still a possibility that the $\bsmm$ branching ratio
can be large enough to be found at the Tevatron Run II, if we allow the 
$3\sigma$ range for the $a_\mu^{\rm SUSY}$ (see Fig,~2 (a), for example). 

%\vspace{-1.5cm}

%\vspace{-1.5cm}

\clearpage
\textwidth=12cm

\vspace{-2cm}

\clearpage

\begin{figure}
\includegraphics[width=0.8\textwidth]{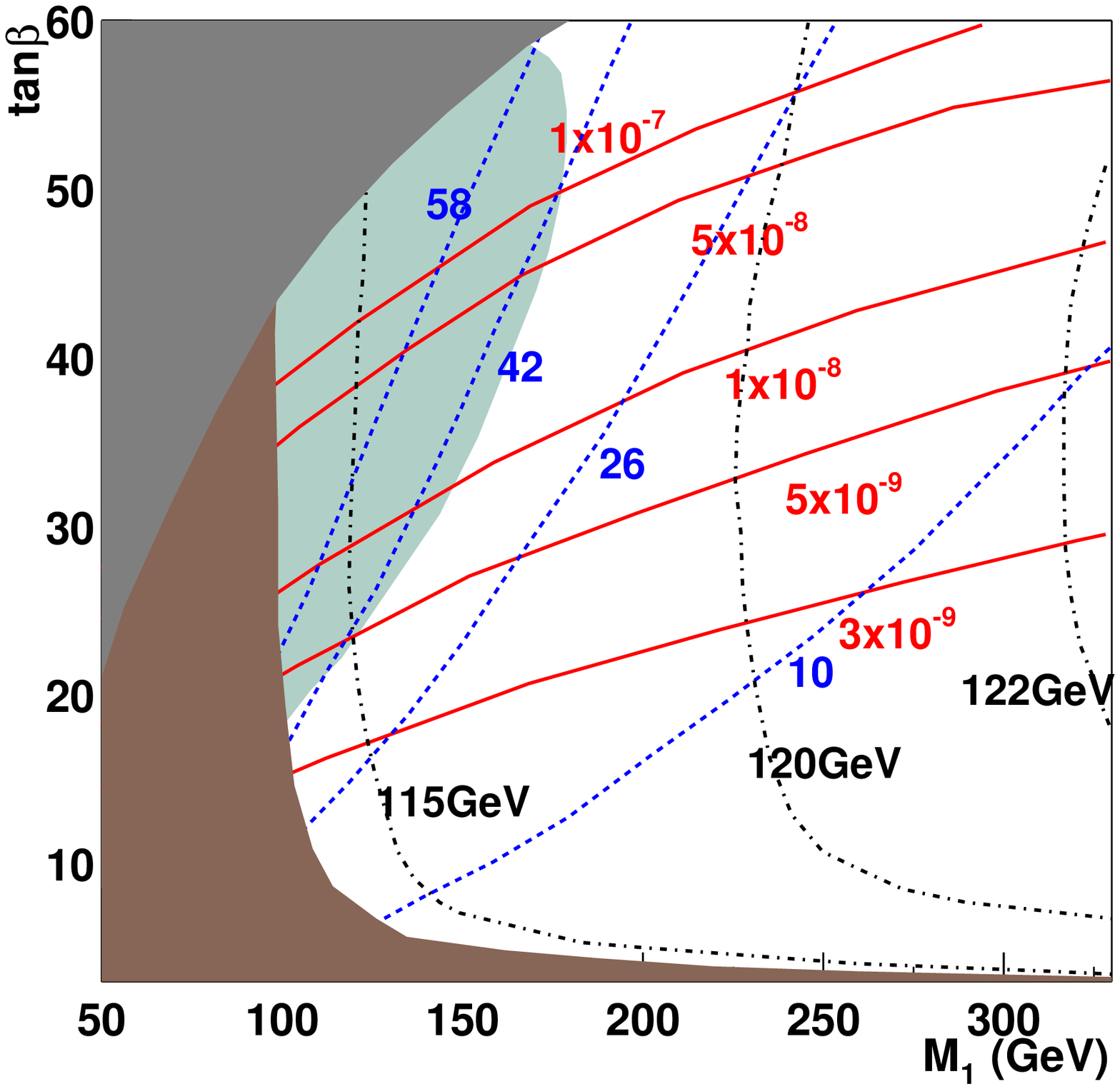}
\caption{\label{fig:heterop1}
The contour plots for $a_\mu^{\rm SUSY}$ in unit of $10^{-10}$
(in the short dashed curves) and the Br ($B_s \rightarrow \mu^+ \mu^- $)
(in the solid curves) in the $( M_1 , \tan\beta)$ plane for the heterotic
M theory with $\epsilon = +0.5$ and $\theta = 0.15 \pi$. }
%\label{fig:heterop1}
\end{figure}

\begin{figure}
\centering
\hspace{-1.5cm}
\subfigure[]{
\includegraphics[width=0.5\textwidth]%{figs/hetero_M_05/bsll-b2sr.ps}}
{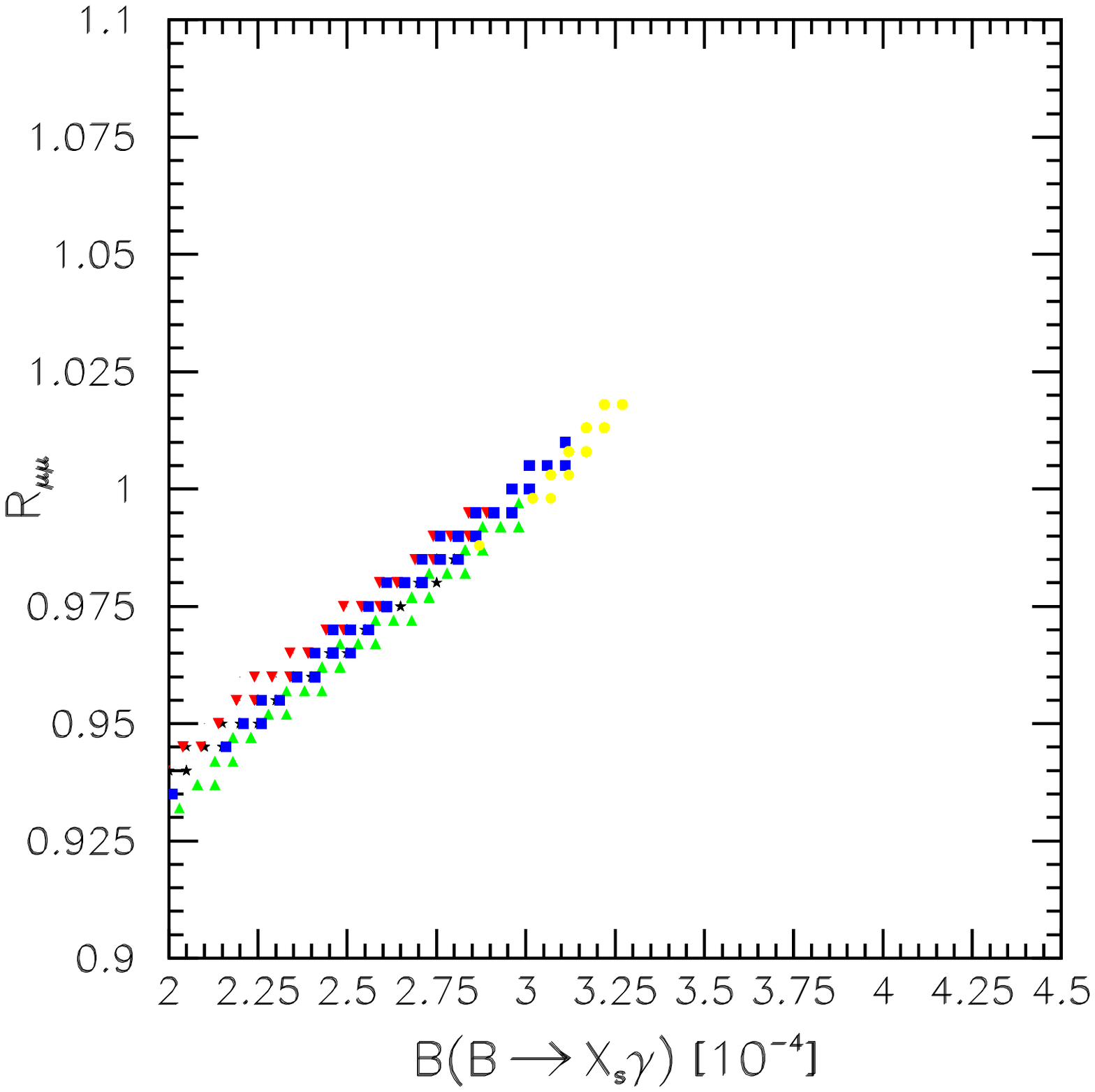}}
%\hspace{-1.5cm}
\subfigure[]{
\includegraphics[width=0.5\textwidth]%{figs/hetero_M_05/br_bs-mug23.ps}}
{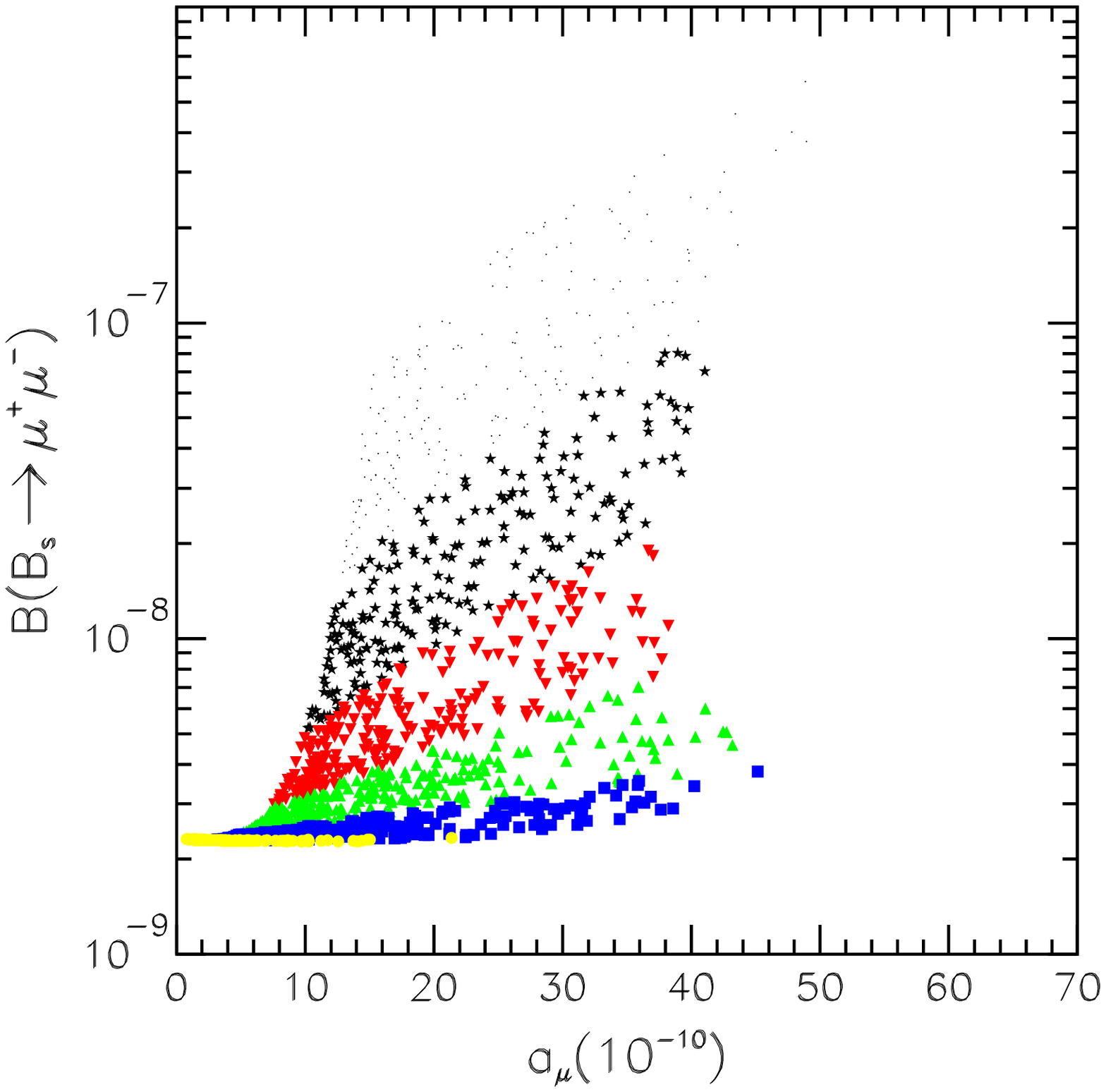}}
\caption{\label{fig:heterop2}
The correlations of the $Br ( B \rightarrow X_s \gamma)$ with
(a) $R_{\mu\mu}$, and (b) $a_\mu^{\rm SUSY}$ for the heterotic
M theory with $\epsilon = +0.5$ and $\theta = 0.15 \pi$. 
The legends are the same as Fig.~\ref{fig:msugra2}.
}
%\label{fig:heterop2}
\end{figure}

\begin{figure}
\includegraphics[width=0.8\textwidth]{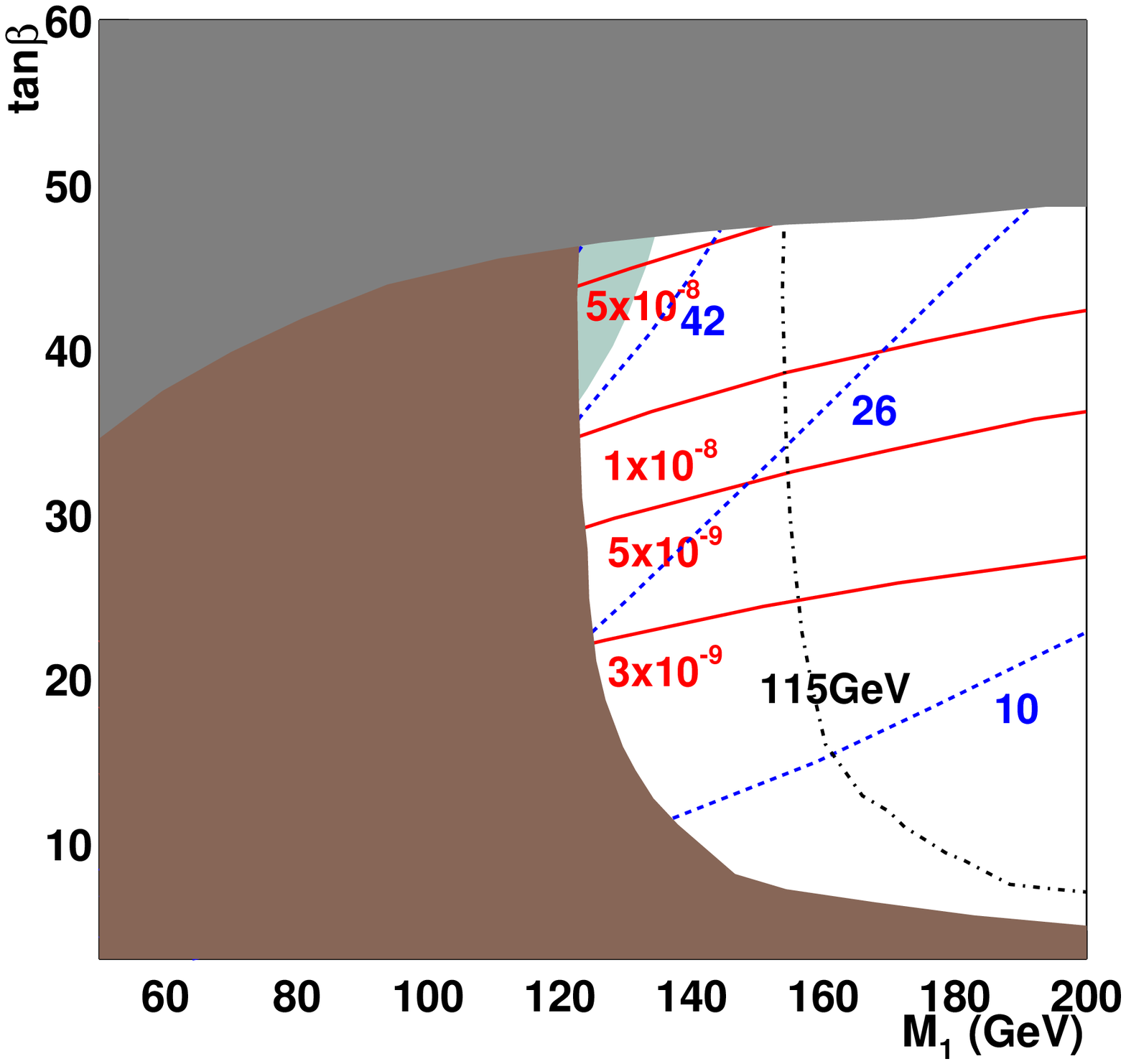}
\caption{\label{fig:heterom1}
The contour plots for $a_\mu^{\rm SUSY}$ in unit of $10^{-10}$
(in the short dashed curves) and the Br ($B_s \rightarrow \mu^+ \mu^- $)
(in the solid curves) in the $( M_1 , \tan\beta)$ plane for the heterotic
M theory with $\epsilon = -0.8$ and $\theta = 0.15 \pi$. }
%\label{fig:heterom1}
\end{figure}

\begin{figure}
\centering
\hspace{-1.5cm}
\subfigure[]{
\includegraphics[width=0.5\textwidth]%{figs/hetero_M_m08/bsll-b2sr.ps}}
{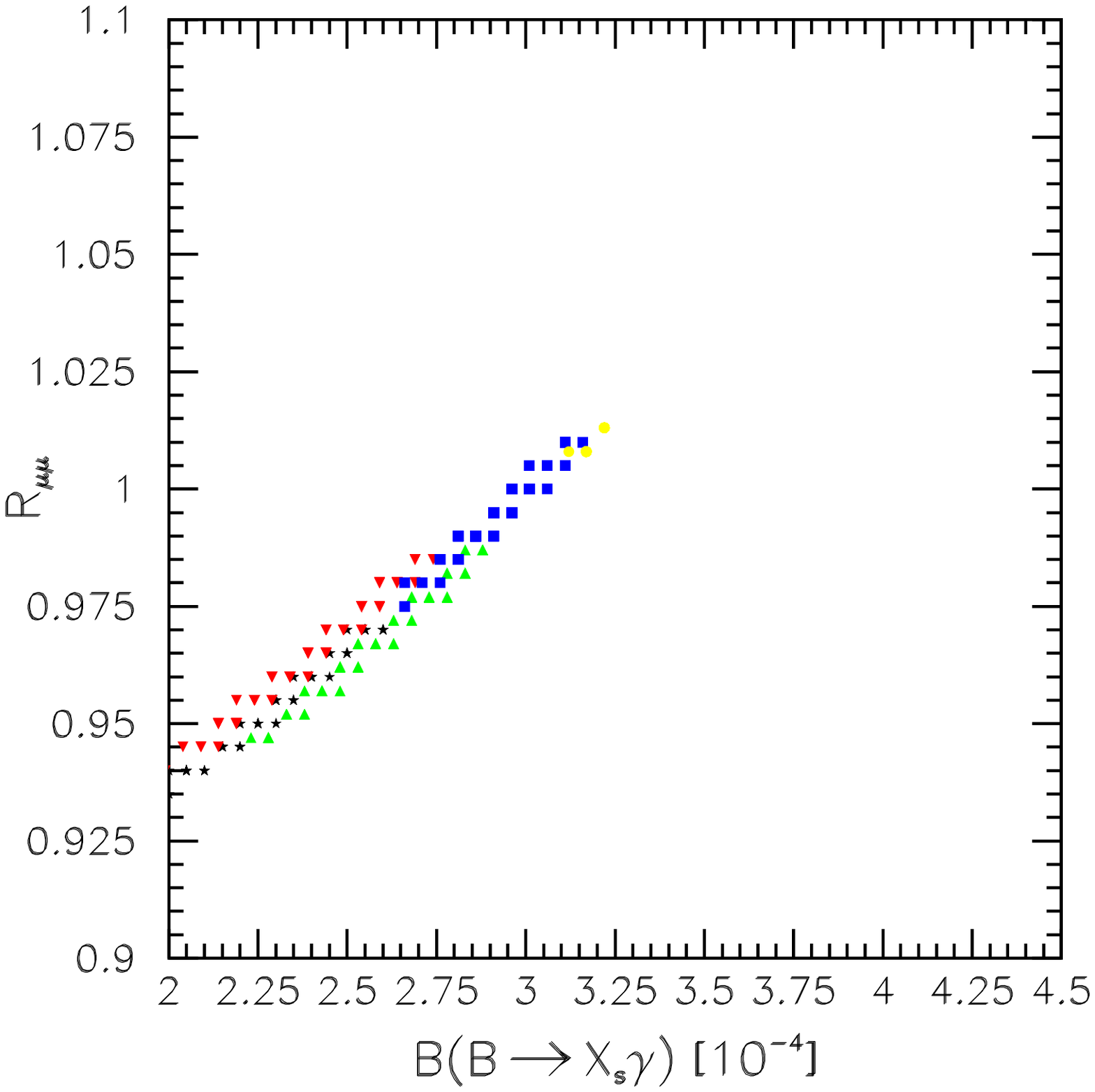}}
%\hspace{-1.5cm}
\subfigure[]{
\includegraphics[width=0.5\textwidth]%{figs/hetero_M_m08/br_bs-mug23.ps}}
{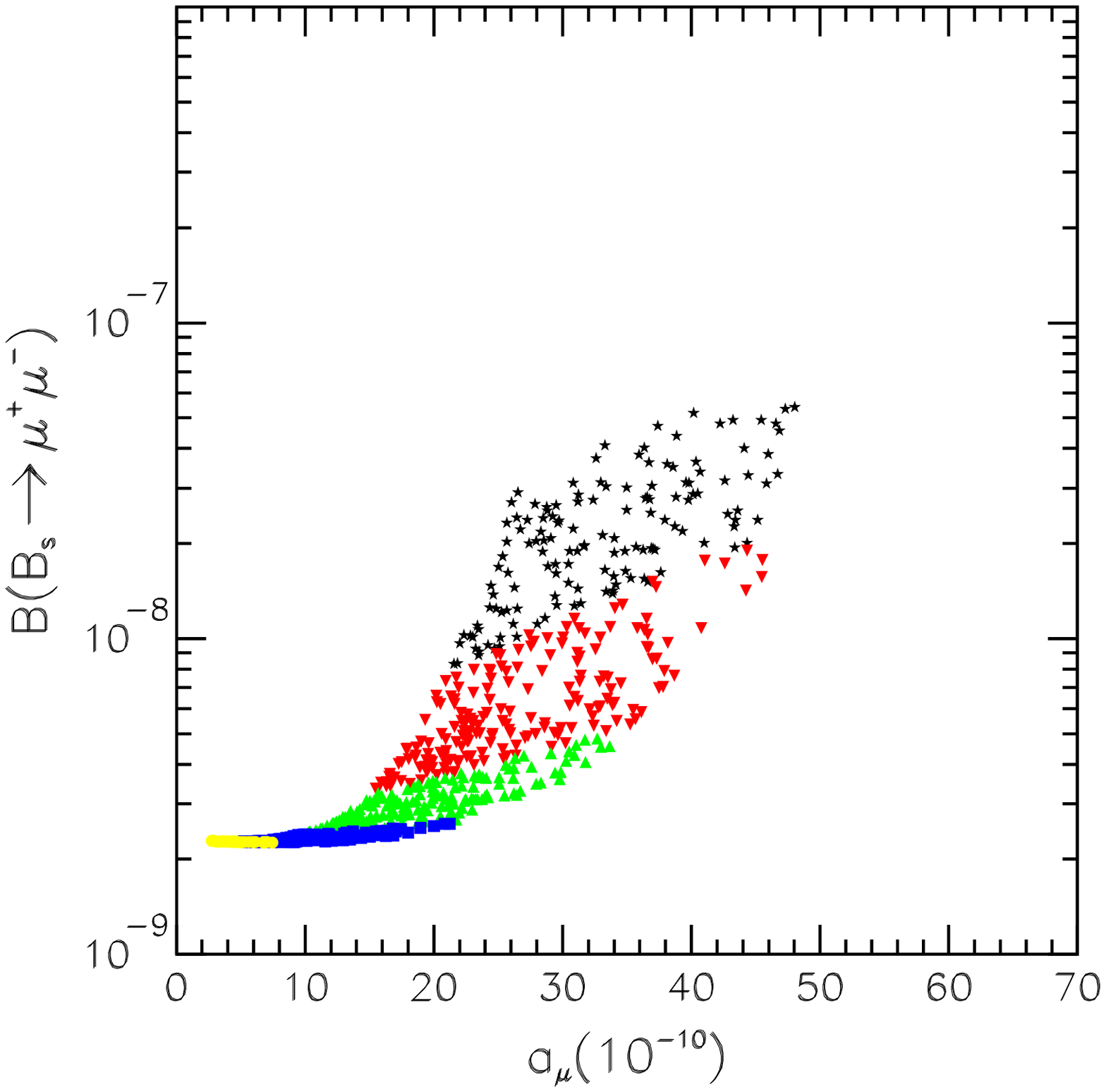}}
\caption{\label{fig:heterom2}
The correlations of the $Br ( B \rightarrow X_s \gamma)$ with
(a) $R_{\mu\mu}$, and (b) $a_\mu^{\rm SUSY}$ for the heterotic
M theory with $\epsilon = -0.8$ and $\theta = 0.15 \pi$. 
The legends are the same as Fig.~\ref{fig:msugra2}.}
%\label{fig:heterom2}
\end{figure}

\begin{figure}
\centering
\hspace{-1.5cm}
\subfigure[]{
\includegraphics[width=0.5\textwidth]%{figs/d-brane/br_bs-mug23.ps}}
{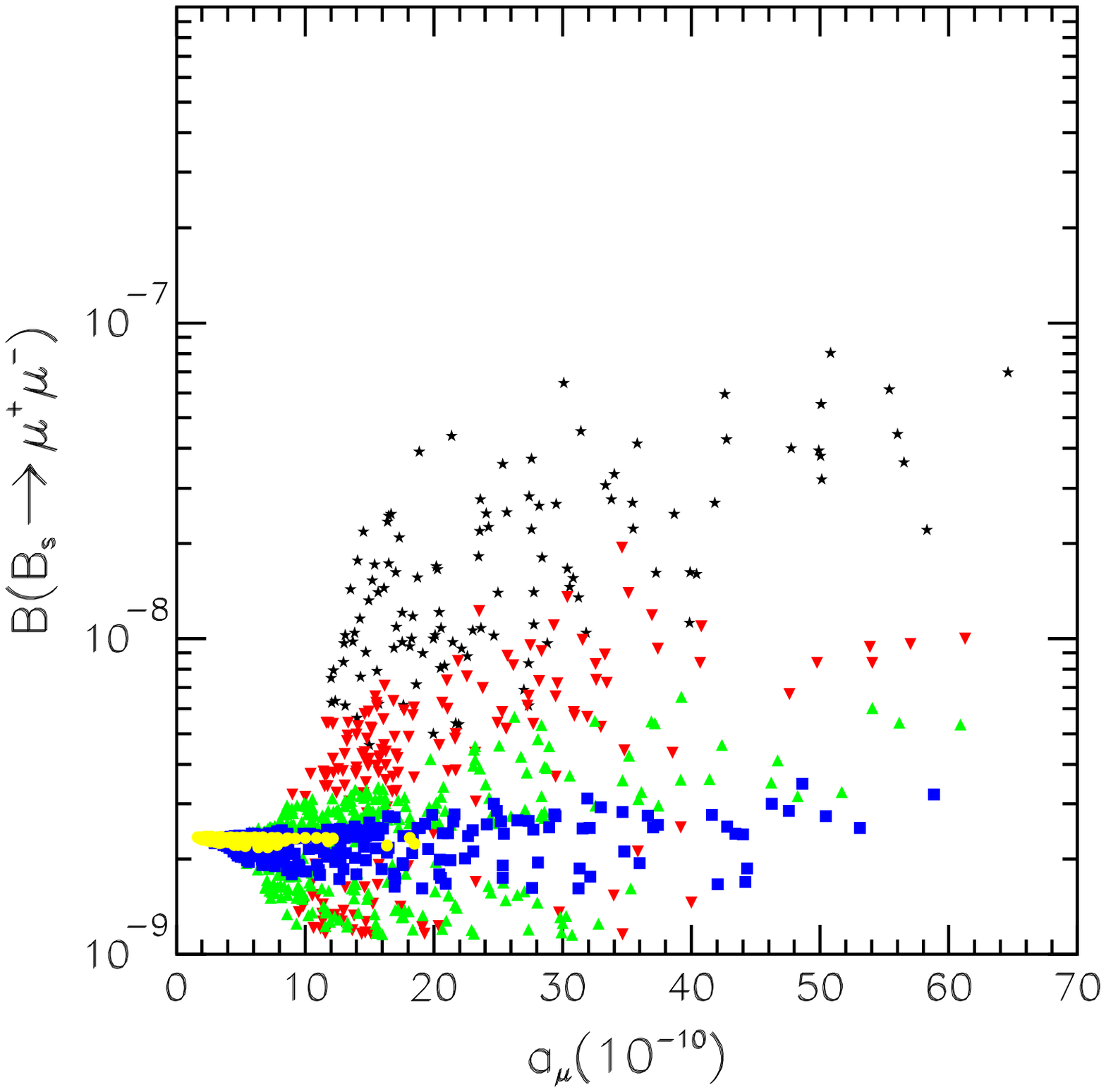}}
\subfigure[]{
\includegraphics[width=0.5\textwidth]%{figs/d-brane/bsll-b2sr.ps}}
{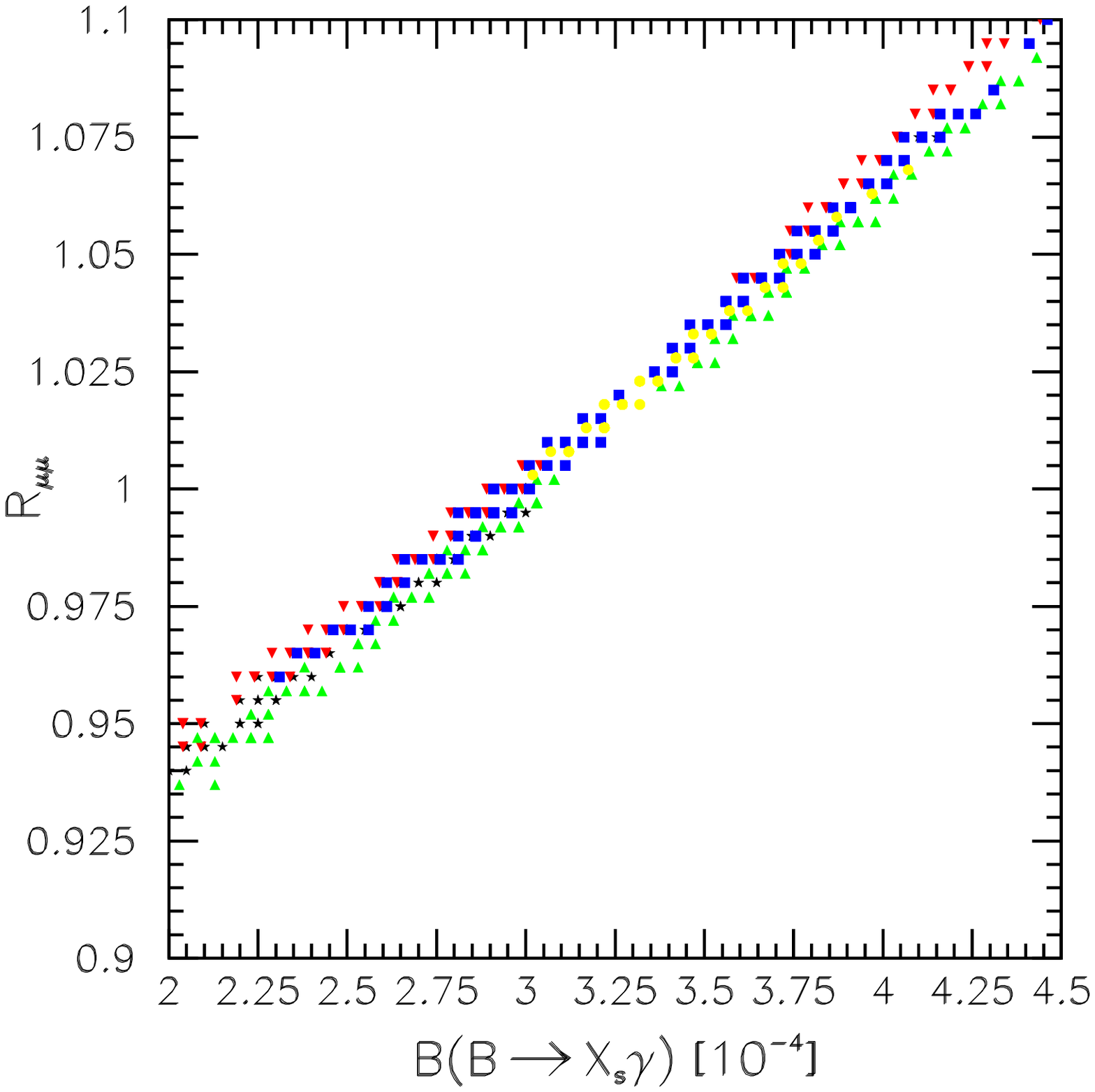}}
%\hspace{-1.5cm}
%\includegraphics[width=0.3\textwidth]%{figs/d-brane/br_bs-mug23.ps}}
%{figs2/d-brane/bsmm-mug2-2.ps}}
\caption{\label{fig:dbrane}
The correlations between
(a) $a_\mu^{\rm SUSY}$ and $B( \bsmm )$, and
(b) $R_{\mu\mu}$ and $B (B \rightarrow X_s \gamma )$ in the $D-$brane model
considered in Sec.~III G. We fix $\Theta_i = 1/\sqrt{3}$ for all $i=1,2,3$,
(the overall modulus limit) and we scan over the following parameter space :
$-\pi/4 \leq \theta \leq 4/ \pi$, $m_{3/2} \leq 300$ GeV, and
$\tan\beta \leq 50$.
}
%\label{fig:dbrane}
\end{figure}

\vfil\eject
\end{document}